\documentclass[preliminary,copyright,creativecommons]{eptcs} 
\usepackage{listings}
\usepackage{xspace}
\usepackage{hyperref}
\def\Coq{\textsc{Coq}\xspace}
\def\Equations{\textsc{Equations}\xspace}

\def\ltmul{$<_{m}$}

\lstdefinestyle{customcaml}{
  belowcaptionskip=1\baselineskip,
  breaklines=true,
  frame=L,
  xleftmargin=\parindent,
  language=C,
  showstringspaces=false,
  basicstyle=\footnotesize\ttfamily,
  keywordstyle=\bfseries\color{green!40!black},
  commentstyle=\itshape\color{purple!40!black},
  identifierstyle=\color{blue},
  stringstyle=\color{orange},
}
\usepackage{utf}
\usepackage[color]{coqdoc}

\def\coqdocbaseindent{1em}

\providecommand{\event}{LFMTP 2015} 
\def\thetitle{Equations for Hereditary Substitution in Leivant's Predicative
  System F: a case study}
\title{\thetitle}
\author{Cyprien Mangin
\institute{Univ Paris Diderot \& \'Ecole Polytechnique\\
Paris, France}
\email{\quad cyprien.mangin@m4x.org}
\and
Matthieu Sozeau
\institute{Inria Paris \& PPS, Univ Paris Diderot\\ Paris, France}
\email{matthieu.sozeau@inria.fr}
}
\def\titlerunning{Equations for Predicative System F}
\def\authorrunning{C. Mangin \& M. Sozeau}
\begin{document}
\maketitle

\begin{abstract}
  This paper presents a case study of formalizing a normalization proof
  for Leivant's Predicative System F \cite{leivant90} using the \Equations
  package. Leivant's Predicative System F is a stratified version of
  System F, where type quantification is annotated with kinds
  representing universe levels. A weaker variant of this system was
  studied by Stump \& Eades \cite{eadespstt10,eades2014semantic},
  employing the hereditary substitution method to show normalization. We
  improve on this result by showing normalization for Leivant's original
  system using hereditary substitutions and a novel multiset ordering on
  types. Our development is done in the \Coq proof assistant using the
  \Equations package, which provides an interface to define
  dependently-typed programs with well-founded recursion and full
  dependent pattern-matching. \Equations allows us to define explicitly
  the hereditary substitution function, clarifying its algorithmic
  behavior in presence of term and type substitutions. From this definition,
  consistency can easily be derived. The algorithmic nature of our
  development is crucial to reflect languages with type quantification,
  enlarging the class of languages on which reflection methods can be
  used in the proof assistant.
\end{abstract}

\section{Introduction}

\Equations \cite{sozeau10} is a toolbox built as a plugin on top of the
\Coq proof assistant for writing dependently-typed programs in
\Coq. Given a high-level specification of a function, using dependent
pattern-matching and complex recursion schemes, its purpose is to
compile it to pure \Coq terms. This compilation scheme builds on the
work of Goguen et al \cite{goguen2006}, which explains dependent
pattern-matching \cite{coquand92baastad} in terms of manipulations of propositional
equalities. In essence, dependent pattern-matching is compiled away
using a reduction-preserving encoding with eliminators for the equality type
between datatypes. In addition to this compilation scheme, \Equations
also automatically derives the resulting equations as propositional
equalities, abstracting entirely from the encoding of pattern-matching found
in the actual compiled definition, and an elimination scheme
corresponding to the graph of the function.  This elimination scheme can
then be used to simplify proofs that directly follow the case-analysis
and recursion behavior of the function without repeating it. \Equations
supports definitions using arbitrarily complex well-founded recursion
schemes, including the nested kind of recursion found in hereditary
substitution functions, and the generation of unfolding lemmas and
elimination schemes for those as well. Additionally, \Equations plays
well with the Program extension of \Coq to manipulate subset types (also
known as refinement types). 

The purpose of this paper is to present a case study of using \Equations
to show the normalization of Predicative System F, in an algorithmic way
such that the normalization function can actually be run inside the
proof assistant. 

Predicative System F was introduced by Leivant \cite{leivant90} to study
the logical strength of different extensions of arithmetic. Using kinds
to represent levels of allowed predicative quantification, he can show
that super-elementary functions can be represented in this system. He
employs a Tait-Girard logical relation proof to argue normalization.

Stump and Eades \cite{eadespstt10} took the system and studied a
normalization proof using hereditary substitution. However, they needed
to define a variant of the system where the kinding rule of universal
quantifications is more restrictive, which makes level $n+1$ not closed
by quantifications over types in level $n$. They claim that the same
derivations can be done in their system but give no proof of such
fact. We remedy this situation by giving a simple normalization proof
still based on hereditary substitution using a novel ordering on types
based on multisets of kinds.

The hereditary substitution function ends up being defined as one would
do in e.g. ML, but combining Program and \Equations, it can be shown to
inhabit a richer type, providing a proof that the function indeed
computes normal forms given inputs in normal form. The pre and
postconditions of the hereditary subsitution function, which will be explicited later,
are actually necessary to justify the termination of this
function. From this it is easy to derive normalization and show that the
system is consistent (relatively to Coq's theory, \emph{with the K
  axiom} currently, although we hope to have an axiom-free version
working by the time of the workshop).

The paper is organized as follows: in \S \ref{sec:equations} we present
a gentle introduction to the \Equations package and its features and
quickly explain the main differences with the original presentation from
\cite{sozeau10}. Then we summarize the standard definitions and
metatheoretical results on Predicative System F that we proved and
highlight the main differences with the presentation of
\cite{eadespstt10}. We provide the \Coq development of our proof
supplemented with some commentary to help follow along. First in \S
\ref{sec:typing-reduction} we present the language definition, with its
typing and reduction rules. Then we show in \S \ref{sec:metatheory} some
metatheoretical properties on this language, such as substitution lemmas
and regularity. Section \ref{sec:normalization} is dedicated to
showing strong normalization of the calculus, which includes defining a
well-founded ordering to justify the termination of the hereditary
substitution function, and its definition itself using the \Equations
package. We also provide as an appendix the code which is produced from
this definition by \Coq's extraction mechanism. Finally, we compare
with related work and conclude in \S \ref{sec:relat-work-concl}.

\section{Equations}
\label{sec:equations}
\begin{coqdoccode}
\coqdocemptyline
\end{coqdoccode}
\Equations allows one to write recursive functions by specifying a
    list of clauses with a pattern on the left and a term on the right,
    \`a la Agda \cite{norell:thesis} and Epigram
    \cite{epigram:pratprog}. Here is an example recursive definition on
    lists, where wildcards corresponds to arbitrary fresh variables in patterns: \begin{coqdoccode}
\coqdocemptyline
\coqdocnoindent
\coqdockw{Equations} \coqdef{Equations.length}{length}{\coqdocdefinition{length}} \{\coqdocvar{A}\} (\coqdocvar{l} : \coqexternalref{list}{http://coq.inria.fr/distrib/8.5beta2/stdlib/Coq.Init.Datatypes}{\coqdocinductive{list}} \coqdocvariable{A}) : \coqexternalref{nat}{http://coq.inria.fr/distrib/8.5beta2/stdlib/Coq.Init.Datatypes}{\coqdocinductive{nat}} :=\coqdoceol
\coqdocnoindent
\coqref{Equations.length}{\coqdocdefinition{length}} \coqdocvar{\_} \coqexternalref{ListNotations.:list scope:'[' ']'}{http://coq.inria.fr/distrib/8.5beta2/stdlib/Coq.Lists.List}{\coqdocnotation{[]}} \ensuremath{\Rightarrow} 0;\coqdoceol
\coqdocnoindent
\coqref{Equations.length}{\coqdocdefinition{length}} \coqdocvar{\_} (\coqexternalref{cons}{http://coq.inria.fr/distrib/8.5beta2/stdlib/Coq.Init.Datatypes}{\coqdocconstructor{cons}} \coqexternalref{cons}{http://coq.inria.fr/distrib/8.5beta2/stdlib/Coq.Init.Datatypes}{\coqdocconstructor{\_}} \coqexternalref{cons}{http://coq.inria.fr/distrib/8.5beta2/stdlib/Coq.Init.Datatypes}{\coqdocconstructor{t}}) \ensuremath{\Rightarrow} \coqexternalref{S}{http://coq.inria.fr/distrib/8.5beta2/stdlib/Coq.Init.Datatypes}{\coqdocconstructor{S}} (\coqref{Equations.length}{\coqdocdefinition{length}} \coqdocvar{t}).\coqdoceol
\coqdocemptyline
\end{coqdoccode}
The package starts by building a splitting tree for the definition
  and then compiles it to a pure \Coq term. From the splitting tree,
  it also derives the equations as propositional equalities, which can
  be more robust to use than reduction when writing proofs about the
  constant, although in this particular case the compiled definition is
  the same as the one from the standard library. Here we have two
  leaves in the computation tree hence two equations: \begin{coqdoccode}
\coqdocemptyline
\coqdocnoindent
\coqdockw{Check} \coqref{Equations.length equation 1}{\coqdocdefinition{length\_equation\_1}} : \coqexternalref{:type scope:'xE2x88x80' x '..' x ',' x}{http://coq.inria.fr/distrib/8.5beta2/stdlib/Coq.Unicode.Utf8\_core}{\coqdocnotation{∀}} \coqdocvar{A} : \coqdockw{Type}\coqexternalref{:type scope:'xE2x88x80' x '..' x ',' x}{http://coq.inria.fr/distrib/8.5beta2/stdlib/Coq.Unicode.Utf8\_core}{\coqdocnotation{,}} \coqref{Equations.length}{\coqdocdefinition{length}} \coqexternalref{ListNotations.:list scope:'[' ']'}{http://coq.inria.fr/distrib/8.5beta2/stdlib/Coq.Lists.List}{\coqdocnotation{[]}} \coqexternalref{:type scope:x '=' x}{http://coq.inria.fr/distrib/8.5beta2/stdlib/Coq.Init.Logic}{\coqdocnotation{=}} 0.\coqdoceol
\coqdocnoindent
\coqdockw{Check} \coqref{Equations.length equation 2}{\coqdocdefinition{length\_equation\_2}} : \coqexternalref{:type scope:'xE2x88x80' x '..' x ',' x}{http://coq.inria.fr/distrib/8.5beta2/stdlib/Coq.Unicode.Utf8\_core}{\coqdocnotation{∀}} \coqexternalref{:type scope:'xE2x88x80' x '..' x ',' x}{http://coq.inria.fr/distrib/8.5beta2/stdlib/Coq.Unicode.Utf8\_core}{\coqdocnotation{(}}\coqdocvar{A} : \coqdockw{Type}) (\coqdocvar{a} : \coqdocvariable{A}) (\coqdocvar{l} : \coqexternalref{list}{http://coq.inria.fr/distrib/8.5beta2/stdlib/Coq.Init.Datatypes}{\coqdocinductive{list}} \coqdocvariable{A}\coqexternalref{:type scope:'xE2x88x80' x '..' x ',' x}{http://coq.inria.fr/distrib/8.5beta2/stdlib/Coq.Unicode.Utf8\_core}{\coqdocnotation{),}} \coqref{Equations.length}{\coqdocdefinition{length}} (\coqexternalref{cons}{http://coq.inria.fr/distrib/8.5beta2/stdlib/Coq.Init.Datatypes}{\coqdocconstructor{cons}} \coqdocvariable{a} \coqdocvariable{l}) \coqexternalref{:type scope:x '=' x}{http://coq.inria.fr/distrib/8.5beta2/stdlib/Coq.Init.Logic}{\coqdocnotation{=}} \coqexternalref{S}{http://coq.inria.fr/distrib/8.5beta2/stdlib/Coq.Init.Datatypes}{\coqdocconstructor{S}} (\coqref{Equations.length}{\coqdocdefinition{length}} \coqdocvariable{l}).\coqdoceol
\coqdocemptyline
\end{coqdoccode}
These two equations are automatically added to a rewrite hint
  database named \coqref{Equations.length}{\coqdocdefinition{length}} and can be used during proofs using the \coqdocvar{simp}
  \coqref{Equations.length}{\coqdocdefinition{length}} tactic.  In addition, an elimination principle for \coqref{Equations.length}{\coqdocdefinition{length}} is
  derived. Note that \coqref{Equations.length comp}{\coqdocdefinition{length\_comp}} is just a definition of the return type of \coqref{Equations.length}{\coqdocdefinition{length}} in
  terms of its arguments, i.e. it is \coqdocvar{\ensuremath{\lambda}} \coqdocvariable{A} \coqdocvariable{l}, \coqexternalref{nat}{http://coq.inria.fr/distrib/8.5beta2/stdlib/Coq.Init.Datatypes}{\coqdocinductive{nat}} here: \begin{coqdoccode}
\coqdocemptyline
\coqdocnoindent
\coqdockw{Check} \coqref{Equations.length elim}{\coqdocdefinition{length\_elim}} : \coqexternalref{:type scope:'xE2x88x80' x '..' x ',' x}{http://coq.inria.fr/distrib/8.5beta2/stdlib/Coq.Unicode.Utf8\_core}{\coqdocnotation{∀}} \coqdocvar{P} : \coqexternalref{:type scope:'xE2x88x80' x '..' x ',' x}{http://coq.inria.fr/distrib/8.5beta2/stdlib/Coq.Unicode.Utf8\_core}{\coqdocnotation{∀}} \coqexternalref{:type scope:'xE2x88x80' x '..' x ',' x}{http://coq.inria.fr/distrib/8.5beta2/stdlib/Coq.Unicode.Utf8\_core}{\coqdocnotation{(}}\coqdocvar{A} : \coqdockw{Type}) (\coqdocvar{l} : \coqexternalref{list}{http://coq.inria.fr/distrib/8.5beta2/stdlib/Coq.Init.Datatypes}{\coqdocinductive{list}} \coqdocvariable{A}\coqexternalref{:type scope:'xE2x88x80' x '..' x ',' x}{http://coq.inria.fr/distrib/8.5beta2/stdlib/Coq.Unicode.Utf8\_core}{\coqdocnotation{),}} \coqref{Equations.length comp}{\coqdocdefinition{length\_comp}} \coqdocvariable{l} \coqexternalref{:type scope:x 'xE2x86x92' x}{http://coq.inria.fr/distrib/8.5beta2/stdlib/Coq.Unicode.Utf8\_core}{\coqdocnotation{→}} \coqdockw{Prop}\coqexternalref{:type scope:'xE2x88x80' x '..' x ',' x}{http://coq.inria.fr/distrib/8.5beta2/stdlib/Coq.Unicode.Utf8\_core}{\coqdocnotation{,}}\coqdoceol
\coqdocindent{2.00em}
\coqexternalref{:type scope:x 'xE2x86x92' x}{http://coq.inria.fr/distrib/8.5beta2/stdlib/Coq.Unicode.Utf8\_core}{\coqdocnotation{(}}\coqexternalref{:type scope:'xE2x88x80' x '..' x ',' x}{http://coq.inria.fr/distrib/8.5beta2/stdlib/Coq.Unicode.Utf8\_core}{\coqdocnotation{∀}} \coqdocvar{A} : \coqdockw{Type}\coqexternalref{:type scope:'xE2x88x80' x '..' x ',' x}{http://coq.inria.fr/distrib/8.5beta2/stdlib/Coq.Unicode.Utf8\_core}{\coqdocnotation{,}} \coqdocvariable{P} \coqdocvariable{A} \coqexternalref{ListNotations.:list scope:'[' ']'}{http://coq.inria.fr/distrib/8.5beta2/stdlib/Coq.Lists.List}{\coqdocnotation{[]}} 0\coqexternalref{:type scope:x 'xE2x86x92' x}{http://coq.inria.fr/distrib/8.5beta2/stdlib/Coq.Unicode.Utf8\_core}{\coqdocnotation{)}} \coqexternalref{:type scope:x 'xE2x86x92' x}{http://coq.inria.fr/distrib/8.5beta2/stdlib/Coq.Unicode.Utf8\_core}{\coqdocnotation{→}} \coqdoceol
\coqdocindent{2.00em}
\coqexternalref{:type scope:x 'xE2x86x92' x}{http://coq.inria.fr/distrib/8.5beta2/stdlib/Coq.Unicode.Utf8\_core}{\coqdocnotation{(}}\coqexternalref{:type scope:'xE2x88x80' x '..' x ',' x}{http://coq.inria.fr/distrib/8.5beta2/stdlib/Coq.Unicode.Utf8\_core}{\coqdocnotation{∀}} \coqexternalref{:type scope:'xE2x88x80' x '..' x ',' x}{http://coq.inria.fr/distrib/8.5beta2/stdlib/Coq.Unicode.Utf8\_core}{\coqdocnotation{(}}\coqdocvar{A} : \coqdockw{Type}) (\coqdocvar{a} : \coqdocvariable{A}) (\coqdocvar{l} : \coqexternalref{list}{http://coq.inria.fr/distrib/8.5beta2/stdlib/Coq.Init.Datatypes}{\coqdocinductive{list}} \coqdocvariable{A}\coqexternalref{:type scope:'xE2x88x80' x '..' x ',' x}{http://coq.inria.fr/distrib/8.5beta2/stdlib/Coq.Unicode.Utf8\_core}{\coqdocnotation{),}} \coqdocvariable{P} \coqdocvariable{A} \coqdocvariable{l} (\coqref{Equations.length}{\coqdocdefinition{length}} \coqdocvariable{l}) \coqexternalref{:type scope:x 'xE2x86x92' x}{http://coq.inria.fr/distrib/8.5beta2/stdlib/Coq.Unicode.Utf8\_core}{\coqdocnotation{→}} \coqdocvariable{P} \coqdocvariable{A} (\coqexternalref{cons}{http://coq.inria.fr/distrib/8.5beta2/stdlib/Coq.Init.Datatypes}{\coqdocconstructor{cons}} \coqdocvariable{a} \coqdocvariable{l}) (\coqexternalref{S}{http://coq.inria.fr/distrib/8.5beta2/stdlib/Coq.Init.Datatypes}{\coqdocconstructor{S}} (\coqref{Equations.length}{\coqdocdefinition{length}} \coqdocvariable{l}))\coqexternalref{:type scope:x 'xE2x86x92' x}{http://coq.inria.fr/distrib/8.5beta2/stdlib/Coq.Unicode.Utf8\_core}{\coqdocnotation{)}} \coqexternalref{:type scope:x 'xE2x86x92' x}{http://coq.inria.fr/distrib/8.5beta2/stdlib/Coq.Unicode.Utf8\_core}{\coqdocnotation{→}} \coqdoceol
\coqdocindent{2.00em}
\coqexternalref{:type scope:'xE2x88x80' x '..' x ',' x}{http://coq.inria.fr/distrib/8.5beta2/stdlib/Coq.Unicode.Utf8\_core}{\coqdocnotation{∀}} \coqexternalref{:type scope:'xE2x88x80' x '..' x ',' x}{http://coq.inria.fr/distrib/8.5beta2/stdlib/Coq.Unicode.Utf8\_core}{\coqdocnotation{(}}\coqdocvar{A} : \coqdockw{Type}) (\coqdocvar{l} : \coqexternalref{list}{http://coq.inria.fr/distrib/8.5beta2/stdlib/Coq.Init.Datatypes}{\coqdocinductive{list}} \coqdocvariable{A}\coqexternalref{:type scope:'xE2x88x80' x '..' x ',' x}{http://coq.inria.fr/distrib/8.5beta2/stdlib/Coq.Unicode.Utf8\_core}{\coqdocnotation{),}} \coqdocvariable{P} \coqdocvariable{A} \coqdocvariable{l} (\coqref{Equations.length}{\coqdocdefinition{length}} \coqdocvariable{l}).\coqdoceol
\coqdocemptyline
\coqdocemptyline
\end{coqdoccode}
This elimination principle can be used in proofs to eliminate
  \textit{calls} to \coqref{Equations.length}{\coqdocdefinition{length}} and refine at the same time the arguments and
  results of the call in the goal. For example, to prove the following
  lemma, one can apply the functional elimination principle using the
  \coqdocvar{funelim} tactic to eliminate the \coqref{Equations.length}{\coqdocdefinition{length}} \coqdocvariable{l} call: \begin{coqdoccode}
\coqdocemptyline
\coqdocnoindent
\coqdockw{Lemma} \coqdef{Equations.length rev}{length\_rev}{\coqdoclemma{length\_rev}} \{\coqdocvar{A}\} (\coqdocvar{l} : \coqexternalref{list}{http://coq.inria.fr/distrib/8.5beta2/stdlib/Coq.Init.Datatypes}{\coqdocinductive{list}} \coqdocvariable{A}) : \coqref{Equations.length}{\coqdocdefinition{length}} (\coqexternalref{rev}{http://coq.inria.fr/distrib/8.5beta2/stdlib/Coq.Lists.List}{\coqdocdefinition{rev}} \coqdocvariable{l}) \coqexternalref{:type scope:x '=' x}{http://coq.inria.fr/distrib/8.5beta2/stdlib/Coq.Init.Logic}{\coqdocnotation{=}} \coqref{Equations.length}{\coqdocdefinition{length}} \coqdocvariable{l}.\coqdoceol
\coqdocnoindent
\coqdockw{Proof}.\coqdoceol
\coqdocindent{1.00em}
\coqdocvar{funelim} (\coqref{Equations.length}{\coqdocdefinition{length}} \coqdocvar{l}).\coqdoceol
\coqdocemptyline
\end{coqdoccode}
We get two subgoals, easily solved by simplification and arithmetic.
\coqdoceol
\coqdocemptyline
\coqdocnoindent
\coqdoceol
\coqdocindent{1.00em}
\coqdocvariable{A} : \coqdockw{Type}\coqdoceol
\coqdocindent{1.00em}
============================\coqdoceol
\coqdocindent{1.50em}
\coqref{Equations.length}{\coqdocdefinition{length}} (\coqexternalref{rev}{http://coq.inria.fr/distrib/8.5beta2/stdlib/Coq.Lists.List}{\coqdocdefinition{rev}} []) = 0\coqdoceol
\coqdocnoindent
\coqdoceol
\coqdocindent{1.00em}
\coqdocvariable{A} : \coqdockw{Type}\coqdoceol
\coqdocindent{1.00em}
\coqdocvariable{a} : \coqdocvariable{A}\coqdoceol
\coqdocindent{1.00em}
\coqdocvariable{l} : \coqexternalref{list}{http://coq.inria.fr/distrib/8.5beta2/stdlib/Coq.Init.Datatypes}{\coqdocinductive{list}} \coqdocvariable{A}\coqdoceol
\coqdocindent{1.00em}
\coqdocvar{H} : \coqref{Equations.length}{\coqdocdefinition{length}} (\coqexternalref{rev}{http://coq.inria.fr/distrib/8.5beta2/stdlib/Coq.Lists.List}{\coqdocdefinition{rev}} \coqdocvariable{l}) = \coqref{Equations.length}{\coqdocdefinition{length}} \coqdocvariable{l}\coqdoceol
\coqdocindent{1.00em}
============================\coqdoceol
\coqdocindent{1.50em}
\coqref{Equations.length}{\coqdocdefinition{length}} (\coqexternalref{rev}{http://coq.inria.fr/distrib/8.5beta2/stdlib/Coq.Lists.List}{\coqdocdefinition{rev}} (\coqexternalref{cons}{http://coq.inria.fr/distrib/8.5beta2/stdlib/Coq.Init.Datatypes}{\coqdocconstructor{cons}} \coqdocvariable{a} \coqdocvariable{l})) = \coqexternalref{S}{http://coq.inria.fr/distrib/8.5beta2/stdlib/Coq.Init.Datatypes}{\coqdocconstructor{S}} (\coqref{Equations.length}{\coqdocdefinition{length}} \coqdocvariable{l})

\coqdocemptyline
\begin{coqdoccode}
\coqdocemptyline
\coqdocindent{1.00em}
- \coqdocvar{simp} \coqdocvar{length}.\coqdoceol
\coqdocindent{1.00em}
- \coqdoctac{simpl}; \coqdoctac{rewrite} \coqref{Equations.length app}{\coqdoclemma{length\_app}}, \coqdocvar{H}; \coqdocvar{simp} \coqdocvar{length}; \coqdoctac{omega}.\coqdoceol
\coqdocnoindent
\coqdockw{Qed}.\coqdoceol
\coqdocemptyline
\end{coqdoccode}
\subsection{Dependent Pattern-Matching}

  \Equations handles not only simple pattern-matching on inductive
  types, but also dependent pattern-matching on inductive \textit{families}.
  With respect to the standard \Coq match construct, it eases the
  definition of complex pattern-matchings by compiling in the proof term
  all the inversion and unification steps that must be witnessed.
  Here is an example with the \coqref{Equations.le}{\coqdocinductive{le}} relation on natural numbers.
 \begin{coqdoccode}
\coqdocemptyline
\coqdocemptyline
\coqdocnoindent
\coqdockw{Inductive} \coqdef{Equations.le}{le}{\coqdocinductive{le}} : \coqexternalref{nat}{http://coq.inria.fr/distrib/8.5beta2/stdlib/Coq.Init.Datatypes}{\coqdocinductive{nat}} \coqexternalref{:type scope:x '->' x}{http://coq.inria.fr/distrib/8.5beta2/stdlib/Coq.Init.Logic}{\coqdocnotation{\ensuremath{\rightarrow}}} \coqexternalref{nat}{http://coq.inria.fr/distrib/8.5beta2/stdlib/Coq.Init.Datatypes}{\coqdocinductive{nat}} \coqexternalref{:type scope:x '->' x}{http://coq.inria.fr/distrib/8.5beta2/stdlib/Coq.Init.Logic}{\coqdocnotation{\ensuremath{\rightarrow}}} \coqdockw{Set} :=\coqdoceol
\coqdocnoindent
\ensuremath{|} \coqdef{Equations.lz}{lz}{\coqdocconstructor{lz}} : \coqdockw{\ensuremath{\forall}} \{\coqdocvar{n}\}, 0 \coqref{Equations.::x '<=' x}{\coqdocnotation{$\le$}} \coqdocvariable{n}\coqdoceol
\coqdocnoindent
\ensuremath{|} \coqdef{Equations.ls}{ls}{\coqdocconstructor{ls}} : \coqdockw{\ensuremath{\forall}} \{\coqdocvar{m} \coqdocvar{n}\}, \coqdocvariable{m} \coqref{Equations.::x '<=' x}{\coqdocnotation{$\le$}} \coqdocvariable{n} \coqexternalref{:type scope:x '->' x}{http://coq.inria.fr/distrib/8.5beta2/stdlib/Coq.Init.Logic}{\coqdocnotation{\ensuremath{\rightarrow}}} \coqref{Equations.::x '<=' x}{\coqdocnotation{(}}\coqexternalref{S}{http://coq.inria.fr/distrib/8.5beta2/stdlib/Coq.Init.Datatypes}{\coqdocconstructor{S}} \coqdocvariable{m}\coqref{Equations.::x '<=' x}{\coqdocnotation{)}} \coqref{Equations.::x '<=' x}{\coqdocnotation{$\le$}} \coqref{Equations.::x '<=' x}{\coqdocnotation{(}}\coqexternalref{S}{http://coq.inria.fr/distrib/8.5beta2/stdlib/Coq.Init.Datatypes}{\coqdocconstructor{S}} \coqdocvariable{n}\coqref{Equations.::x '<=' x}{\coqdocnotation{)}} \coqdockw{where} \coqdef{Equations.::x '<=' x}{"}{"}m \coqdocnotation{\ensuremath{\le}} n" := (\coqref{Equations.le}{\coqdocinductive{le}} \coqdocvar{m} \coqdocvar{n}).\coqdoceol
\coqdocemptyline
\coqdocemptyline
\end{coqdoccode}
Proving antisymmetry of this relation requires only two cases, 
    because pattern matching on the first argument determines the endpoints
    of the second argument: \begin{coqdoccode}
\coqdocemptyline
\coqdocnoindent
\coqdockw{Equations} \coqdef{Equations.antisym}{antisym}{\coqdocdefinition{antisym}} \{\coqdocvar{m} \coqdocvar{n} : \coqexternalref{nat}{http://coq.inria.fr/distrib/8.5beta2/stdlib/Coq.Init.Datatypes}{\coqdocinductive{nat}}\} (\coqdocvar{x} : \coqdocvariable{m} \coqref{Equations.::x '<=' x}{\coqdocnotation{$\le$}} \coqdocvariable{n}) (\coqdocvar{y} : \coqdocvariable{n} \coqref{Equations.::x '<=' x}{\coqdocnotation{$\le$}} \coqdocvariable{m}) : \coqdocvar{m} \coqexternalref{:type scope:x '=' x}{http://coq.inria.fr/distrib/8.5beta2/stdlib/Coq.Init.Logic}{\coqdocnotation{=}} \coqdocvar{n} :=\coqdoceol
\coqdocnoindent
\coqref{Equations.antisym}{\coqdocdefinition{antisym}} \coqdocvar{\_} \coqdocvar{\_} \coqdocvar{x} \coqdocvar{y} \coqdoctac{by} \coqdocvar{rec} \coqdocvar{x} \ensuremath{\Rightarrow}\coqdoceol
\coqdocnoindent
\coqref{Equations.antisym}{\coqdocdefinition{antisym}} \coqdocvar{\_} \coqdocvar{\_} \coqref{Equations.lz}{\coqdocconstructor{lz}} \coqref{Equations.lz}{\coqdocconstructor{lz}} \ensuremath{\Rightarrow} \coqexternalref{eq refl}{http://coq.inria.fr/distrib/8.5beta2/stdlib/Coq.Init.Logic}{\coqdocconstructor{eq\_refl}};\coqdoceol
\coqdocnoindent
\coqref{Equations.antisym}{\coqdocdefinition{antisym}} \coqdocvar{\_} \coqdocvar{\_} (\coqref{Equations.ls}{\coqdocconstructor{ls}} \coqref{Equations.ls}{\coqdocconstructor{x}}) (\coqref{Equations.ls}{\coqdocconstructor{ls}} \coqref{Equations.ls}{\coqdocconstructor{y}}) \ensuremath{\Rightarrow} \coqexternalref{f equal}{http://coq.inria.fr/distrib/8.5beta2/stdlib/Coq.Init.Logic}{\coqdoclemma{f\_equal}} \coqexternalref{S}{http://coq.inria.fr/distrib/8.5beta2/stdlib/Coq.Init.Datatypes}{\coqdocconstructor{S}} (\coqref{Equations.antisym comp proj}{\coqdocdefinition{antisym}} \coqdocvar{x} \coqdocvar{y}).\coqdoceol
\coqdocemptyline
\end{coqdoccode}
More precisely, in the \coqdocvariable{x} = \coqref{Equations.lz}{\coqdocconstructor{lz}} case, it is possible during the
    translation to deduce that \coqdocvariable{m} must be 0, which implies that \coqdocvariable{y}
    cannot be some application of \coqref{Equations.ls}{\coqdocconstructor{ls}}. These deductions are done
    automatically by \Equations, which allows to reduce this proof to its
    simplest form.  Writing it explicitely in pure \Coq would be
    actually annoying and require explicit mention of impossible cases and
    surgical rewritings with equalities. 

\subsection{Recursion}

  Note that we use a clause \coqdoctac{by} \coqdocvar{rec} \coqdocvariable{x} \ensuremath{\Rightarrow} here in addition to the
  pattern-matching.  This is a different kind of right-hand-side, that
  allows to specify the recursion scheme of the function. We are using
  well-founded recursion on the \coqdocvariable{m} $\le$ \coqdocvariable{n} hypothesis here. The implicit
  ordering used is actually automatically derived using a \coqdocvar{Derive}
  \coqdocvar{Subterm} \coqdockw{for} \coqref{Equations.le}{\coqdocinductive{le}} command, and corresponds to the transitive closure of
  the direct subterm relation, i.e. the deep structural recursion
  ordering.

  The compiled definition cannot be checked using the built-in
  structural guardness check of \Coq, because the equality
  manipulations appearing in the term go outside of the subset of
  recursion schemes recognized by it.  It would have to handle
  commutative cuts and specific constructs on the equality type. Also,
  the syntactic check can be very slow on medium-sized terms.

  The solution here, using the logic to justify the recursive calls,
  means that we are freed from any syntactic restriction, and any
  logical justification for termination is allowed. At each recursive
  call, we must simply provide a proof that the given argument is
  strictly smaller than the initial one in the subterm relation.  An
  automatic proof search using the constructors of the subterm relation
  for \coqref{Equations.le}{\coqdocinductive{le}} solves these subgoals for us here, otherwise they are given
  as obligations for the user to prove.

  As in the case of \coqref{Equations.length}{\coqdocdefinition{length}}, we provide equations and an elimination
  principle for the definition. In case well-founded recursion is used, we
  first prove an unfolding lemma for the definition which allows us to
  remove any reasoning on the termination conditions after the definition.
  The equations are as expected:
\begin{coqdoccode}
\coqdocemptyline
\coqdocnoindent
\coqdockw{Check} \coqref{Equations.antisym equation 1}{\coqdocdefinition{antisym\_equation\_1}} : \coqref{Equations.antisym}{\coqdocdefinition{antisym}} \coqref{Equations.lz}{\coqdocconstructor{lz}} \coqref{Equations.lz}{\coqdocconstructor{lz}} \coqexternalref{:type scope:x '=' x}{http://coq.inria.fr/distrib/8.5beta2/stdlib/Coq.Init.Logic}{\coqdocnotation{=}} \coqexternalref{eq refl}{http://coq.inria.fr/distrib/8.5beta2/stdlib/Coq.Init.Logic}{\coqdocconstructor{eq\_refl}}.\coqdoceol
\coqdocnoindent
\coqdockw{Check} \coqref{Equations.antisym equation 2}{\coqdocdefinition{antisym\_equation\_2}} : \coqexternalref{:type scope:'xE2x88x80' x '..' x ',' x}{http://coq.inria.fr/distrib/8.5beta2/stdlib/Coq.Unicode.Utf8\_core}{\coqdocnotation{∀}} \coqexternalref{:type scope:'xE2x88x80' x '..' x ',' x}{http://coq.inria.fr/distrib/8.5beta2/stdlib/Coq.Unicode.Utf8\_core}{\coqdocnotation{(}}\coqdocvar{n1} \coqdocvar{m0} : \coqexternalref{nat}{http://coq.inria.fr/distrib/8.5beta2/stdlib/Coq.Init.Datatypes}{\coqdocinductive{nat}}) (\coqdocvar{l} : \coqdocvariable{n1} \coqref{Equations.::x '<=' x}{\coqdocnotation{$\le$}} \coqdocvariable{m0}) (\coqdocvar{l0} : \coqdocvariable{m0} \coqref{Equations.::x '<=' x}{\coqdocnotation{$\le$}} \coqdocvariable{n1}\coqexternalref{:type scope:'xE2x88x80' x '..' x ',' x}{http://coq.inria.fr/distrib/8.5beta2/stdlib/Coq.Unicode.Utf8\_core}{\coqdocnotation{),}}\coqdoceol
\coqdocindent{2.00em}
\coqref{Equations.antisym}{\coqdocdefinition{antisym}} (\coqref{Equations.ls}{\coqdocconstructor{ls}} \coqdocvariable{l}) (\coqref{Equations.ls}{\coqdocconstructor{ls}} \coqdocvariable{l0}) \coqexternalref{:type scope:x '=' x}{http://coq.inria.fr/distrib/8.5beta2/stdlib/Coq.Init.Logic}{\coqdocnotation{=}} \coqexternalref{f equal}{http://coq.inria.fr/distrib/8.5beta2/stdlib/Coq.Init.Logic}{\coqdoclemma{f\_equal}} \coqexternalref{S}{http://coq.inria.fr/distrib/8.5beta2/stdlib/Coq.Init.Datatypes}{\coqdocconstructor{S}} (\coqref{Equations.antisym}{\coqdocdefinition{antisym}} \coqdocvariable{l} \coqdocvariable{l0}).\coqdoceol
\coqdocemptyline
\end{coqdoccode}
And the elimination principle, with the correct inductive hypothesis in the 
   recursive case: \begin{coqdoccode}
\coqdocemptyline
\coqdocnoindent
\coqdockw{Check} \coqref{Equations.antisym elim}{\coqdocdefinition{antisym\_elim}} : \coqexternalref{:type scope:'xE2x88x80' x '..' x ',' x}{http://coq.inria.fr/distrib/8.5beta2/stdlib/Coq.Unicode.Utf8\_core}{\coqdocnotation{∀}} \coqdocvar{P} : \coqexternalref{:type scope:'xE2x88x80' x '..' x ',' x}{http://coq.inria.fr/distrib/8.5beta2/stdlib/Coq.Unicode.Utf8\_core}{\coqdocnotation{∀}} \coqexternalref{:type scope:'xE2x88x80' x '..' x ',' x}{http://coq.inria.fr/distrib/8.5beta2/stdlib/Coq.Unicode.Utf8\_core}{\coqdocnotation{(}}\coqdocvar{m} \coqdocvar{n} : \coqexternalref{nat}{http://coq.inria.fr/distrib/8.5beta2/stdlib/Coq.Init.Datatypes}{\coqdocinductive{nat}}) (\coqdocvar{x} : \coqdocvariable{m} \coqref{Equations.::x '<=' x}{\coqdocnotation{$\le$}} \coqdocvariable{n}) (\coqdocvar{y} : \coqdocvariable{n} \coqref{Equations.::x '<=' x}{\coqdocnotation{$\le$}} \coqdocvariable{m}\coqexternalref{:type scope:'xE2x88x80' x '..' x ',' x}{http://coq.inria.fr/distrib/8.5beta2/stdlib/Coq.Unicode.Utf8\_core}{\coqdocnotation{),}} \coqref{Equations.antisym comp}{\coqdocdefinition{antisym\_comp}} \coqdocvariable{x} \coqdocvariable{y} \coqexternalref{:type scope:x 'xE2x86x92' x}{http://coq.inria.fr/distrib/8.5beta2/stdlib/Coq.Unicode.Utf8\_core}{\coqdocnotation{→}} \coqdockw{Prop}\coqexternalref{:type scope:'xE2x88x80' x '..' x ',' x}{http://coq.inria.fr/distrib/8.5beta2/stdlib/Coq.Unicode.Utf8\_core}{\coqdocnotation{,}}\coqdoceol
\coqdocindent{2.00em}
\coqdocvariable{P} 0 0 \coqref{Equations.lz}{\coqdocconstructor{lz}} \coqref{Equations.lz}{\coqdocconstructor{lz}} \coqexternalref{eq refl}{http://coq.inria.fr/distrib/8.5beta2/stdlib/Coq.Init.Logic}{\coqdocconstructor{eq\_refl}}\coqdoceol
\coqdocindent{2.00em}
\coqexternalref{:type scope:x 'xE2x86x92' x}{http://coq.inria.fr/distrib/8.5beta2/stdlib/Coq.Unicode.Utf8\_core}{\coqdocnotation{→}} \coqexternalref{:type scope:x 'xE2x86x92' x}{http://coq.inria.fr/distrib/8.5beta2/stdlib/Coq.Unicode.Utf8\_core}{\coqdocnotation{(}}\coqexternalref{:type scope:'xE2x88x80' x '..' x ',' x}{http://coq.inria.fr/distrib/8.5beta2/stdlib/Coq.Unicode.Utf8\_core}{\coqdocnotation{∀}} \coqexternalref{:type scope:'xE2x88x80' x '..' x ',' x}{http://coq.inria.fr/distrib/8.5beta2/stdlib/Coq.Unicode.Utf8\_core}{\coqdocnotation{(}}\coqdocvar{n1} \coqdocvar{m0} : \coqexternalref{nat}{http://coq.inria.fr/distrib/8.5beta2/stdlib/Coq.Init.Datatypes}{\coqdocinductive{nat}}) (\coqdocvar{l} : \coqdocvariable{n1} \coqref{Equations.::x '<=' x}{\coqdocnotation{$\le$}} \coqdocvariable{m0}) (\coqdocvar{l0} : \coqdocvariable{m0} \coqref{Equations.::x '<=' x}{\coqdocnotation{$\le$}} \coqdocvariable{n1}\coqexternalref{:type scope:'xE2x88x80' x '..' x ',' x}{http://coq.inria.fr/distrib/8.5beta2/stdlib/Coq.Unicode.Utf8\_core}{\coqdocnotation{),}} \coqdocvariable{P} \coqdocvariable{n1} \coqdocvariable{m0} \coqdocvariable{l} \coqdocvariable{l0} (\coqref{Equations.antisym}{\coqdocdefinition{antisym}} \coqdocvariable{l} \coqdocvariable{l0}) \coqexternalref{:type scope:x 'xE2x86x92' x}{http://coq.inria.fr/distrib/8.5beta2/stdlib/Coq.Unicode.Utf8\_core}{\coqdocnotation{→}} \coqdoceol
\coqdocindent{3.50em}
\coqdocvariable{P} (\coqexternalref{S}{http://coq.inria.fr/distrib/8.5beta2/stdlib/Coq.Init.Datatypes}{\coqdocconstructor{S}} \coqdocvariable{n1}) (\coqexternalref{S}{http://coq.inria.fr/distrib/8.5beta2/stdlib/Coq.Init.Datatypes}{\coqdocconstructor{S}} \coqdocvariable{m0}) (\coqref{Equations.ls}{\coqdocconstructor{ls}} \coqdocvariable{l}) (\coqref{Equations.ls}{\coqdocconstructor{ls}} \coqdocvariable{l0}) (\coqexternalref{f equal}{http://coq.inria.fr/distrib/8.5beta2/stdlib/Coq.Init.Logic}{\coqdoclemma{f\_equal}} \coqexternalref{S}{http://coq.inria.fr/distrib/8.5beta2/stdlib/Coq.Init.Datatypes}{\coqdocconstructor{S}} (\coqref{Equations.antisym}{\coqdocdefinition{antisym}} \coqdocvariable{l} \coqdocvariable{l0}))\coqexternalref{:type scope:x 'xE2x86x92' x}{http://coq.inria.fr/distrib/8.5beta2/stdlib/Coq.Unicode.Utf8\_core}{\coqdocnotation{)}} \coqexternalref{:type scope:x 'xE2x86x92' x}{http://coq.inria.fr/distrib/8.5beta2/stdlib/Coq.Unicode.Utf8\_core}{\coqdocnotation{→}} \coqdoceol
\coqdocindent{1.00em}
\coqexternalref{:type scope:'xE2x88x80' x '..' x ',' x}{http://coq.inria.fr/distrib/8.5beta2/stdlib/Coq.Unicode.Utf8\_core}{\coqdocnotation{∀}} \coqexternalref{:type scope:'xE2x88x80' x '..' x ',' x}{http://coq.inria.fr/distrib/8.5beta2/stdlib/Coq.Unicode.Utf8\_core}{\coqdocnotation{(}}\coqdocvar{m} \coqdocvar{n} : \coqexternalref{nat}{http://coq.inria.fr/distrib/8.5beta2/stdlib/Coq.Init.Datatypes}{\coqdocinductive{nat}}) (\coqdocvar{x} : \coqdocvariable{m} \coqref{Equations.::x '<=' x}{\coqdocnotation{$\le$}} \coqdocvariable{n}) (\coqdocvar{y} : \coqdocvariable{n} \coqref{Equations.::x '<=' x}{\coqdocnotation{$\le$}} \coqdocvariable{m}\coqexternalref{:type scope:'xE2x88x80' x '..' x ',' x}{http://coq.inria.fr/distrib/8.5beta2/stdlib/Coq.Unicode.Utf8\_core}{\coqdocnotation{),}} \coqdocvariable{P} \coqdocvariable{m} \coqdocvariable{n} \coqdocvariable{x} \coqdocvariable{y} (\coqref{Equations.antisym}{\coqdocdefinition{antisym}} \coqdocvariable{x} \coqdocvariable{y}).\coqdoceol
\end{coqdoccode}

\begin{coqdoccode}
\coqdocemptyline
\end{coqdoccode}
The last feature of \Equations necessary to write real definitions
  is the \coqdockw{with} construct. This construct allows to do pattern-matching
  on intermediary results in a definition. A typical example is the
  \coqref{EquationsWith.filter}{\coqdocdefinition{filter}} function on lists, which selects all elements of the original
  list respecting some boolean predicate: \begin{coqdoccode}
\coqdocemptyline
\coqdocindent{1.00em}
\coqdockw{Context} \{\coqdocvar{A}\} (\coqdocvar{p} : \coqdocvariable{A} \coqexternalref{:type scope:x '->' x}{http://coq.inria.fr/distrib/8.5beta2/stdlib/Coq.Init.Logic}{\coqdocnotation{\ensuremath{\rightarrow}}} \coqexternalref{bool}{http://coq.inria.fr/distrib/8.5beta2/stdlib/Coq.Init.Datatypes}{\coqdocinductive{bool}}).\coqdoceol
\coqdocemptyline
\coqdocindent{1.00em}
\coqdockw{Equations} \coqdef{EquationsWith.filter}{filter}{\coqdocdefinition{filter}} (\coqdocvar{l} : \coqexternalref{list}{http://coq.inria.fr/distrib/8.5beta2/stdlib/Coq.Init.Datatypes}{\coqdocinductive{list}} \coqdocvariable{A}) : \coqexternalref{list}{http://coq.inria.fr/distrib/8.5beta2/stdlib/Coq.Init.Datatypes}{\coqdocinductive{list}} \coqdocvariable{A} :=\coqdoceol
\coqdocindent{1.00em}
\coqref{EquationsWith.filter}{\coqdocdefinition{filter}} \coqexternalref{ListNotations.:list scope:'[' ']'}{http://coq.inria.fr/distrib/8.5beta2/stdlib/Coq.Lists.List}{\coqdocnotation{[]}} := \coqexternalref{ListNotations.:list scope:'[' ']'}{http://coq.inria.fr/distrib/8.5beta2/stdlib/Coq.Lists.List}{\coqdocnotation{[]}} ;\coqdoceol
\coqdocindent{1.00em}
\coqref{EquationsWith.filter}{\coqdocdefinition{filter}} (\coqexternalref{cons}{http://coq.inria.fr/distrib/8.5beta2/stdlib/Coq.Init.Datatypes}{\coqdocconstructor{cons}} \coqexternalref{cons}{http://coq.inria.fr/distrib/8.5beta2/stdlib/Coq.Init.Datatypes}{\coqdocconstructor{a}} \coqexternalref{cons}{http://coq.inria.fr/distrib/8.5beta2/stdlib/Coq.Init.Datatypes}{\coqdocconstructor{l}}) $\Leftarrow$ \coqdocvariable{p} \coqdocvar{a} \ensuremath{\Rightarrow} \{ \ensuremath{|} \coqexternalref{true}{http://coq.inria.fr/distrib/8.5beta2/stdlib/Coq.Init.Datatypes}{\coqdocconstructor{true}} := \coqexternalref{cons}{http://coq.inria.fr/distrib/8.5beta2/stdlib/Coq.Init.Datatypes}{\coqdocconstructor{cons}} \coqdocvar{a} (\coqref{EquationsWith.filter}{\coqdocdefinition{filter}} \coqdocvar{l}) ; \ensuremath{|} \coqexternalref{false}{http://coq.inria.fr/distrib/8.5beta2/stdlib/Coq.Init.Datatypes}{\coqdocconstructor{false}} := \coqref{EquationsWith.filter}{\coqdocdefinition{filter}} \coqdocvar{l} \}.\coqdoceol
\end{coqdoccode}
The $\Leftarrow$ \coqdocvariable{p} \coqdocvariable{a} \ensuremath{\Rightarrow} right-hand side adds a new pattern to the left-hand
  side of its subprogram, for an object of type \coqexternalref{bool}{http://coq.inria.fr/distrib/8.5beta2/stdlib/Coq.Init.Datatypes}{\coqdocinductive{bool}} here. The
  subprogram is actually defined as another proxy constant, which takes
  as arguments the variables \coqdocvariable{a}, \coqdocvariable{p} and a new variable of type \coqexternalref{bool}{http://coq.inria.fr/distrib/8.5beta2/stdlib/Coq.Init.Datatypes}{\coqdocinductive{bool}}.
  The clauses of the subprogram can shortcut the \coqref{EquationsWith.filter}{\coqdocdefinition{filter}} (\coqexternalref{cons}{http://coq.inria.fr/distrib/8.5beta2/stdlib/Coq.Init.Datatypes}{\coqdocconstructor{cons}} \coqdocvariable{a} \coqdocvariable{l})
  part of the pattern which is automatically inferred from the enclosing
  left-hand side.

  The generated equations for such definitions go through the proxy
  constant, hence we have two equations for \coqref{EquationsWith.filter}{\coqdocdefinition{filter}} and two for
  \coqdocvar{filter\_helper\_1}, which is the name of the proxy constant. To
  generate the elimination principle, a mutually inductive graph is
  generated, and the predicate applying to the subprogram is defined in
  terms of the original one, adding an equality between the new variable
  and the exact term it is applied to in the enclosing program. This
  way, we cannot forget during proofs that the \coqexternalref{true}{http://coq.inria.fr/distrib/8.5beta2/stdlib/Coq.Init.Datatypes}{\coqdocconstructor{true}} or \coqexternalref{false}{http://coq.inria.fr/distrib/8.5beta2/stdlib/Coq.Init.Datatypes}{\coqdocconstructor{false}} cases
  are actually results of a call to \coqdocvariable{p} \coqdocvariable{a}. Note that there are three leaves in
  the original program (and splitting tree) hence three cases to consider here.
 \begin{coqdoccode}
\coqdocemptyline
\coqdocnoindent
\coqdockw{Check} $\star$\coqref{EquationsWith.filter elim}{\coqdocdefinition{filter\_elim}} : \coqexternalref{:type scope:'xE2x88x80' x '..' x ',' x}{http://coq.inria.fr/distrib/8.5beta2/stdlib/Coq.Unicode.Utf8\_core}{\coqdocnotation{∀}} \coqexternalref{:type scope:'xE2x88x80' x '..' x ',' x}{http://coq.inria.fr/distrib/8.5beta2/stdlib/Coq.Unicode.Utf8\_core}{\coqdocnotation{(}}\coqdocvar{A} : \coqdockw{Type}) (\coqdocvar{p} : \coqdocvariable{A} \coqexternalref{:type scope:x 'xE2x86x92' x}{http://coq.inria.fr/distrib/8.5beta2/stdlib/Coq.Unicode.Utf8\_core}{\coqdocnotation{→}} \coqexternalref{bool}{http://coq.inria.fr/distrib/8.5beta2/stdlib/Coq.Init.Datatypes}{\coqdocinductive{bool}}) (\coqdocvar{P} : \coqexternalref{list}{http://coq.inria.fr/distrib/8.5beta2/stdlib/Coq.Init.Datatypes}{\coqdocinductive{list}} \coqdocvariable{A} \coqexternalref{:type scope:x 'xE2x86x92' x}{http://coq.inria.fr/distrib/8.5beta2/stdlib/Coq.Unicode.Utf8\_core}{\coqdocnotation{→}} \coqexternalref{list}{http://coq.inria.fr/distrib/8.5beta2/stdlib/Coq.Init.Datatypes}{\coqdocinductive{list}} \coqdocvariable{A} \coqexternalref{:type scope:x 'xE2x86x92' x}{http://coq.inria.fr/distrib/8.5beta2/stdlib/Coq.Unicode.Utf8\_core}{\coqdocnotation{→}} \coqdockw{Prop})\coqdoceol
\coqdocindent{0.50em}
(\coqdocvar{P0}:=\coqexternalref{::'xCExBB' x '..' x ',' x}{http://coq.inria.fr/distrib/8.5beta2/stdlib/Coq.Unicode.Utf8\_core}{\coqdocnotation{\ensuremath{\lambda}}} \coqexternalref{::'xCExBB' x '..' x ',' x}{http://coq.inria.fr/distrib/8.5beta2/stdlib/Coq.Unicode.Utf8\_core}{\coqdocnotation{(}}\coqdocvar{a} : \coqdocvariable{A}) (\coqdoctac{refine} : \coqexternalref{bool}{http://coq.inria.fr/distrib/8.5beta2/stdlib/Coq.Init.Datatypes}{\coqdocinductive{bool}}) (\coqdocvar{l} \coqdocvar{H} : \coqexternalref{list}{http://coq.inria.fr/distrib/8.5beta2/stdlib/Coq.Init.Datatypes}{\coqdocinductive{list}} \coqdocvariable{A}\coqexternalref{::'xCExBB' x '..' x ',' x}{http://coq.inria.fr/distrib/8.5beta2/stdlib/Coq.Unicode.Utf8\_core}{\coqdocnotation{),}} \coqdocvariable{p} \coqdocvariable{a} \coqexternalref{:type scope:x '=' x}{http://coq.inria.fr/distrib/8.5beta2/stdlib/Coq.Init.Logic}{\coqdocnotation{=}} \coqdocvariable{refine} \coqexternalref{:type scope:x 'xE2x86x92' x}{http://coq.inria.fr/distrib/8.5beta2/stdlib/Coq.Unicode.Utf8\_core}{\coqdocnotation{→}} \coqdocvariable{P} (\coqexternalref{cons}{http://coq.inria.fr/distrib/8.5beta2/stdlib/Coq.Init.Datatypes}{\coqdocconstructor{cons}} \coqdocvariable{a} \coqdocvariable{l}) \coqdocvariable{H}\coqexternalref{:type scope:'xE2x88x80' x '..' x ',' x}{http://coq.inria.fr/distrib/8.5beta2/stdlib/Coq.Unicode.Utf8\_core}{\coqdocnotation{),}}\coqdoceol
\coqdocindent{1.00em}
\coqdocvariable{P} \coqexternalref{ListNotations.:list scope:'[' ']'}{http://coq.inria.fr/distrib/8.5beta2/stdlib/Coq.Lists.List}{\coqdocnotation{[]}} \coqexternalref{ListNotations.:list scope:'[' ']'}{http://coq.inria.fr/distrib/8.5beta2/stdlib/Coq.Lists.List}{\coqdocnotation{[]}} \coqexternalref{:type scope:x 'xE2x86x92' x}{http://coq.inria.fr/distrib/8.5beta2/stdlib/Coq.Unicode.Utf8\_core}{\coqdocnotation{→}} \coqexternalref{:type scope:x 'xE2x86x92' x}{http://coq.inria.fr/distrib/8.5beta2/stdlib/Coq.Unicode.Utf8\_core}{\coqdocnotation{(}}\coqexternalref{:type scope:'xE2x88x80' x '..' x ',' x}{http://coq.inria.fr/distrib/8.5beta2/stdlib/Coq.Unicode.Utf8\_core}{\coqdocnotation{∀}} \coqexternalref{:type scope:'xE2x88x80' x '..' x ',' x}{http://coq.inria.fr/distrib/8.5beta2/stdlib/Coq.Unicode.Utf8\_core}{\coqdocnotation{(}}\coqdocvar{a} : \coqdocvariable{A}) (\coqdocvar{l} : \coqexternalref{list}{http://coq.inria.fr/distrib/8.5beta2/stdlib/Coq.Init.Datatypes}{\coqdocinductive{list}} \coqdocvariable{A}\coqexternalref{:type scope:'xE2x88x80' x '..' x ',' x}{http://coq.inria.fr/distrib/8.5beta2/stdlib/Coq.Unicode.Utf8\_core}{\coqdocnotation{),}} \coqdocvariable{P} \coqdocvariable{l} (\coqref{EquationsWith.filter}{\coqdocdefinition{filter}} \coqdocvariable{p} \coqdocvariable{l}) \coqexternalref{:type scope:x 'xE2x86x92' x}{http://coq.inria.fr/distrib/8.5beta2/stdlib/Coq.Unicode.Utf8\_core}{\coqdocnotation{→}} \coqdocvariable{P0} \coqdocvariable{a} \coqexternalref{true}{http://coq.inria.fr/distrib/8.5beta2/stdlib/Coq.Init.Datatypes}{\coqdocconstructor{true}} \coqdocvariable{l} (\coqexternalref{cons}{http://coq.inria.fr/distrib/8.5beta2/stdlib/Coq.Init.Datatypes}{\coqdocconstructor{cons}} \coqdocvariable{a} (\coqref{EquationsWith.filter}{\coqdocdefinition{filter}} \coqdocvariable{p} \coqdocvariable{l}))\coqexternalref{:type scope:x 'xE2x86x92' x}{http://coq.inria.fr/distrib/8.5beta2/stdlib/Coq.Unicode.Utf8\_core}{\coqdocnotation{)}} \coqexternalref{:type scope:x 'xE2x86x92' x}{http://coq.inria.fr/distrib/8.5beta2/stdlib/Coq.Unicode.Utf8\_core}{\coqdocnotation{→}} \coqdoceol
\coqdocindent{1.00em}
\coqexternalref{:type scope:x 'xE2x86x92' x}{http://coq.inria.fr/distrib/8.5beta2/stdlib/Coq.Unicode.Utf8\_core}{\coqdocnotation{(}}\coqexternalref{:type scope:'xE2x88x80' x '..' x ',' x}{http://coq.inria.fr/distrib/8.5beta2/stdlib/Coq.Unicode.Utf8\_core}{\coqdocnotation{∀}} \coqexternalref{:type scope:'xE2x88x80' x '..' x ',' x}{http://coq.inria.fr/distrib/8.5beta2/stdlib/Coq.Unicode.Utf8\_core}{\coqdocnotation{(}}\coqdocvar{a} : \coqdocvariable{A}) (\coqdocvar{l} : \coqexternalref{list}{http://coq.inria.fr/distrib/8.5beta2/stdlib/Coq.Init.Datatypes}{\coqdocinductive{list}} \coqdocvariable{A}\coqexternalref{:type scope:'xE2x88x80' x '..' x ',' x}{http://coq.inria.fr/distrib/8.5beta2/stdlib/Coq.Unicode.Utf8\_core}{\coqdocnotation{),}} \coqdocvariable{P} \coqdocvariable{l} (\coqref{EquationsWith.filter}{\coqdocdefinition{filter}} \coqdocvariable{p} \coqdocvariable{l}) \coqexternalref{:type scope:x 'xE2x86x92' x}{http://coq.inria.fr/distrib/8.5beta2/stdlib/Coq.Unicode.Utf8\_core}{\coqdocnotation{→}} \coqdocvariable{P0} \coqdocvariable{a} \coqexternalref{false}{http://coq.inria.fr/distrib/8.5beta2/stdlib/Coq.Init.Datatypes}{\coqdocconstructor{false}} \coqdocvariable{l} (\coqref{EquationsWith.filter}{\coqdocdefinition{filter}} \coqdocvariable{p} \coqdocvariable{l})\coqexternalref{:type scope:x 'xE2x86x92' x}{http://coq.inria.fr/distrib/8.5beta2/stdlib/Coq.Unicode.Utf8\_core}{\coqdocnotation{)}} \coqexternalref{:type scope:x 'xE2x86x92' x}{http://coq.inria.fr/distrib/8.5beta2/stdlib/Coq.Unicode.Utf8\_core}{\coqdocnotation{→}} \coqdoceol
\coqdocindent{1.00em}
\coqexternalref{:type scope:'xE2x88x80' x '..' x ',' x}{http://coq.inria.fr/distrib/8.5beta2/stdlib/Coq.Unicode.Utf8\_core}{\coqdocnotation{∀}} \coqdocvar{l} : \coqexternalref{list}{http://coq.inria.fr/distrib/8.5beta2/stdlib/Coq.Init.Datatypes}{\coqdocinductive{list}} \coqdocvariable{A}\coqexternalref{:type scope:'xE2x88x80' x '..' x ',' x}{http://coq.inria.fr/distrib/8.5beta2/stdlib/Coq.Unicode.Utf8\_core}{\coqdocnotation{,}} \coqdocvariable{P} \coqdocvariable{l} (\coqref{EquationsWith.filter}{\coqdocdefinition{filter}} \coqdocvariable{p} \coqdocvariable{l}).\coqdoceol
\coqdocemptyline
\end{coqdoccode}
In general, the term used as a new discriminee is abstracted from
  the context and return type at this point of the program before checking the 
  subprogram. In that case the eliminator predicate for the subprogram has
  a dependent binding for the \coqdocvar{t} = \coqdoctac{refine} hypothesis that is used to rewrite 
  in the type of hypotheses and results. This is examplified in the following
  classical example: \begin{coqdoccode}
\coqdocemptyline
\coqdocnoindent
\coqdockw{Inductive} \coqdef{EquationsWith.incl}{incl}{\coqdocinductive{incl}} \{\coqdocvar{A}\} : \coqexternalref{list}{http://coq.inria.fr/distrib/8.5beta2/stdlib/Coq.Init.Datatypes}{\coqdocinductive{list}} \coqdocvar{A} \coqexternalref{:type scope:x '->' x}{http://coq.inria.fr/distrib/8.5beta2/stdlib/Coq.Init.Logic}{\coqdocnotation{\ensuremath{\rightarrow}}} \coqexternalref{list}{http://coq.inria.fr/distrib/8.5beta2/stdlib/Coq.Init.Datatypes}{\coqdocinductive{list}} \coqdocvar{A} \coqexternalref{:type scope:x '->' x}{http://coq.inria.fr/distrib/8.5beta2/stdlib/Coq.Init.Logic}{\coqdocnotation{\ensuremath{\rightarrow}}} \coqdockw{Prop} :=\coqdoceol
\coqdocindent{1.00em}
\coqdef{EquationsWith.stop}{stop}{\coqdocconstructor{stop}} : \coqref{EquationsWith.incl}{\coqdocinductive{incl}} \coqexternalref{nil}{http://coq.inria.fr/distrib/8.5beta2/stdlib/Coq.Init.Datatypes}{\coqdocconstructor{nil}} \coqexternalref{nil}{http://coq.inria.fr/distrib/8.5beta2/stdlib/Coq.Init.Datatypes}{\coqdocconstructor{nil}} \coqdoceol
\coqdocnoindent
\ensuremath{|} \coqdef{EquationsWith.keep}{keep}{\coqdocconstructor{keep}} \{\coqdocvar{x} : \coqdocvar{A}\} \{\coqdocvar{xs} \coqdocvar{ys} : \coqexternalref{list}{http://coq.inria.fr/distrib/8.5beta2/stdlib/Coq.Init.Datatypes}{\coqdocinductive{list}} \coqdocvar{A}\} : \coqref{EquationsWith.incl}{\coqdocinductive{incl}} \coqdocvariable{xs} \coqdocvariable{ys} \coqexternalref{:type scope:x '->' x}{http://coq.inria.fr/distrib/8.5beta2/stdlib/Coq.Init.Logic}{\coqdocnotation{\ensuremath{\rightarrow}}} \coqref{EquationsWith.incl}{\coqdocinductive{incl}} (\coqexternalref{cons}{http://coq.inria.fr/distrib/8.5beta2/stdlib/Coq.Init.Datatypes}{\coqdocconstructor{cons}} \coqdocvariable{x} \coqdocvariable{xs}) (\coqexternalref{cons}{http://coq.inria.fr/distrib/8.5beta2/stdlib/Coq.Init.Datatypes}{\coqdocconstructor{cons}} \coqdocvariable{x} \coqdocvariable{ys})\coqdoceol
\coqdocnoindent
\ensuremath{|} \coqdef{EquationsWith.skip}{skip}{\coqdocconstructor{skip}} \{\coqdocvar{x} : \coqdocvar{A}\} \{\coqdocvar{xs} \coqdocvar{ys} : \coqexternalref{list}{http://coq.inria.fr/distrib/8.5beta2/stdlib/Coq.Init.Datatypes}{\coqdocinductive{list}} \coqdocvar{A}\} : \coqref{EquationsWith.incl}{\coqdocinductive{incl}} \coqdocvariable{xs} \coqdocvariable{ys} \coqexternalref{:type scope:x '->' x}{http://coq.inria.fr/distrib/8.5beta2/stdlib/Coq.Init.Logic}{\coqdocnotation{\ensuremath{\rightarrow}}} \coqref{EquationsWith.incl}{\coqdocinductive{incl}} (\coqdocvariable{xs}) (\coqexternalref{cons}{http://coq.inria.fr/distrib/8.5beta2/stdlib/Coq.Init.Datatypes}{\coqdocconstructor{cons}} \coqdocvariable{x} \coqdocvariable{ys}).\coqdoceol
\coqdocemptyline
\end{coqdoccode}
We define list inclusion inductively and show that filtering out some elements 
  from a list xs results in a included list. \begin{coqdoccode}
\coqdocemptyline
\coqdocnoindent
\coqdockw{Equations}(\coqdocvar{nocomp}) \coqdef{EquationsWith.sublist}{sublist}{\coqdocdefinition{sublist}} \{\coqdocvar{A}\} (\coqdocvar{p} : \coqdocvariable{A} \coqexternalref{:type scope:x '->' x}{http://coq.inria.fr/distrib/8.5beta2/stdlib/Coq.Init.Logic}{\coqdocnotation{\ensuremath{\rightarrow}}} \coqexternalref{bool}{http://coq.inria.fr/distrib/8.5beta2/stdlib/Coq.Init.Datatypes}{\coqdocinductive{bool}}) (\coqdocvar{xs} : \coqexternalref{list}{http://coq.inria.fr/distrib/8.5beta2/stdlib/Coq.Init.Datatypes}{\coqdocinductive{list}} \coqdocvariable{A}) : \coqref{EquationsWith.incl}{\coqdocinductive{incl}} (\coqref{EquationsWith.filter}{\coqdocdefinition{filter}} \coqdocvar{p} \coqdocvar{xs}) \coqdocvar{xs} :=\coqdoceol
\coqdocnoindent
\coqref{EquationsWith.sublist}{\coqdocdefinition{sublist}} \coqdocvar{A} \coqdocvar{p} \coqexternalref{nil}{http://coq.inria.fr/distrib/8.5beta2/stdlib/Coq.Init.Datatypes}{\coqdocconstructor{nil}} := \coqref{EquationsWith.stop}{\coqdocconstructor{stop}} ;\coqdoceol
\coqdocnoindent
\coqref{EquationsWith.sublist}{\coqdocdefinition{sublist}} \coqdocvar{A} \coqdocvar{p} (\coqexternalref{cons}{http://coq.inria.fr/distrib/8.5beta2/stdlib/Coq.Init.Datatypes}{\coqdocconstructor{cons}} \coqexternalref{cons}{http://coq.inria.fr/distrib/8.5beta2/stdlib/Coq.Init.Datatypes}{\coqdocconstructor{x}} \coqexternalref{cons}{http://coq.inria.fr/distrib/8.5beta2/stdlib/Coq.Init.Datatypes}{\coqdocconstructor{xs}}) \coqdockw{with} \coqdocvar{p} \coqdocvar{x} := \{\coqdoceol
\coqdocindent{1.00em}
\ensuremath{|} \coqexternalref{true}{http://coq.inria.fr/distrib/8.5beta2/stdlib/Coq.Init.Datatypes}{\coqdocconstructor{true}} := \coqref{EquationsWith.keep}{\coqdocconstructor{keep}} (\coqref{EquationsWith.sublist}{\coqdocdefinition{sublist}} \coqdocvar{p} \coqdocvar{xs}) ; \ensuremath{|} \coqexternalref{false}{http://coq.inria.fr/distrib/8.5beta2/stdlib/Coq.Init.Datatypes}{\coqdocconstructor{false}} := \coqref{EquationsWith.skip}{\coqdocconstructor{skip}} (\coqref{EquationsWith.sublist}{\coqdocdefinition{sublist}} \coqdocvar{p} \coqdocvar{xs}) \}.\coqdoceol
\coqdocemptyline
\end{coqdoccode}
Here at the \coqdockw{with} node, the return type is 
\coqref{EquationsWith.incl}{\coqdocinductive{incl}} (\coqdockw{if} \coqdocvariable{p} \coqdocvariable{x} \coqdockw{then} \coqexternalref{cons}{http://coq.inria.fr/distrib/8.5beta2/stdlib/Coq.Init.Datatypes}{\coqdocconstructor{cons}} \coqdocvariable{x} (\coqref{EquationsWith.filter}{\coqdocdefinition{filter}} \coqdocvariable{p} \coqdocvariable{xs}) \coqdockw{else} \coqref{EquationsWith.filter}{\coqdocdefinition{filter}} \coqdocvariable{p} \coqdocvariable{xs}) (\coqexternalref{cons}{http://coq.inria.fr/distrib/8.5beta2/stdlib/Coq.Init.Datatypes}{\coqdocconstructor{cons}} \coqdocvariable{x} \coqdocvariable{xs}).
  We abstract \coqdocvariable{p} \coqdocvariable{x} from the return type and check the new subprogram 
  in context \coqdocvariable{A} \coqdocvariable{P} \coqdocvariable{x} \coqdocvariable{xs} (\coqdoctac{refine} : \coqexternalref{bool}{http://coq.inria.fr/distrib/8.5beta2/stdlib/Coq.Init.Datatypes}{\coqdocinductive{bool}}) with return type:
  \coqref{EquationsWith.incl}{\coqdocinductive{incl}} (\coqdockw{if} \coqdoctac{refine} \coqdockw{then} \coqexternalref{cons}{http://coq.inria.fr/distrib/8.5beta2/stdlib/Coq.Init.Datatypes}{\coqdocconstructor{cons}} \coqdocvariable{x} (\coqref{EquationsWith.filter}{\coqdocdefinition{filter}} \coqdocvariable{p}) \coqdocvariable{xs} \coqdockw{else} \coqref{EquationsWith.filter}{\coqdocdefinition{filter}} \coqdocvariable{p} \coqdocvariable{xs}) (\coqexternalref{cons}{http://coq.inria.fr/distrib/8.5beta2/stdlib/Coq.Init.Datatypes}{\coqdocconstructor{cons}} \coqdocvariable{x} \coqdocvariable{xs}).

  Each of the patterns instantiates \coqdoctac{refine} to a constructor, so the
  return type reduces to the two expected cases matching with conclusions
  of the \coqref{EquationsWith.incl}{\coqdocinductive{incl}} relation. The (\coqdocvar{nocomp}) option indicates that we do not
  want the return type to be defined using a \coqdocvar{\_comp} constant. Indeed,
  the term \coqref{EquationsWith.keep}{\coqdocconstructor{keep}} (\coqref{EquationsWith.sublist}{\coqdocdefinition{sublist}} \coqdocvariable{p} \coqdocvariable{xs}) would not be well-typed, as it is expected
  to have type \coqdocvar{sublist\_comp} \coqdocvariable{p} (\coqdocvariable{x} :: \coqdocvariable{xs}) which is \coqref{EquationsWith.incl}{\coqdocinductive{incl}} (\coqref{EquationsWith.filter}{\coqdocdefinition{filter}} \coqdocvariable{p} (\coqdocvariable{x} :: \coqdocvariable{xs})) (\coqdocvariable{x} :: \coqdocvariable{xs}),
  but has type \coqref{EquationsWith.incl}{\coqdocinductive{incl}} (\coqdocvariable{x} :: \coqref{EquationsWith.filter}{\coqdocdefinition{filter}} \coqdocvariable{p} \coqdocvariable{xs}) (\coqdocvariable{x} :: \coqdocvariable{xs}).
  This is just a technical limitation we hope to remove in the future. \begin{coqdoccode}
\coqdocemptyline
\coqdocnoindent
\coqdockw{Check} $\star$\coqref{EquationsWith.sublist elim}{\coqdocdefinition{sublist\_elim}} : \coqexternalref{:type scope:'xE2x88x80' x '..' x ',' x}{http://coq.inria.fr/distrib/8.5beta2/stdlib/Coq.Unicode.Utf8\_core}{\coqdocnotation{∀}} \coqexternalref{:type scope:'xE2x88x80' x '..' x ',' x}{http://coq.inria.fr/distrib/8.5beta2/stdlib/Coq.Unicode.Utf8\_core}{\coqdocnotation{(}}\coqdocvar{P} : \coqexternalref{:type scope:'xE2x88x80' x '..' x ',' x}{http://coq.inria.fr/distrib/8.5beta2/stdlib/Coq.Unicode.Utf8\_core}{\coqdocnotation{∀}} \coqexternalref{:type scope:'xE2x88x80' x '..' x ',' x}{http://coq.inria.fr/distrib/8.5beta2/stdlib/Coq.Unicode.Utf8\_core}{\coqdocnotation{(}}\coqdocvar{A} : \coqdockw{Type}) (\coqdocvar{p} : \coqdocvariable{A} \coqexternalref{:type scope:x 'xE2x86x92' x}{http://coq.inria.fr/distrib/8.5beta2/stdlib/Coq.Unicode.Utf8\_core}{\coqdocnotation{→}} \coqexternalref{bool}{http://coq.inria.fr/distrib/8.5beta2/stdlib/Coq.Init.Datatypes}{\coqdocinductive{bool}}) (\coqdocvar{xs} : \coqexternalref{list}{http://coq.inria.fr/distrib/8.5beta2/stdlib/Coq.Init.Datatypes}{\coqdocinductive{list}} \coqdocvariable{A}\coqexternalref{:type scope:'xE2x88x80' x '..' x ',' x}{http://coq.inria.fr/distrib/8.5beta2/stdlib/Coq.Unicode.Utf8\_core}{\coqdocnotation{),}} \coqref{EquationsWith.incl}{\coqdocinductive{incl}} (\coqref{EquationsWith.filter}{\coqdocdefinition{filter}} \coqdocvariable{p} \coqdocvariable{xs}) \coqdocvariable{xs} \coqexternalref{:type scope:x 'xE2x86x92' x}{http://coq.inria.fr/distrib/8.5beta2/stdlib/Coq.Unicode.Utf8\_core}{\coqdocnotation{→}} \coqdockw{Prop})\coqdoceol
\coqdocnoindent
(\coqdocvar{P0}:= \coqexternalref{::'xCExBB' x '..' x ',' x}{http://coq.inria.fr/distrib/8.5beta2/stdlib/Coq.Unicode.Utf8\_core}{\coqdocnotation{\ensuremath{\lambda}}} \coqexternalref{::'xCExBB' x '..' x ',' x}{http://coq.inria.fr/distrib/8.5beta2/stdlib/Coq.Unicode.Utf8\_core}{\coqdocnotation{(}}\coqdocvar{A} : \coqdockw{Type}) (\coqdocvar{p} : \coqdocvariable{A} \coqexternalref{:type scope:x 'xE2x86x92' x}{http://coq.inria.fr/distrib/8.5beta2/stdlib/Coq.Unicode.Utf8\_core}{\coqdocnotation{→}} \coqexternalref{bool}{http://coq.inria.fr/distrib/8.5beta2/stdlib/Coq.Init.Datatypes}{\coqdocinductive{bool}}) (\coqdocvar{a} : \coqdocvariable{A}) (\coqdoctac{refine} : \coqexternalref{bool}{http://coq.inria.fr/distrib/8.5beta2/stdlib/Coq.Init.Datatypes}{\coqdocinductive{bool}}) (\coqdocvar{l} : \coqexternalref{list}{http://coq.inria.fr/distrib/8.5beta2/stdlib/Coq.Init.Datatypes}{\coqdocinductive{list}} \coqdocvariable{A})\coqdoceol
\coqdocindent{0.50em}
(\coqdocvar{H} : \coqref{EquationsWith.incl}{\coqdocinductive{incl}} (\coqref{EquationsWith.filter obligation 2}{\coqdoclemma{filter\_obligation\_2}} (\coqref{EquationsWith.filter}{\coqdocdefinition{filter}} \coqdocvariable{p}) \coqdocvariable{a} \coqdocvariable{refine} \coqdocvariable{l}) (\coqexternalref{cons}{http://coq.inria.fr/distrib/8.5beta2/stdlib/Coq.Init.Datatypes}{\coqdocconstructor{cons}} \coqdocvariable{a} \coqdocvariable{l})\coqexternalref{::'xCExBB' x '..' x ',' x}{http://coq.inria.fr/distrib/8.5beta2/stdlib/Coq.Unicode.Utf8\_core}{\coqdocnotation{),}}\coqdoceol
\coqdocindent{0.50em}
\coqexternalref{:type scope:'xE2x88x80' x '..' x ',' x}{http://coq.inria.fr/distrib/8.5beta2/stdlib/Coq.Unicode.Utf8\_core}{\coqdocnotation{∀}} \coqdocvar{Heq} : \coqdocvariable{p} \coqdocvariable{a} \coqexternalref{:type scope:x '=' x}{http://coq.inria.fr/distrib/8.5beta2/stdlib/Coq.Init.Logic}{\coqdocnotation{=}} \coqdocvariable{refine}\coqexternalref{:type scope:'xE2x88x80' x '..' x ',' x}{http://coq.inria.fr/distrib/8.5beta2/stdlib/Coq.Unicode.Utf8\_core}{\coqdocnotation{,}} \coqdocvariable{P} \coqdocvariable{A} \coqdocvariable{p} (\coqexternalref{cons}{http://coq.inria.fr/distrib/8.5beta2/stdlib/Coq.Init.Datatypes}{\coqdocconstructor{cons}} \coqdocvariable{a} \coqdocvariable{l}) \coqdoceol
\coqdocindent{1.50em}
(\coqexternalref{eq rect r}{http://coq.inria.fr/distrib/8.5beta2/stdlib/Coq.Init.Logic}{\coqdocdefinition{eq\_rect\_r}} (\coqexternalref{::'xCExBB' x '..' x ',' x}{http://coq.inria.fr/distrib/8.5beta2/stdlib/Coq.Unicode.Utf8\_core}{\coqdocnotation{\ensuremath{\lambda}}} \coqdocvar{r} : \coqexternalref{bool}{http://coq.inria.fr/distrib/8.5beta2/stdlib/Coq.Init.Datatypes}{\coqdocinductive{bool}}\coqexternalref{::'xCExBB' x '..' x ',' x}{http://coq.inria.fr/distrib/8.5beta2/stdlib/Coq.Unicode.Utf8\_core}{\coqdocnotation{,}} \coqref{EquationsWith.incl}{\coqdocinductive{incl}} (\coqref{EquationsWith.filter obligation 2}{\coqdoclemma{filter\_obligation\_2}} (\coqref{EquationsWith.filter}{\coqdocdefinition{filter}} \coqdocvariable{p}) \coqdocvariable{a} \coqdocvariable{r} \coqdocvariable{l}) (\coqexternalref{cons}{http://coq.inria.fr/distrib/8.5beta2/stdlib/Coq.Init.Datatypes}{\coqdocconstructor{cons}} \coqdocvariable{a} \coqdocvariable{l})) \coqdocvariable{H} \coqdocvariable{Heq})\coqexternalref{:type scope:'xE2x88x80' x '..' x ',' x}{http://coq.inria.fr/distrib/8.5beta2/stdlib/Coq.Unicode.Utf8\_core}{\coqdocnotation{),}}\coqdoceol
\coqdocindent{1.00em}
\coqexternalref{:type scope:x 'xE2x86x92' x}{http://coq.inria.fr/distrib/8.5beta2/stdlib/Coq.Unicode.Utf8\_core}{\coqdocnotation{(}}\coqexternalref{:type scope:'xE2x88x80' x '..' x ',' x}{http://coq.inria.fr/distrib/8.5beta2/stdlib/Coq.Unicode.Utf8\_core}{\coqdocnotation{∀}} \coqexternalref{:type scope:'xE2x88x80' x '..' x ',' x}{http://coq.inria.fr/distrib/8.5beta2/stdlib/Coq.Unicode.Utf8\_core}{\coqdocnotation{(}}\coqdocvar{A} : \coqdockw{Type}) (\coqdocvar{p} : \coqdocvariable{A} \coqexternalref{:type scope:x 'xE2x86x92' x}{http://coq.inria.fr/distrib/8.5beta2/stdlib/Coq.Unicode.Utf8\_core}{\coqdocnotation{→}} \coqexternalref{bool}{http://coq.inria.fr/distrib/8.5beta2/stdlib/Coq.Init.Datatypes}{\coqdocinductive{bool}}\coqexternalref{:type scope:'xE2x88x80' x '..' x ',' x}{http://coq.inria.fr/distrib/8.5beta2/stdlib/Coq.Unicode.Utf8\_core}{\coqdocnotation{),}} \coqdocvariable{P} \coqdocvariable{A} \coqdocvariable{p} \coqexternalref{ListNotations.:list scope:'[' ']'}{http://coq.inria.fr/distrib/8.5beta2/stdlib/Coq.Lists.List}{\coqdocnotation{[]}} \coqref{EquationsWith.stop}{\coqdocconstructor{stop}}\coqexternalref{:type scope:x 'xE2x86x92' x}{http://coq.inria.fr/distrib/8.5beta2/stdlib/Coq.Unicode.Utf8\_core}{\coqdocnotation{)}} \coqexternalref{:type scope:x 'xE2x86x92' x}{http://coq.inria.fr/distrib/8.5beta2/stdlib/Coq.Unicode.Utf8\_core}{\coqdocnotation{→}} \coqdoceol
\coqdocindent{1.00em}
\coqexternalref{:type scope:x 'xE2x86x92' x}{http://coq.inria.fr/distrib/8.5beta2/stdlib/Coq.Unicode.Utf8\_core}{\coqdocnotation{(}}\coqexternalref{:type scope:'xE2x88x80' x '..' x ',' x}{http://coq.inria.fr/distrib/8.5beta2/stdlib/Coq.Unicode.Utf8\_core}{\coqdocnotation{∀}} \coqexternalref{:type scope:'xE2x88x80' x '..' x ',' x}{http://coq.inria.fr/distrib/8.5beta2/stdlib/Coq.Unicode.Utf8\_core}{\coqdocnotation{(}}\coqdocvar{A} : \coqdockw{Type}) (\coqdocvar{p} : \coqdocvariable{A} \coqexternalref{:type scope:x 'xE2x86x92' x}{http://coq.inria.fr/distrib/8.5beta2/stdlib/Coq.Unicode.Utf8\_core}{\coqdocnotation{→}} \coqexternalref{bool}{http://coq.inria.fr/distrib/8.5beta2/stdlib/Coq.Init.Datatypes}{\coqdocinductive{bool}}) (\coqdocvar{a} : \coqdocvariable{A}) (\coqdocvar{l} : \coqexternalref{list}{http://coq.inria.fr/distrib/8.5beta2/stdlib/Coq.Init.Datatypes}{\coqdocinductive{list}} \coqdocvariable{A}\coqexternalref{:type scope:'xE2x88x80' x '..' x ',' x}{http://coq.inria.fr/distrib/8.5beta2/stdlib/Coq.Unicode.Utf8\_core}{\coqdocnotation{),}}\coqdoceol
\coqdocindent{2.50em}
\coqdocvariable{P} \coqdocvariable{A} \coqdocvariable{p} \coqdocvariable{l} (\coqref{EquationsWith.sublist}{\coqdocdefinition{sublist}} \coqdocvariable{p} \coqdocvariable{l}) \coqexternalref{:type scope:x 'xE2x86x92' x}{http://coq.inria.fr/distrib/8.5beta2/stdlib/Coq.Unicode.Utf8\_core}{\coqdocnotation{→}} \coqdocvariable{P0} \coqdocvariable{A} \coqdocvariable{p} \coqdocvariable{a} \coqexternalref{true}{http://coq.inria.fr/distrib/8.5beta2/stdlib/Coq.Init.Datatypes}{\coqdocconstructor{true}} \coqdocvariable{l} (\coqref{EquationsWith.keep}{\coqdocconstructor{keep}} (\coqref{EquationsWith.sublist}{\coqdocdefinition{sublist}} \coqdocvariable{p} \coqdocvariable{l}))\coqexternalref{:type scope:x 'xE2x86x92' x}{http://coq.inria.fr/distrib/8.5beta2/stdlib/Coq.Unicode.Utf8\_core}{\coqdocnotation{)}} \coqexternalref{:type scope:x 'xE2x86x92' x}{http://coq.inria.fr/distrib/8.5beta2/stdlib/Coq.Unicode.Utf8\_core}{\coqdocnotation{→}} \coqdoceol
\coqdocindent{1.00em}
\coqexternalref{:type scope:x 'xE2x86x92' x}{http://coq.inria.fr/distrib/8.5beta2/stdlib/Coq.Unicode.Utf8\_core}{\coqdocnotation{(}}\coqexternalref{:type scope:'xE2x88x80' x '..' x ',' x}{http://coq.inria.fr/distrib/8.5beta2/stdlib/Coq.Unicode.Utf8\_core}{\coqdocnotation{∀}} \coqexternalref{:type scope:'xE2x88x80' x '..' x ',' x}{http://coq.inria.fr/distrib/8.5beta2/stdlib/Coq.Unicode.Utf8\_core}{\coqdocnotation{(}}\coqdocvar{A} : \coqdockw{Type}) (\coqdocvar{p} : \coqdocvariable{A} \coqexternalref{:type scope:x 'xE2x86x92' x}{http://coq.inria.fr/distrib/8.5beta2/stdlib/Coq.Unicode.Utf8\_core}{\coqdocnotation{→}} \coqexternalref{bool}{http://coq.inria.fr/distrib/8.5beta2/stdlib/Coq.Init.Datatypes}{\coqdocinductive{bool}}) (\coqdocvar{a} : \coqdocvariable{A}) (\coqdocvar{l} : \coqexternalref{list}{http://coq.inria.fr/distrib/8.5beta2/stdlib/Coq.Init.Datatypes}{\coqdocinductive{list}} \coqdocvariable{A}\coqexternalref{:type scope:'xE2x88x80' x '..' x ',' x}{http://coq.inria.fr/distrib/8.5beta2/stdlib/Coq.Unicode.Utf8\_core}{\coqdocnotation{),}}\coqdoceol
\coqdocindent{2.50em}
\coqdocvariable{P} \coqdocvariable{A} \coqdocvariable{p} \coqdocvariable{l} (\coqref{EquationsWith.sublist}{\coqdocdefinition{sublist}} \coqdocvariable{p} \coqdocvariable{l}) \coqexternalref{:type scope:x 'xE2x86x92' x}{http://coq.inria.fr/distrib/8.5beta2/stdlib/Coq.Unicode.Utf8\_core}{\coqdocnotation{→}} \coqdocvariable{P0} \coqdocvariable{A} \coqdocvariable{p} \coqdocvariable{a} \coqexternalref{false}{http://coq.inria.fr/distrib/8.5beta2/stdlib/Coq.Init.Datatypes}{\coqdocconstructor{false}} \coqdocvariable{l} (\coqref{EquationsWith.skip}{\coqdocconstructor{skip}} (\coqref{EquationsWith.sublist}{\coqdocdefinition{sublist}} \coqdocvariable{p} \coqdocvariable{l}))\coqexternalref{:type scope:x 'xE2x86x92' x}{http://coq.inria.fr/distrib/8.5beta2/stdlib/Coq.Unicode.Utf8\_core}{\coqdocnotation{)}} \coqexternalref{:type scope:x 'xE2x86x92' x}{http://coq.inria.fr/distrib/8.5beta2/stdlib/Coq.Unicode.Utf8\_core}{\coqdocnotation{→}}\coqdoceol
\coqdocindent{1.50em}
\coqexternalref{:type scope:'xE2x88x80' x '..' x ',' x}{http://coq.inria.fr/distrib/8.5beta2/stdlib/Coq.Unicode.Utf8\_core}{\coqdocnotation{∀}} \coqexternalref{:type scope:'xE2x88x80' x '..' x ',' x}{http://coq.inria.fr/distrib/8.5beta2/stdlib/Coq.Unicode.Utf8\_core}{\coqdocnotation{(}}\coqdocvar{A} : \coqdockw{Type}) (\coqdocvar{p} : \coqdocvariable{A} \coqexternalref{:type scope:x 'xE2x86x92' x}{http://coq.inria.fr/distrib/8.5beta2/stdlib/Coq.Unicode.Utf8\_core}{\coqdocnotation{→}} \coqexternalref{bool}{http://coq.inria.fr/distrib/8.5beta2/stdlib/Coq.Init.Datatypes}{\coqdocinductive{bool}}) (\coqdocvar{xs} : \coqexternalref{list}{http://coq.inria.fr/distrib/8.5beta2/stdlib/Coq.Init.Datatypes}{\coqdocinductive{list}} \coqdocvariable{A}\coqexternalref{:type scope:'xE2x88x80' x '..' x ',' x}{http://coq.inria.fr/distrib/8.5beta2/stdlib/Coq.Unicode.Utf8\_core}{\coqdocnotation{),}} \coqdocvariable{P} \coqdocvariable{A} \coqdocvariable{p} \coqdocvariable{xs} (\coqref{EquationsWith.sublist}{\coqdocdefinition{sublist}} \coqdocvariable{p} \coqdocvariable{xs}).\coqdoceol
\coqdocemptyline
\end{coqdoccode}
The resulting elimination principle, while maybe not so useful in
   that case as this program constructs a \textit{proof}, shows the explicit
   rewriting needed in the definition of the subpredicate \coqdocvariable{P0}. \begin{coqdoccode}
\end{coqdoccode}

This concludes our exposition of \Equations and we now turn to the
formalization of Predicative System F.

\section{Typing and reduction}
\label{sec:typing-reduction}
\begin{coqdoccode}
\coqdocemptyline
\coqdocemptyline
\coqdocemptyline
\end{coqdoccode}
\subsection{Definition of terms}

 Recall that Predicative System F is a typed lambda calculus with
  type abstractions and applications. Our type structure is very simple
  here, with just the function space and universal quantification on
  kinded type variables. We use an absolutely standard de Bruijn
  encoding for type and term variables. The kinds (a.k.a universe
  levels) are represented using natural numbers. Our development is
  based on Jérôme Vouillon's solution to the POPLmark challenge for
  System $F^{sub}$ \cite{vouillonpoplmark}. \begin{coqdoccode}
\coqdocemptyline
\coqdocnoindent
\coqdockw{Definition} \coqdef{Fpred.Definitions.kind}{kind}{\coqdocdefinition{kind}} := \coqexternalref{nat}{http://coq.inria.fr/distrib/8.5beta2/stdlib/Coq.Init.Datatypes}{\coqdocinductive{nat}}.\coqdoceol
\coqdocemptyline
\coqdocnoindent
\coqdockw{Inductive} \coqdef{Fpred.Definitions.typ}{typ}{\coqdocinductive{typ}} : \coqdockw{Set} :=\coqdoceol
\coqdocindent{1.00em}
\ensuremath{|} \coqdef{Fpred.Definitions.tvar}{tvar}{\coqdocconstructor{tvar}} : \coqexternalref{nat}{http://coq.inria.fr/distrib/8.5beta2/stdlib/Coq.Init.Datatypes}{\coqdocinductive{nat}} \coqexternalref{:type scope:x '->' x}{http://coq.inria.fr/distrib/8.5beta2/stdlib/Coq.Init.Logic}{\coqdocnotation{\ensuremath{\rightarrow}}} \coqref{Fpred.Definitions.typ}{\coqdocinductive{typ}} \ensuremath{|} \coqdef{Fpred.Definitions.arrow}{arrow}{\coqdocconstructor{arrow}} : \coqref{Fpred.Definitions.typ}{\coqdocinductive{typ}} \coqexternalref{:type scope:x '->' x}{http://coq.inria.fr/distrib/8.5beta2/stdlib/Coq.Init.Logic}{\coqdocnotation{\ensuremath{\rightarrow}}} \coqref{Fpred.Definitions.typ}{\coqdocinductive{typ}} \coqexternalref{:type scope:x '->' x}{http://coq.inria.fr/distrib/8.5beta2/stdlib/Coq.Init.Logic}{\coqdocnotation{\ensuremath{\rightarrow}}} \coqref{Fpred.Definitions.typ}{\coqdocinductive{typ}} \ensuremath{|} \coqdef{Fpred.Definitions.all}{all}{\coqdocconstructor{all}} : \coqref{Fpred.Definitions.kind}{\coqdocdefinition{kind}} \coqexternalref{:type scope:x '->' x}{http://coq.inria.fr/distrib/8.5beta2/stdlib/Coq.Init.Logic}{\coqdocnotation{\ensuremath{\rightarrow}}} \coqref{Fpred.Definitions.typ}{\coqdocinductive{typ}} \coqexternalref{:type scope:x '->' x}{http://coq.inria.fr/distrib/8.5beta2/stdlib/Coq.Init.Logic}{\coqdocnotation{\ensuremath{\rightarrow}}} \coqref{Fpred.Definitions.typ}{\coqdocinductive{typ}}.\coqdoceol
\end{coqdoccode}
We will write ∀ \coqdocvariable{X} :* \coqdocvariable{k}. \coqdocvariable{T} for the quantification over types of
    kind \coqdocvariable{k}. \begin{coqdoccode}
\coqdocemptyline
\coqdocnoindent
\coqdockw{Inductive} \coqdef{Fpred.Definitions.term}{term}{\coqdocinductive{term}} : \coqdockw{Set} :=\coqdoceol
\coqdocindent{1.00em}
\ensuremath{|} \coqdef{Fpred.Definitions.var}{var}{\coqdocconstructor{var}} : \coqexternalref{nat}{http://coq.inria.fr/distrib/8.5beta2/stdlib/Coq.Init.Datatypes}{\coqdocinductive{nat}} \coqexternalref{:type scope:x '->' x}{http://coq.inria.fr/distrib/8.5beta2/stdlib/Coq.Init.Logic}{\coqdocnotation{\ensuremath{\rightarrow}}} \coqref{Fpred.Definitions.term}{\coqdocinductive{term}}\coqdoceol
\coqdocindent{1.00em}
\ensuremath{|} \coqdef{Fpred.Definitions.abs}{abs}{\coqdocconstructor{abs}} : \coqref{Fpred.Definitions.typ}{\coqdocinductive{typ}} \coqexternalref{:type scope:x '->' x}{http://coq.inria.fr/distrib/8.5beta2/stdlib/Coq.Init.Logic}{\coqdocnotation{\ensuremath{\rightarrow}}} \coqref{Fpred.Definitions.term}{\coqdocinductive{term}} \coqexternalref{:type scope:x '->' x}{http://coq.inria.fr/distrib/8.5beta2/stdlib/Coq.Init.Logic}{\coqdocnotation{\ensuremath{\rightarrow}}} \coqref{Fpred.Definitions.term}{\coqdocinductive{term}}\coqdoceol
\coqdocindent{1.00em}
\ensuremath{|} \coqdef{Fpred.Definitions.app}{app}{\coqdocconstructor{app}} : \coqref{Fpred.Definitions.term}{\coqdocinductive{term}} \coqexternalref{:type scope:x '->' x}{http://coq.inria.fr/distrib/8.5beta2/stdlib/Coq.Init.Logic}{\coqdocnotation{\ensuremath{\rightarrow}}} \coqref{Fpred.Definitions.term}{\coqdocinductive{term}} \coqexternalref{:type scope:x '->' x}{http://coq.inria.fr/distrib/8.5beta2/stdlib/Coq.Init.Logic}{\coqdocnotation{\ensuremath{\rightarrow}}} \coqref{Fpred.Definitions.term}{\coqdocinductive{term}}\coqdoceol
\coqdocindent{1.00em}
\ensuremath{|} \coqdef{Fpred.Definitions.tabs}{tabs}{\coqdocconstructor{tabs}} : \coqref{Fpred.Definitions.kind}{\coqdocdefinition{kind}} \coqexternalref{:type scope:x '->' x}{http://coq.inria.fr/distrib/8.5beta2/stdlib/Coq.Init.Logic}{\coqdocnotation{\ensuremath{\rightarrow}}} \coqref{Fpred.Definitions.term}{\coqdocinductive{term}} \coqexternalref{:type scope:x '->' x}{http://coq.inria.fr/distrib/8.5beta2/stdlib/Coq.Init.Logic}{\coqdocnotation{\ensuremath{\rightarrow}}} \coqref{Fpred.Definitions.term}{\coqdocinductive{term}}\coqdoceol
\coqdocindent{1.00em}
\ensuremath{|} \coqdef{Fpred.Definitions.tapp}{tapp}{\coqdocconstructor{tapp}} : \coqref{Fpred.Definitions.term}{\coqdocinductive{term}} \coqexternalref{:type scope:x '->' x}{http://coq.inria.fr/distrib/8.5beta2/stdlib/Coq.Init.Logic}{\coqdocnotation{\ensuremath{\rightarrow}}} \coqref{Fpred.Definitions.typ}{\coqdocinductive{typ}} \coqexternalref{:type scope:x '->' x}{http://coq.inria.fr/distrib/8.5beta2/stdlib/Coq.Init.Logic}{\coqdocnotation{\ensuremath{\rightarrow}}} \coqref{Fpred.Definitions.term}{\coqdocinductive{term}}.\coqdoceol
\coqdocemptyline
\end{coqdoccode}
Our raw terms are simply the abstract syntax trees. \subsection{Shiftings and substitutions}

 We define the different operations of shifting and substitutions
    with the \Equations package, we only show the substitution function
    here which uses a \coqdockw{with} right-hand side. All the development can be downloaded
    or browsed at \url{http://equations-fpred.gforge.inria.fr}. \begin{coqdoccode}
\coqdocemptyline
\coqdocemptyline
\coqdocnoindent
\coqdockw{Check} \coqref{Fpred.Definitions.shift typ}{\coqdocdefinition{shift\_typ}} : \coqexternalref{:type scope:'xE2x88x80' x '..' x ',' x}{http://coq.inria.fr/distrib/8.5beta2/stdlib/Coq.Unicode.Utf8\_core}{\coqdocnotation{∀}} \coqexternalref{:type scope:'xE2x88x80' x '..' x ',' x}{http://coq.inria.fr/distrib/8.5beta2/stdlib/Coq.Unicode.Utf8\_core}{\coqdocnotation{(}}\coqdocvar{X} : \coqexternalref{nat}{http://coq.inria.fr/distrib/8.5beta2/stdlib/Coq.Init.Datatypes}{\coqdocinductive{nat}}) (\coqdocvar{t} : \coqref{Fpred.Definitions.term}{\coqdocinductive{term}}\coqexternalref{:type scope:'xE2x88x80' x '..' x ',' x}{http://coq.inria.fr/distrib/8.5beta2/stdlib/Coq.Unicode.Utf8\_core}{\coqdocnotation{),}} \coqref{Fpred.Definitions.term}{\coqdocinductive{term}}.\coqdoceol
\coqdocnoindent
\coqdockw{Check} \coqref{Fpred.Definitions.tsubst}{\coqdocdefinition{tsubst}} : \coqref{Fpred.Definitions.typ}{\coqdocinductive{typ}} \coqexternalref{:type scope:x '->' x}{http://coq.inria.fr/distrib/8.5beta2/stdlib/Coq.Init.Logic}{\coqdocnotation{\ensuremath{\rightarrow}}} \coqexternalref{nat}{http://coq.inria.fr/distrib/8.5beta2/stdlib/Coq.Init.Datatypes}{\coqdocinductive{nat}} \coqexternalref{:type scope:x '->' x}{http://coq.inria.fr/distrib/8.5beta2/stdlib/Coq.Init.Logic}{\coqdocnotation{\ensuremath{\rightarrow}}} \coqref{Fpred.Definitions.typ}{\coqdocinductive{typ}} \coqexternalref{:type scope:x '->' x}{http://coq.inria.fr/distrib/8.5beta2/stdlib/Coq.Init.Logic}{\coqdocnotation{\ensuremath{\rightarrow}}} \coqref{Fpred.Definitions.typ}{\coqdocinductive{typ}}.\coqdoceol
\coqdocemptyline
\coqdocnoindent
\coqdockw{Equations} \coqdef{Fpred.Definitions.subst}{subst}{\coqdocdefinition{subst}} (\coqdocvar{t} : \coqref{Fpred.Definitions.term}{\coqdocinductive{term}}) (\coqdocvar{x} : \coqexternalref{nat}{http://coq.inria.fr/distrib/8.5beta2/stdlib/Coq.Init.Datatypes}{\coqdocinductive{nat}}) (\coqdocvar{t'} : \coqref{Fpred.Definitions.term}{\coqdocinductive{term}}) : \coqref{Fpred.Definitions.term}{\coqdocinductive{term}} :=\coqdoceol
\coqdocnoindent
\coqref{Fpred.Definitions.subst}{\coqdocdefinition{subst}} (\coqref{Fpred.Definitions.var}{\coqdocconstructor{var}} \coqref{Fpred.Definitions.var}{\coqdocconstructor{y}}) \coqdocvar{x} \coqdocvar{t'} $\Leftarrow$ \coqexternalref{lt eq lt dec}{http://coq.inria.fr/distrib/8.5beta2/stdlib/Coq.Arith.Compare\_dec}{\coqdocdefinition{lt\_eq\_lt\_dec}} \coqdocvar{y} \coqdocvar{x} \ensuremath{\Rightarrow} \{\coqdoceol
\coqdocindent{1.00em}
\ensuremath{|} \coqexternalref{inleft}{http://coq.inria.fr/distrib/8.5beta2/stdlib/Coq.Init.Specif}{\coqdocconstructor{inleft}} \coqexternalref{inleft}{http://coq.inria.fr/distrib/8.5beta2/stdlib/Coq.Init.Specif}{\coqdocconstructor{(}}\coqexternalref{inleft}{http://coq.inria.fr/distrib/8.5beta2/stdlib/Coq.Init.Specif}{\coqdocconstructor{left}}  \coqexternalref{inleft}{http://coq.inria.fr/distrib/8.5beta2/stdlib/Coq.Init.Specif}{\coqdocconstructor{\_}}\coqexternalref{inleft}{http://coq.inria.fr/distrib/8.5beta2/stdlib/Coq.Init.Specif}{\coqdocconstructor{)}} \ensuremath{\Rightarrow} \coqref{Fpred.Definitions.var}{\coqdocconstructor{var}} \coqdocvar{y};\coqdoceol
\coqdocindent{1.00em}
\ensuremath{|} \coqexternalref{inleft}{http://coq.inria.fr/distrib/8.5beta2/stdlib/Coq.Init.Specif}{\coqdocconstructor{inleft}} \coqexternalref{inleft}{http://coq.inria.fr/distrib/8.5beta2/stdlib/Coq.Init.Specif}{\coqdocconstructor{(}}\coqexternalref{inleft}{http://coq.inria.fr/distrib/8.5beta2/stdlib/Coq.Init.Specif}{\coqdocconstructor{right}} \coqexternalref{inleft}{http://coq.inria.fr/distrib/8.5beta2/stdlib/Coq.Init.Specif}{\coqdocconstructor{\_}}\coqexternalref{inleft}{http://coq.inria.fr/distrib/8.5beta2/stdlib/Coq.Init.Specif}{\coqdocconstructor{)}} \ensuremath{\Rightarrow} \coqdocvar{t'};\coqdoceol
\coqdocindent{1.00em}
\ensuremath{|} \coqexternalref{inright}{http://coq.inria.fr/distrib/8.5beta2/stdlib/Coq.Init.Specif}{\coqdocconstructor{inright}} \coqexternalref{inright}{http://coq.inria.fr/distrib/8.5beta2/stdlib/Coq.Init.Specif}{\coqdocconstructor{\_}}        \ensuremath{\Rightarrow} \coqref{Fpred.Definitions.var}{\coqdocconstructor{var}} (\coqdocvar{y} \coqexternalref{:nat scope:x '-' x}{http://coq.inria.fr/distrib/8.5beta2/stdlib/Coq.Init.Peano}{\coqdocnotation{-}} 1) \};\coqdoceol
\coqdocnoindent
\coqref{Fpred.Definitions.subst}{\coqdocdefinition{subst}} (\coqref{Fpred.Definitions.abs}{\coqdocconstructor{abs}} \coqref{Fpred.Definitions.abs}{\coqdocconstructor{T1}} \coqref{Fpred.Definitions.abs}{\coqdocconstructor{t2}})  \coqdocvar{x} \coqdocvar{t'} \ensuremath{\Rightarrow} \coqref{Fpred.Definitions.abs}{\coqdocconstructor{abs}} \coqdocvar{T1} (\coqref{Fpred.Definitions.subst}{\coqdocdefinition{subst}} \coqdocvar{t2} (1 \coqexternalref{:nat scope:x '+' x}{http://coq.inria.fr/distrib/8.5beta2/stdlib/Coq.Init.Peano}{\coqdocnotation{+}} \coqdocvar{x}) (\coqref{Fpred.Definitions.shift}{\coqdocdefinition{shift}} 0 \coqdocvar{t'}));\coqdoceol
\coqdocnoindent
\coqref{Fpred.Definitions.subst}{\coqdocdefinition{subst}} (\coqref{Fpred.Definitions.app}{\coqdocconstructor{app}} \coqref{Fpred.Definitions.app}{\coqdocconstructor{t1}} \coqref{Fpred.Definitions.app}{\coqdocconstructor{t2}})  \coqdocvar{x} \coqdocvar{t'} \ensuremath{\Rightarrow} \coqref{Fpred.Definitions.app}{\coqdocconstructor{app}} (\coqref{Fpred.Definitions.subst}{\coqdocdefinition{subst}} \coqdocvar{t1} \coqdocvar{x} \coqdocvar{t'}) (\coqref{Fpred.Definitions.subst}{\coqdocdefinition{subst}} \coqdocvar{t2} \coqdocvar{x} \coqdocvar{t'});\coqdoceol
\coqdocnoindent
\coqref{Fpred.Definitions.subst}{\coqdocdefinition{subst}} (\coqref{Fpred.Definitions.tabs}{\coqdocconstructor{tabs}} \coqref{Fpred.Definitions.tabs}{\coqdocconstructor{k}} \coqref{Fpred.Definitions.tabs}{\coqdocconstructor{t2}})  \coqdocvar{x} \coqdocvar{t'} \ensuremath{\Rightarrow} \coqref{Fpred.Definitions.tabs}{\coqdocconstructor{tabs}} \coqdocvar{k} (\coqref{Fpred.Definitions.subst}{\coqdocdefinition{subst}} \coqdocvar{t2} \coqdocvar{x} (\coqref{Fpred.Definitions.shift typ}{\coqdocdefinition{shift\_typ}} 0 \coqdocvar{t'}));\coqdoceol
\coqdocnoindent
\coqref{Fpred.Definitions.subst}{\coqdocdefinition{subst}} (\coqref{Fpred.Definitions.tapp}{\coqdocconstructor{tapp}} \coqref{Fpred.Definitions.tapp}{\coqdocconstructor{t1}} \coqref{Fpred.Definitions.tapp}{\coqdocconstructor{T2}}) \coqdocvar{x} \coqdocvar{t'} \ensuremath{\Rightarrow} \coqref{Fpred.Definitions.tapp}{\coqdocconstructor{tapp}} (\coqref{Fpred.Definitions.subst}{\coqdocdefinition{subst}} \coqdocvar{t1} \coqdocvar{x} \coqdocvar{t'}) \coqdocvar{T2}.\coqdoceol
\coqdocemptyline
\coqdocemptyline
\coqdocnoindent
\coqdockw{Check} \coqref{Fpred.Definitions.subst typ}{\coqdocdefinition{subst\_typ}} : \coqref{Fpred.Definitions.term}{\coqdocinductive{term}} \coqexternalref{:type scope:x '->' x}{http://coq.inria.fr/distrib/8.5beta2/stdlib/Coq.Init.Logic}{\coqdocnotation{\ensuremath{\rightarrow}}} \coqexternalref{nat}{http://coq.inria.fr/distrib/8.5beta2/stdlib/Coq.Init.Datatypes}{\coqdocinductive{nat}} \coqexternalref{:type scope:x '->' x}{http://coq.inria.fr/distrib/8.5beta2/stdlib/Coq.Init.Logic}{\coqdocnotation{\ensuremath{\rightarrow}}} \coqref{Fpred.Definitions.typ}{\coqdocinductive{typ}} \coqexternalref{:type scope:x '->' x}{http://coq.inria.fr/distrib/8.5beta2/stdlib/Coq.Init.Logic}{\coqdocnotation{\ensuremath{\rightarrow}}} \coqref{Fpred.Definitions.term}{\coqdocinductive{term}}.\coqdoceol
\coqdocemptyline
\coqdocemptyline
\end{coqdoccode}
\subsection{Contexts}

 We define the contexts \coqref{Fpred.Definitions.env}{\coqdocinductive{env}} and the two functions \coqref{Fpred.Definitions.get kind}{\coqdocdefinition{get\_kind}} and
   \coqref{Fpred.Definitions.get var}{\coqdocdefinition{get\_var}} which access the context. A context is an interleaving of
   types and terms contexts. Vouillon's great idea is to have parallel
   de Bruijn indexings for type and term variables, which means separate
   indices for type and term variables. That way, shifting and
   substitution of one kind does not influence the other, making weakening
   and substitution lemmas much simpler, we follow this idea here. \begin{coqdoccode}
\coqdocemptyline
\coqdocnoindent
\coqdockw{Inductive} \coqdef{Fpred.Definitions.env}{env}{\coqdocinductive{env}} : \coqdockw{Set} :=\coqdoceol
\coqdocindent{1.00em}
\ensuremath{|} \coqdef{Fpred.Definitions.empty}{empty}{\coqdocconstructor{empty}} : \coqref{Fpred.Definitions.env}{\coqdocinductive{env}} \ensuremath{|} \coqdef{Fpred.Definitions.evar}{evar}{\coqdocconstructor{evar}} : \coqref{Fpred.Definitions.env}{\coqdocinductive{env}} \coqexternalref{:type scope:x '->' x}{http://coq.inria.fr/distrib/8.5beta2/stdlib/Coq.Init.Logic}{\coqdocnotation{\ensuremath{\rightarrow}}} \coqref{Fpred.Definitions.typ}{\coqdocinductive{typ}} \coqexternalref{:type scope:x '->' x}{http://coq.inria.fr/distrib/8.5beta2/stdlib/Coq.Init.Logic}{\coqdocnotation{\ensuremath{\rightarrow}}} \coqref{Fpred.Definitions.env}{\coqdocinductive{env}} \ensuremath{|} \coqdef{Fpred.Definitions.etvar}{etvar}{\coqdocconstructor{etvar}} : \coqref{Fpred.Definitions.env}{\coqdocinductive{env}} \coqexternalref{:type scope:x '->' x}{http://coq.inria.fr/distrib/8.5beta2/stdlib/Coq.Init.Logic}{\coqdocnotation{\ensuremath{\rightarrow}}} \coqref{Fpred.Definitions.kind}{\coqdocdefinition{kind}} \coqexternalref{:type scope:x '->' x}{http://coq.inria.fr/distrib/8.5beta2/stdlib/Coq.Init.Logic}{\coqdocnotation{\ensuremath{\rightarrow}}} \coqref{Fpred.Definitions.env}{\coqdocinductive{env}}.\coqdoceol
\coqdocemptyline
\end{coqdoccode}
Note that \Equations allows wildcards and overlapping clauses with a
  first match semantics, as usual. \begin{coqdoccode}
\coqdocemptyline
\coqdocnoindent
\coqdockw{Equations}(\coqdocvar{nocomp}) \coqdef{Fpred.Definitions.get kind}{get\_kind}{\coqdocdefinition{get\_kind}} (\coqdocvar{e} : \coqref{Fpred.Definitions.env}{\coqdocinductive{env}}) (\coqdocvar{X} : \coqexternalref{nat}{http://coq.inria.fr/distrib/8.5beta2/stdlib/Coq.Init.Datatypes}{\coqdocinductive{nat}}) : \coqexternalref{option}{http://coq.inria.fr/distrib/8.5beta2/stdlib/Coq.Init.Datatypes}{\coqdocinductive{option}} \coqref{Fpred.Definitions.kind}{\coqdocdefinition{kind}} :=\coqdoceol
\coqdocnoindent
\coqref{Fpred.Definitions.get kind}{\coqdocdefinition{get\_kind}}  \coqref{Fpred.Definitions.empty}{\coqdocconstructor{empty}}       \coqdocvar{\_}    \ensuremath{\Rightarrow} \coqexternalref{None}{http://coq.inria.fr/distrib/8.5beta2/stdlib/Coq.Init.Datatypes}{\coqdocconstructor{None}};\coqdoceol
\coqdocnoindent
\coqref{Fpred.Definitions.get kind}{\coqdocdefinition{get\_kind}} (\coqref{Fpred.Definitions.evar}{\coqdocconstructor{evar}} \coqref{Fpred.Definitions.evar}{\coqdocconstructor{e}} \coqref{Fpred.Definitions.evar}{\coqdocconstructor{\_}})   \coqdocvar{X}    \ensuremath{\Rightarrow} \coqref{Fpred.Definitions.get kind}{\coqdocdefinition{get\_kind}} \coqdocvar{e} \coqdocvar{X};\coqdoceol
\coqdocnoindent
\coqref{Fpred.Definitions.get kind}{\coqdocdefinition{get\_kind}} (\coqref{Fpred.Definitions.etvar}{\coqdocconstructor{etvar}} \coqref{Fpred.Definitions.etvar}{\coqdocconstructor{\_}} \coqref{Fpred.Definitions.etvar}{\coqdocconstructor{T}})  \coqexternalref{O}{http://coq.inria.fr/distrib/8.5beta2/stdlib/Coq.Init.Datatypes}{\coqdocconstructor{O}}    \ensuremath{\Rightarrow} \coqexternalref{Some}{http://coq.inria.fr/distrib/8.5beta2/stdlib/Coq.Init.Datatypes}{\coqdocconstructor{Some}} \coqdocvar{T};\coqdoceol
\coqdocnoindent
\coqref{Fpred.Definitions.get kind}{\coqdocdefinition{get\_kind}} (\coqref{Fpred.Definitions.etvar}{\coqdocconstructor{etvar}} \coqref{Fpred.Definitions.etvar}{\coqdocconstructor{e}} \coqref{Fpred.Definitions.etvar}{\coqdocconstructor{\_}}) (\coqexternalref{S}{http://coq.inria.fr/distrib/8.5beta2/stdlib/Coq.Init.Datatypes}{\coqdocconstructor{S}} \coqexternalref{S}{http://coq.inria.fr/distrib/8.5beta2/stdlib/Coq.Init.Datatypes}{\coqdocconstructor{X}}) \ensuremath{\Rightarrow} \coqref{Fpred.Definitions.get kind}{\coqdocdefinition{get\_kind}} \coqdocvar{e} \coqdocvar{X}.\coqdoceol
\coqdocemptyline
\end{coqdoccode}
We need the functorial map on \coqexternalref{option}{http://coq.inria.fr/distrib/8.5beta2/stdlib/Coq.Init.Datatypes}{\coqdocinductive{option}} types to ease writing these
  partial lookup functions. \begin{coqdoccode}
\coqdocemptyline
\coqdocnoindent
\coqdockw{Equations} \coqdef{Fpred.Definitions.opt map}{opt\_map}{\coqdocdefinition{opt\_map}} (\coqdocvar{A} \coqdocvar{B} : \coqdockw{Set}) (\coqdocvar{f} : \coqdocvariable{A} \coqexternalref{:type scope:x '->' x}{http://coq.inria.fr/distrib/8.5beta2/stdlib/Coq.Init.Logic}{\coqdocnotation{\ensuremath{\rightarrow}}} \coqdocvariable{B}) (\coqdocvar{x} : \coqexternalref{option}{http://coq.inria.fr/distrib/8.5beta2/stdlib/Coq.Init.Datatypes}{\coqdocinductive{option}} \coqdocvariable{A}) : \coqexternalref{option}{http://coq.inria.fr/distrib/8.5beta2/stdlib/Coq.Init.Datatypes}{\coqdocinductive{option}} \coqdocvar{B} :=\coqdoceol
\coqdocnoindent
\coqref{Fpred.Definitions.opt map}{\coqdocdefinition{opt\_map}} \coqdocvar{\_} \coqdocvar{\_} \coqdocvar{f} (\coqexternalref{Some}{http://coq.inria.fr/distrib/8.5beta2/stdlib/Coq.Init.Datatypes}{\coqdocconstructor{Some}} \coqexternalref{Some}{http://coq.inria.fr/distrib/8.5beta2/stdlib/Coq.Init.Datatypes}{\coqdocconstructor{x}}) \ensuremath{\Rightarrow} \coqexternalref{Some}{http://coq.inria.fr/distrib/8.5beta2/stdlib/Coq.Init.Datatypes}{\coqdocconstructor{Some}} (\coqdocvar{f} \coqdocvar{x});\coqdoceol
\coqdocnoindent
\coqref{Fpred.Definitions.opt map}{\coqdocdefinition{opt\_map}} \coqdocvar{\_} \coqdocvar{\_} \coqdocvar{\_}  \coqexternalref{None}{http://coq.inria.fr/distrib/8.5beta2/stdlib/Coq.Init.Datatypes}{\coqdocconstructor{None}}    \ensuremath{\Rightarrow} \coqexternalref{None}{http://coq.inria.fr/distrib/8.5beta2/stdlib/Coq.Init.Datatypes}{\coqdocconstructor{None}}.\coqdoceol
\coqdocemptyline
\coqdocnoindent
\coqdockw{Equations}(\coqdocvar{nocomp}) \coqdef{Fpred.Definitions.get var}{get\_var}{\coqdocdefinition{get\_var}} (\coqdocvar{e} : \coqref{Fpred.Definitions.env}{\coqdocinductive{env}}) (\coqdocvar{x} : \coqexternalref{nat}{http://coq.inria.fr/distrib/8.5beta2/stdlib/Coq.Init.Datatypes}{\coqdocinductive{nat}}) : \coqexternalref{option}{http://coq.inria.fr/distrib/8.5beta2/stdlib/Coq.Init.Datatypes}{\coqdocinductive{option}} \coqref{Fpred.Definitions.typ}{\coqdocinductive{typ}} :=\coqdoceol
\coqdocnoindent
\coqref{Fpred.Definitions.get var}{\coqdocdefinition{get\_var}}  \coqref{Fpred.Definitions.empty}{\coqdocconstructor{empty}}       \coqdocvar{\_}    \ensuremath{\Rightarrow} \coqexternalref{None}{http://coq.inria.fr/distrib/8.5beta2/stdlib/Coq.Init.Datatypes}{\coqdocconstructor{None}};\coqdoceol
\coqdocnoindent
\coqref{Fpred.Definitions.get var}{\coqdocdefinition{get\_var}} (\coqref{Fpred.Definitions.etvar}{\coqdocconstructor{etvar}} \coqref{Fpred.Definitions.etvar}{\coqdocconstructor{e}} \coqref{Fpred.Definitions.etvar}{\coqdocconstructor{\_}})  \coqdocvar{x}    \ensuremath{\Rightarrow} \coqref{Fpred.Definitions.opt map}{\coqdocdefinition{opt\_map}} (\coqref{Fpred.Definitions.tshift}{\coqdocdefinition{tshift}} 0) (\coqref{Fpred.Definitions.get var}{\coqdocdefinition{get\_var}} \coqdocvar{e} \coqdocvar{x});\coqdoceol
\coqdocnoindent
\coqref{Fpred.Definitions.get var}{\coqdocdefinition{get\_var}} (\coqref{Fpred.Definitions.evar}{\coqdocconstructor{evar}}  \coqref{Fpred.Definitions.evar}{\coqdocconstructor{\_}} \coqref{Fpred.Definitions.evar}{\coqdocconstructor{T}})  \coqexternalref{O}{http://coq.inria.fr/distrib/8.5beta2/stdlib/Coq.Init.Datatypes}{\coqdocconstructor{O}}    \ensuremath{\Rightarrow} \coqexternalref{Some}{http://coq.inria.fr/distrib/8.5beta2/stdlib/Coq.Init.Datatypes}{\coqdocconstructor{Some}} \coqdocvar{T};\coqdoceol
\coqdocnoindent
\coqref{Fpred.Definitions.get var}{\coqdocdefinition{get\_var}} (\coqref{Fpred.Definitions.evar}{\coqdocconstructor{evar}}  \coqref{Fpred.Definitions.evar}{\coqdocconstructor{e}} \coqref{Fpred.Definitions.evar}{\coqdocconstructor{\_}}) (\coqexternalref{S}{http://coq.inria.fr/distrib/8.5beta2/stdlib/Coq.Init.Datatypes}{\coqdocconstructor{S}} \coqexternalref{S}{http://coq.inria.fr/distrib/8.5beta2/stdlib/Coq.Init.Datatypes}{\coqdocconstructor{x}}) \ensuremath{\Rightarrow} \coqref{Fpred.Definitions.get var}{\coqdocdefinition{get\_var}} \coqdocvar{e} \coqdocvar{x}.\coqdoceol
\coqdocemptyline
\end{coqdoccode}
\subsection{Well-formedness conditions}

  We also define some well-formedness conditions for types, terms
  and contexts. Namely, in a type (resp. in a term), the variables
  must all be kinded (resp. typed). We just show the \coqref{Fpred.Definitions.wf typ}{\coqdocdefinition{wf\_typ}} definition here,
  those follow Stump and Haye's work. \begin{coqdoccode}
\coqdocemptyline
\coqdocnoindent
\coqdockw{Equations} \coqdef{Fpred.Definitions.wf typ}{wf\_typ}{\coqdocdefinition{wf\_typ}} (\coqdocvar{e} : \coqref{Fpred.Definitions.env}{\coqdocinductive{env}}) (\coqdocvar{T} : \coqref{Fpred.Definitions.typ}{\coqdocinductive{typ}}) : \coqdockw{Prop} :=\coqdoceol
\coqdocnoindent
\coqref{Fpred.Definitions.wf typ}{\coqdocdefinition{wf\_typ}} \coqdocvar{e} (\coqref{Fpred.Definitions.tvar}{\coqdocconstructor{tvar}} \coqref{Fpred.Definitions.tvar}{\coqdocconstructor{X}})      \ensuremath{\Rightarrow} \coqref{Fpred.Definitions.get kind}{\coqdocdefinition{get\_kind}} \coqdocvar{e} \coqdocvar{X} \coqexternalref{:type scope:x '<>' x}{http://coq.inria.fr/distrib/8.5beta2/stdlib/Coq.Init.Logic}{\coqdocnotation{\ensuremath{\not=}}} \coqexternalref{None}{http://coq.inria.fr/distrib/8.5beta2/stdlib/Coq.Init.Datatypes}{\coqdocconstructor{None}};\coqdoceol
\coqdocnoindent
\coqref{Fpred.Definitions.wf typ}{\coqdocdefinition{wf\_typ}} \coqdocvar{e} (\coqref{Fpred.Definitions.arrow}{\coqdocconstructor{arrow}} \coqref{Fpred.Definitions.arrow}{\coqdocconstructor{T1}} \coqref{Fpred.Definitions.arrow}{\coqdocconstructor{T2}}) \ensuremath{\Rightarrow} \coqref{Fpred.Definitions.wf typ}{\coqdocdefinition{wf\_typ}} \coqdocvar{e} \coqdocvar{T1} \coqexternalref{:type scope:x '/x5C' x}{http://coq.inria.fr/distrib/8.5beta2/stdlib/Coq.Init.Logic}{\coqdocnotation{\ensuremath{\land}}} \coqref{Fpred.Definitions.wf typ}{\coqdocdefinition{wf\_typ}} \coqdocvar{e} \coqdocvar{T2};\coqdoceol
\coqdocnoindent
\coqref{Fpred.Definitions.wf typ}{\coqdocdefinition{wf\_typ}} \coqdocvar{e} (\coqref{Fpred.Definitions.all}{\coqdocconstructor{all}} \coqref{Fpred.Definitions.all}{\coqdocconstructor{k}} \coqref{Fpred.Definitions.all}{\coqdocconstructor{T2}})    \ensuremath{\Rightarrow} \coqref{Fpred.Definitions.wf typ}{\coqdocdefinition{wf\_typ}} (\coqref{Fpred.Definitions.etvar}{\coqdocconstructor{etvar}} \coqdocvar{e} \coqdocvar{k}) \coqdocvar{T2}.\coqdoceol
\coqdocemptyline
\coqdocemptyline
\end{coqdoccode}
\subsection{Kinding and typing rules}

  The kinding rules are the main difference between Leivant's and
  Stump's presentations. We refer to these works for pen and paper
  presentations of these systems, due to lack of space, we cannot
  include them here. The case for universal quantification sets the
  level of a universal type at 1 + \coqexternalref{max}{http://coq.inria.fr/distrib/8.5beta2/stdlib/Coq.Init.Peano}{\coqdocabbreviation{max}} \coqdocvariable{k} \coqdocvariable{k'}, where \coqdocvariable{k} and \coqdocvariable{k'} are
  respectively the domain and codomain kinds, in Stump's case, which
  allows for a straightforward order on types based on levels, but this
  means that each level is not closed under products from lower levels
  anymore. In other words, multiple quantifications at the same level
  raise the overall level. For example (∀ \coqdocvariable{X} :* 0. \coqdocvariable{X}) :* 1 as expected
  but ∀ \coqdocvariable{X} :* 0. ∀ \coqdocvar{Y} :* 0. \coqdocvariable{X} :* (1 + \coqexternalref{max}{http://coq.inria.fr/distrib/8.5beta2/stdlib/Coq.Init.Peano}{\coqdocabbreviation{max}} 0 (1 + \coqexternalref{max}{http://coq.inria.fr/distrib/8.5beta2/stdlib/Coq.Init.Peano}{\coqdocabbreviation{max}} 0 0)) = 2. This is
  a very strange behavior.

  We use the standard predicative product rule which sets the product
  level to \coqexternalref{max}{http://coq.inria.fr/distrib/8.5beta2/stdlib/Coq.Init.Peano}{\coqdocabbreviation{max}} (\coqdocvariable{k}+1) \coqdocvariable{k'}, which directly corresponds to Martin-Löf's
  Predicative Type Theory. Note that the system includes cumulativity
  through the Var rule which allows to lift a type variable declared at
  level \coqdocvariable{k} into any higher level \coqdocvariable{k'}.  \begin{coqdoccode}
\coqdocemptyline
\coqdocnoindent
\coqdockw{Inductive} \coqdef{Fpred.Definitions.kinding}{kinding}{\coqdocinductive{kinding}} : \coqref{Fpred.Definitions.env}{\coqdocinductive{env}} \coqexternalref{:type scope:x '->' x}{http://coq.inria.fr/distrib/8.5beta2/stdlib/Coq.Init.Logic}{\coqdocnotation{\ensuremath{\rightarrow}}} \coqref{Fpred.Definitions.typ}{\coqdocinductive{typ}} \coqexternalref{:type scope:x '->' x}{http://coq.inria.fr/distrib/8.5beta2/stdlib/Coq.Init.Logic}{\coqdocnotation{\ensuremath{\rightarrow}}} \coqref{Fpred.Definitions.kind}{\coqdocdefinition{kind}} \coqexternalref{:type scope:x '->' x}{http://coq.inria.fr/distrib/8.5beta2/stdlib/Coq.Init.Logic}{\coqdocnotation{\ensuremath{\rightarrow}}} \coqdockw{Prop} :=\coqdoceol
\coqdocnoindent
\ensuremath{|} \coqdef{Fpred.Definitions.T TVar}{T\_TVar}{\coqdocconstructor{T\_TVar}} : \coqdockw{\ensuremath{\forall}} (\coqdocvar{e} : \coqref{Fpred.Definitions.env}{\coqdocinductive{env}}) (\coqdocvar{X} : \coqexternalref{nat}{http://coq.inria.fr/distrib/8.5beta2/stdlib/Coq.Init.Datatypes}{\coqdocinductive{nat}}) (\coqdocvar{k} \coqdocvar{k'} : \coqref{Fpred.Definitions.kind}{\coqdocdefinition{kind}}), \coqref{Fpred.Definitions.wf env}{\coqdocdefinition{wf\_env}} \coqdocvariable{e} \coqexternalref{:type scope:x '->' x}{http://coq.inria.fr/distrib/8.5beta2/stdlib/Coq.Init.Logic}{\coqdocnotation{\ensuremath{\rightarrow}}} \coqdoceol
\coqdocindent{1.50em}
\coqref{Fpred.Definitions.get kind}{\coqdocdefinition{get\_kind}} \coqdocvariable{e} \coqdocvariable{X} \coqexternalref{:type scope:x '=' x}{http://coq.inria.fr/distrib/8.5beta2/stdlib/Coq.Init.Logic}{\coqdocnotation{=}} \coqexternalref{Some}{http://coq.inria.fr/distrib/8.5beta2/stdlib/Coq.Init.Datatypes}{\coqdocconstructor{Some}} \coqdocvariable{k} \coqexternalref{:type scope:x '->' x}{http://coq.inria.fr/distrib/8.5beta2/stdlib/Coq.Init.Logic}{\coqdocnotation{\ensuremath{\rightarrow}}} \coqdocvariable{k} \coqexternalref{:nat scope:x '<=' x}{http://coq.inria.fr/distrib/8.5beta2/stdlib/Coq.Init.Peano}{\coqdocnotation{$\Leftarrow$}} \coqdocvariable{k'} \coqexternalref{:type scope:x '->' x}{http://coq.inria.fr/distrib/8.5beta2/stdlib/Coq.Init.Logic}{\coqdocnotation{\ensuremath{\rightarrow}}} \coqref{Fpred.Definitions.kinding}{\coqdocinductive{kinding}} \coqdocvariable{e} (\coqref{Fpred.Definitions.tvar}{\coqdocconstructor{tvar}} \coqdocvariable{X}) \coqdocvariable{k'}\coqdoceol
\coqdocnoindent
\ensuremath{|} \coqdef{Fpred.Definitions.T Arrow}{T\_Arrow}{\coqdocconstructor{T\_Arrow}} : \coqdockw{\ensuremath{\forall}} \coqdocvar{e} \coqdocvar{T} \coqdocvar{U} \coqdocvar{k} \coqdocvar{k'}, \coqref{Fpred.Definitions.kinding}{\coqdocinductive{kinding}} \coqdocvariable{e} \coqdocvariable{T} \coqdocvariable{k} \coqexternalref{:type scope:x '->' x}{http://coq.inria.fr/distrib/8.5beta2/stdlib/Coq.Init.Logic}{\coqdocnotation{\ensuremath{\rightarrow}}} \coqref{Fpred.Definitions.kinding}{\coqdocinductive{kinding}} \coqdocvariable{e} \coqdocvariable{U} \coqdocvariable{k'} \coqexternalref{:type scope:x '->' x}{http://coq.inria.fr/distrib/8.5beta2/stdlib/Coq.Init.Logic}{\coqdocnotation{\ensuremath{\rightarrow}}} \coqref{Fpred.Definitions.kinding}{\coqdocinductive{kinding}} \coqdocvariable{e} (\coqref{Fpred.Definitions.arrow}{\coqdocconstructor{arrow}} \coqdocvariable{T} \coqdocvariable{U}) (\coqexternalref{max}{http://coq.inria.fr/distrib/8.5beta2/stdlib/Coq.Init.Peano}{\coqdocabbreviation{max}} \coqdocvariable{k} \coqdocvariable{k'})\coqdoceol
\coqdocnoindent
\ensuremath{|} \coqdef{Fpred.Definitions.T All}{T\_All}{\coqdocconstructor{T\_All}} : \coqdockw{\ensuremath{\forall}} \coqdocvar{e} \coqdocvar{T} \coqdocvar{k} \coqdocvar{k'}, \coqref{Fpred.Definitions.kinding}{\coqdocinductive{kinding}} (\coqref{Fpred.Definitions.etvar}{\coqdocconstructor{etvar}} \coqdocvariable{e} \coqdocvariable{k}) \coqdocvariable{T} \coqdocvariable{k'} \coqexternalref{:type scope:x '->' x}{http://coq.inria.fr/distrib/8.5beta2/stdlib/Coq.Init.Logic}{\coqdocnotation{\ensuremath{\rightarrow}}} \coqref{Fpred.Definitions.kinding}{\coqdocinductive{kinding}} \coqdocvariable{e} (\coqref{Fpred.Definitions.all}{\coqdocconstructor{all}} \coqdocvariable{k} \coqdocvariable{T}) (\coqexternalref{max}{http://coq.inria.fr/distrib/8.5beta2/stdlib/Coq.Init.Peano}{\coqdocabbreviation{max}} (\coqdocvariable{k}\coqexternalref{:nat scope:x '+' x}{http://coq.inria.fr/distrib/8.5beta2/stdlib/Coq.Init.Peano}{\coqdocnotation{+}}1) \coqdocvariable{k'}).\coqdoceol
\coqdocemptyline
\end{coqdoccode}
The typing relation is straightforward. Just note that we check 
  for well-formedness of environments at the variable case, so typing
  derivations are always well-formed. \begin{coqdoccode}
\coqdocemptyline
\coqdocnoindent
\coqdockw{Inductive} \coqdef{Fpred.Definitions.typing}{typing}{\coqdocinductive{typing}} : \coqref{Fpred.Definitions.env}{\coqdocinductive{env}} \coqexternalref{:type scope:x '->' x}{http://coq.inria.fr/distrib/8.5beta2/stdlib/Coq.Init.Logic}{\coqdocnotation{\ensuremath{\rightarrow}}} \coqref{Fpred.Definitions.term}{\coqdocinductive{term}} \coqexternalref{:type scope:x '->' x}{http://coq.inria.fr/distrib/8.5beta2/stdlib/Coq.Init.Logic}{\coqdocnotation{\ensuremath{\rightarrow}}} \coqref{Fpred.Definitions.typ}{\coqdocinductive{typ}} \coqexternalref{:type scope:x '->' x}{http://coq.inria.fr/distrib/8.5beta2/stdlib/Coq.Init.Logic}{\coqdocnotation{\ensuremath{\rightarrow}}} \coqdockw{Prop} :=\coqdoceol
\coqdocindent{1.00em}
\ensuremath{|} \coqdef{Fpred.Definitions.T Var}{T\_Var}{\coqdocconstructor{T\_Var}} (\coqdocvar{e} : \coqref{Fpred.Definitions.env}{\coqdocinductive{env}}) (\coqdocvar{x} : \coqexternalref{nat}{http://coq.inria.fr/distrib/8.5beta2/stdlib/Coq.Init.Datatypes}{\coqdocinductive{nat}}) (\coqdocvar{T} : \coqref{Fpred.Definitions.typ}{\coqdocinductive{typ}}) : \coqref{Fpred.Definitions.wf env}{\coqdocdefinition{wf\_env}} \coqdocvariable{e} \coqexternalref{:type scope:x '->' x}{http://coq.inria.fr/distrib/8.5beta2/stdlib/Coq.Init.Logic}{\coqdocnotation{\ensuremath{\rightarrow}}} \coqref{Fpred.Definitions.get var}{\coqdocdefinition{get\_var}} \coqdocvariable{e} \coqdocvariable{x} \coqexternalref{:type scope:x '=' x}{http://coq.inria.fr/distrib/8.5beta2/stdlib/Coq.Init.Logic}{\coqdocnotation{=}} \coqexternalref{Some}{http://coq.inria.fr/distrib/8.5beta2/stdlib/Coq.Init.Datatypes}{\coqdocconstructor{Some}} \coqdocvariable{T} \coqexternalref{:type scope:x '->' x}{http://coq.inria.fr/distrib/8.5beta2/stdlib/Coq.Init.Logic}{\coqdocnotation{\ensuremath{\rightarrow}}} \coqref{Fpred.Definitions.typing}{\coqdocinductive{typing}} \coqdocvariable{e} (\coqref{Fpred.Definitions.var}{\coqdocconstructor{var}} \coqdocvariable{x}) \coqdocvariable{T}\coqdoceol
\coqdocindent{1.00em}
\ensuremath{|} \coqdef{Fpred.Definitions.T Abs}{T\_Abs}{\coqdocconstructor{T\_Abs}} (\coqdocvar{e} : \coqref{Fpred.Definitions.env}{\coqdocinductive{env}}) (\coqdocvar{t} : \coqref{Fpred.Definitions.term}{\coqdocinductive{term}}) (\coqdocvar{T1} \coqdocvar{T2} : \coqref{Fpred.Definitions.typ}{\coqdocinductive{typ}}) : \coqdoceol
\coqdocindent{3.00em}
\coqref{Fpred.Definitions.typing}{\coqdocinductive{typing}} (\coqref{Fpred.Definitions.evar}{\coqdocconstructor{evar}} \coqdocvariable{e} \coqdocvariable{T1}) \coqdocvariable{t} \coqdocvariable{T2} \coqexternalref{:type scope:x '->' x}{http://coq.inria.fr/distrib/8.5beta2/stdlib/Coq.Init.Logic}{\coqdocnotation{\ensuremath{\rightarrow}}} \coqref{Fpred.Definitions.typing}{\coqdocinductive{typing}} \coqdocvariable{e} (\coqref{Fpred.Definitions.abs}{\coqdocconstructor{abs}} \coqdocvariable{T1} \coqdocvariable{t}) (\coqref{Fpred.Definitions.arrow}{\coqdocconstructor{arrow}} \coqdocvariable{T1} \coqdocvariable{T2})\coqdoceol
\coqdocindent{1.00em}
\ensuremath{|} \coqdef{Fpred.Definitions.T App}{T\_App}{\coqdocconstructor{T\_App}} (\coqdocvar{e} : \coqref{Fpred.Definitions.env}{\coqdocinductive{env}}) (\coqdocvar{t1} \coqdocvar{t2} : \coqref{Fpred.Definitions.term}{\coqdocinductive{term}}) (\coqdocvar{T11} \coqdocvar{T12} : \coqref{Fpred.Definitions.typ}{\coqdocinductive{typ}}) :\coqdoceol
\coqdocindent{3.00em}
\coqref{Fpred.Definitions.typing}{\coqdocinductive{typing}} \coqdocvariable{e} \coqdocvariable{t1} (\coqref{Fpred.Definitions.arrow}{\coqdocconstructor{arrow}} \coqdocvariable{T11} \coqdocvariable{T12}) \coqexternalref{:type scope:x '->' x}{http://coq.inria.fr/distrib/8.5beta2/stdlib/Coq.Init.Logic}{\coqdocnotation{\ensuremath{\rightarrow}}} \coqref{Fpred.Definitions.typing}{\coqdocinductive{typing}} \coqdocvariable{e} \coqdocvariable{t2} \coqdocvariable{T11} \coqexternalref{:type scope:x '->' x}{http://coq.inria.fr/distrib/8.5beta2/stdlib/Coq.Init.Logic}{\coqdocnotation{\ensuremath{\rightarrow}}} \coqref{Fpred.Definitions.typing}{\coqdocinductive{typing}} \coqdocvariable{e} (\coqref{Fpred.Definitions.app}{\coqdocconstructor{app}} \coqdocvariable{t1} \coqdocvariable{t2}) \coqdocvariable{T12}\coqdoceol
\coqdocindent{1.00em}
\ensuremath{|} \coqdef{Fpred.Definitions.T Tabs}{T\_Tabs}{\coqdocconstructor{T\_Tabs}} (\coqdocvar{e} : \coqref{Fpred.Definitions.env}{\coqdocinductive{env}}) (\coqdocvar{t} : \coqref{Fpred.Definitions.term}{\coqdocinductive{term}}) (\coqdocvar{k} : \coqref{Fpred.Definitions.kind}{\coqdocdefinition{kind}}) (\coqdocvar{T} : \coqref{Fpred.Definitions.typ}{\coqdocinductive{typ}}) :\coqdoceol
\coqdocindent{3.00em}
\coqref{Fpred.Definitions.typing}{\coqdocinductive{typing}} (\coqref{Fpred.Definitions.etvar}{\coqdocconstructor{etvar}} \coqdocvariable{e} \coqdocvariable{k}) \coqdocvariable{t} \coqdocvariable{T} \coqexternalref{:type scope:x '->' x}{http://coq.inria.fr/distrib/8.5beta2/stdlib/Coq.Init.Logic}{\coqdocnotation{\ensuremath{\rightarrow}}} \coqref{Fpred.Definitions.typing}{\coqdocinductive{typing}} \coqdocvariable{e} (\coqref{Fpred.Definitions.tabs}{\coqdocconstructor{tabs}} \coqdocvariable{k} \coqdocvariable{t}) (\coqref{Fpred.Definitions.all}{\coqdocconstructor{all}} \coqdocvariable{k} \coqdocvariable{T})\coqdoceol
\coqdocindent{1.00em}
\ensuremath{|} \coqdef{Fpred.Definitions.T Tapp}{T\_Tapp}{\coqdocconstructor{T\_Tapp}} (\coqdocvar{e} : \coqref{Fpred.Definitions.env}{\coqdocinductive{env}}) (\coqdocvar{t} : \coqref{Fpred.Definitions.term}{\coqdocinductive{term}}) \coqdocvar{k} (\coqdocvar{T1} \coqdocvar{T2} : \coqref{Fpred.Definitions.typ}{\coqdocinductive{typ}}) :\coqdoceol
\coqdocindent{3.00em}
\coqref{Fpred.Definitions.typing}{\coqdocinductive{typing}} \coqdocvariable{e} \coqdocvariable{t} (\coqref{Fpred.Definitions.all}{\coqdocconstructor{all}} \coqdocvariable{k} \coqdocvariable{T1}) \coqexternalref{:type scope:x '->' x}{http://coq.inria.fr/distrib/8.5beta2/stdlib/Coq.Init.Logic}{\coqdocnotation{\ensuremath{\rightarrow}}} \coqref{Fpred.Definitions.kinding}{\coqdocinductive{kinding}} \coqdocvariable{e} \coqdocvariable{T2} \coqdocvariable{k} \coqexternalref{:type scope:x '->' x}{http://coq.inria.fr/distrib/8.5beta2/stdlib/Coq.Init.Logic}{\coqdocnotation{\ensuremath{\rightarrow}}} \coqref{Fpred.Definitions.typing}{\coqdocinductive{typing}} \coqdocvariable{e} (\coqref{Fpred.Definitions.tapp}{\coqdocconstructor{tapp}} \coqdocvariable{t} \coqdocvariable{T2}) (\coqref{Fpred.Definitions.tsubst}{\coqdocdefinition{tsubst}} \coqdocvariable{T1} 0 \coqdocvariable{T2}).\coqdoceol
\end{coqdoccode}

\begin{coqdoccode}
\coqdocemptyline
\end{coqdoccode}
\subsection{Reduction rules}

 To define normalization we must formalize the reduction relation of the calculus.
    Beta-redexes for this calculus are the application of an abstraction
    to a term and the application of a type abstraction to a type. \begin{coqdoccode}
\coqdocemptyline
\coqdocnoindent
\coqdockw{Inductive} \coqdef{Fpred.Reduction.red}{red}{\coqdocinductive{red}} : \coqdocinductive{term} \coqexternalref{:type scope:x '->' x}{http://coq.inria.fr/distrib/8.5beta2/stdlib/Coq.Init.Logic}{\coqdocnotation{\ensuremath{\rightarrow}}} \coqdocinductive{term} \coqexternalref{:type scope:x '->' x}{http://coq.inria.fr/distrib/8.5beta2/stdlib/Coq.Init.Logic}{\coqdocnotation{\ensuremath{\rightarrow}}} \coqdockw{Prop} :=\coqdoceol
\coqdocindent{1.00em}
\ensuremath{|} \coqdef{Fpred.Reduction.E AppAbs}{E\_AppAbs}{\coqdocconstructor{E\_AppAbs}} (\coqdocvar{T} : \coqdocinductive{typ}) (\coqdocvar{t1} \coqdocvar{t2} : \coqdocinductive{term}) : \coqref{Fpred.Reduction.red}{\coqdocinductive{red}} (\coqdocconstructor{app} (\coqdocconstructor{abs} \coqdocvariable{T} \coqdocvariable{t1}) \coqdocvariable{t2}) (\coqdocdefinition{subst} \coqdocvariable{t1} 0 \coqdocvariable{t2})\coqdoceol
\coqdocindent{1.00em}
\ensuremath{|} \coqdef{Fpred.Reduction.E TappTabs}{E\_TappTabs}{\coqdocconstructor{E\_TappTabs}} \coqdocvar{k} (\coqdocvar{T} : \coqdocinductive{typ}) (\coqdocvar{t} : \coqdocinductive{term}) : \coqref{Fpred.Reduction.red}{\coqdocinductive{red}} (\coqdocconstructor{tapp} (\coqdocconstructor{tabs} \coqdocvariable{k} \coqdocvariable{t}) \coqdocvariable{T}) (\coqdocdefinition{subst\_typ} \coqdocvariable{t} 0 \coqdocvariable{T}).\coqdoceol
\coqdocemptyline
\end{coqdoccode}
We define the transitive closure of reduction on terms by closing
  \coqdoctac{red} by context. \begin{coqdoccode}
\coqdocemptyline
\coqdocnoindent
\coqdockw{Inductive} \coqdef{Fpred.Reduction.sred}{sred}{\coqdocinductive{sred}} : \coqdocinductive{term} \coqexternalref{:type scope:x '->' x}{http://coq.inria.fr/distrib/8.5beta2/stdlib/Coq.Init.Logic}{\coqdocnotation{\ensuremath{\rightarrow}}} \coqdocinductive{term} \coqexternalref{:type scope:x '->' x}{http://coq.inria.fr/distrib/8.5beta2/stdlib/Coq.Init.Logic}{\coqdocnotation{\ensuremath{\rightarrow}}} \coqdockw{Prop} :=\coqdoceol
\coqdocindent{1.00em}
\ensuremath{|} \coqdef{Fpred.Reduction.Red sred}{Red\_sred}{\coqdocconstructor{Red\_sred}} \coqdocvar{t} \coqdocvar{t'} : \coqref{Fpred.Reduction.red}{\coqdocinductive{red}} \coqdocvariable{t} \coqdocvariable{t'} \coqexternalref{:type scope:x '->' x}{http://coq.inria.fr/distrib/8.5beta2/stdlib/Coq.Init.Logic}{\coqdocnotation{\ensuremath{\rightarrow}}} \coqref{Fpred.Reduction.sred}{\coqdocinductive{sred}} \coqdocvariable{t} \coqdocvariable{t'}\coqdoceol
\coqdocindent{1.00em}
\ensuremath{|} \coqdef{Fpred.Reduction.sred trans}{sred\_trans}{\coqdocconstructor{sred\_trans}} \coqdocvar{t1} \coqdocvar{t2} \coqdocvar{t3} : \coqref{Fpred.Reduction.sred}{\coqdocinductive{sred}} \coqdocvariable{t1} \coqdocvariable{t2} \coqexternalref{:type scope:x '->' x}{http://coq.inria.fr/distrib/8.5beta2/stdlib/Coq.Init.Logic}{\coqdocnotation{\ensuremath{\rightarrow}}} \coqref{Fpred.Reduction.sred}{\coqdocinductive{sred}} \coqdocvariable{t2} \coqdocvariable{t3} \coqexternalref{:type scope:x '->' x}{http://coq.inria.fr/distrib/8.5beta2/stdlib/Coq.Init.Logic}{\coqdocnotation{\ensuremath{\rightarrow}}} \coqref{Fpred.Reduction.sred}{\coqdocinductive{sred}} \coqdocvariable{t1} \coqdocvariable{t3}\coqdoceol
\coqdocindent{1.00em}
\ensuremath{|} \coqdef{Fpred.Reduction.Par app left}{Par\_app\_left}{\coqdocconstructor{Par\_app\_left}} \coqdocvar{t1} \coqdocvar{t1'} \coqdocvar{t2} : \coqref{Fpred.Reduction.sred}{\coqdocinductive{sred}} \coqdocvariable{t1} \coqdocvariable{t1'} \coqexternalref{:type scope:x '->' x}{http://coq.inria.fr/distrib/8.5beta2/stdlib/Coq.Init.Logic}{\coqdocnotation{\ensuremath{\rightarrow}}} \coqref{Fpred.Reduction.sred}{\coqdocinductive{sred}} (\coqdocconstructor{app} \coqdocvariable{t1} \coqdocvariable{t2}) (\coqdocconstructor{app} \coqdocvariable{t1'} \coqdocvariable{t2})\coqdoceol
\coqdocindent{1.00em}
\ensuremath{|} \coqdef{Fpred.Reduction.Par app right}{Par\_app\_right}{\coqdocconstructor{Par\_app\_right}} \coqdocvar{t1} \coqdocvar{t2} \coqdocvar{t2'} : \coqref{Fpred.Reduction.sred}{\coqdocinductive{sred}} \coqdocvariable{t2} \coqdocvariable{t2'} \coqexternalref{:type scope:x '->' x}{http://coq.inria.fr/distrib/8.5beta2/stdlib/Coq.Init.Logic}{\coqdocnotation{\ensuremath{\rightarrow}}} \coqref{Fpred.Reduction.sred}{\coqdocinductive{sred}} (\coqdocconstructor{app} \coqdocvariable{t1} \coqdocvariable{t2}) (\coqdocconstructor{app} \coqdocvariable{t1} \coqdocvariable{t2'})\coqdoceol
\coqdocindent{1.00em}
\ensuremath{|} \coqdef{Fpred.Reduction.Par abs}{Par\_abs}{\coqdocconstructor{Par\_abs}} \coqdocvar{T} \coqdocvar{t} \coqdocvar{t'}  : \coqref{Fpred.Reduction.sred}{\coqdocinductive{sred}} \coqdocvariable{t} \coqdocvariable{t'} \coqexternalref{:type scope:x '->' x}{http://coq.inria.fr/distrib/8.5beta2/stdlib/Coq.Init.Logic}{\coqdocnotation{\ensuremath{\rightarrow}}} \coqref{Fpred.Reduction.sred}{\coqdocinductive{sred}} (\coqdocconstructor{abs} \coqdocvariable{T} \coqdocvariable{t}) (\coqdocconstructor{abs} \coqdocvariable{T} \coqdocvariable{t'})\coqdoceol
\coqdocindent{1.00em}
\ensuremath{|} \coqdef{Fpred.Reduction.Par tapp}{Par\_tapp}{\coqdocconstructor{Par\_tapp}} \coqdocvar{t} \coqdocvar{t'} \coqdocvar{T} : \coqref{Fpred.Reduction.sred}{\coqdocinductive{sred}} \coqdocvariable{t} \coqdocvariable{t'} \coqexternalref{:type scope:x '->' x}{http://coq.inria.fr/distrib/8.5beta2/stdlib/Coq.Init.Logic}{\coqdocnotation{\ensuremath{\rightarrow}}} \coqref{Fpred.Reduction.sred}{\coqdocinductive{sred}} (\coqdocconstructor{tapp} \coqdocvariable{t} \coqdocvariable{T}) (\coqdocconstructor{tapp} \coqdocvariable{t'} \coqdocvariable{T})\coqdoceol
\coqdocindent{1.00em}
\ensuremath{|} \coqdef{Fpred.Reduction.Par tabs}{Par\_tabs}{\coqdocconstructor{Par\_tabs}} \coqdocvar{k} \coqdocvar{t} \coqdocvar{t'} : \coqref{Fpred.Reduction.sred}{\coqdocinductive{sred}} \coqdocvariable{t} \coqdocvariable{t'} \coqexternalref{:type scope:x '->' x}{http://coq.inria.fr/distrib/8.5beta2/stdlib/Coq.Init.Logic}{\coqdocnotation{\ensuremath{\rightarrow}}} \coqref{Fpred.Reduction.sred}{\coqdocinductive{sred}} (\coqdocconstructor{tabs} \coqdocvariable{k} \coqdocvariable{t}) (\coqdocconstructor{tabs} \coqdocvariable{k} \coqdocvariable{t'}).\coqdoceol
\coqdocemptyline
\coqdocemptyline
\coqdocnoindent
\coqdockw{Definition} \coqdef{Fpred.Reduction.reds}{reds}{\coqdocdefinition{reds}} \coqdocvar{t} \coqdocvar{n} := \coqexternalref{clos refl}{http://coq.inria.fr/distrib/8.5beta2/stdlib/Coq.Relations.Relation\_Operators}{\coqdocinductive{clos\_refl}} \coqref{Fpred.Reduction.sred}{\coqdocinductive{sred}} \coqdocvariable{t} \coqdocvariable{n}.\coqdoceol
\coqdocemptyline
\end{coqdoccode}
We can prove usual congruence lemmas on \coqref{Fpred.Reduction.reds}{\coqdocdefinition{reds}} showing that it indeed
  formalizes parallel reduction. \begin{coqdoccode}
\end{coqdoccode}

\section{Metatheory}
\label{sec:metatheory}
\begin{coqdoccode}
\end{coqdoccode}
The metatheory of the system is pretty straightforward and follows
  the one of $F^{sub}$ closely. We only mention the main idea for type 
  substitution and the statements of the main metatheoretical lemmas. 

 To formalize type substitution, we use a proposition \coqref{Fpred.Metatheory.env subst}{\coqdocinductive{env\_subst}}
    that corresponds to the environment operation: \coqdocvar{E}, \coqdocvariable{X} :* \coqdocvariable{k}, \coqdocvar{E'}
    $\Rightarrow$ \coqdocvar{E}, (\coqdocvariable{X} $\mapsto$ \coqdocvariable{T'}) \coqdocvar{E'} assuming \coqdocvar{E} \ensuremath{\vdash} \coqdocvariable{T'} :*
    \coqdocvariable{k}.  In other words, \coqref{Fpred.Metatheory.env subst}{\coqdocinductive{env\_subst}} \coqdocvariable{X} \coqdocvariable{T} \coqdocvariable{e} \coqdocvariable{e'} holds whenever we can find
    environments \coqdocvar{E}, \coqdocvar{E'} and a kind \coqdocvariable{k} such that \coqdocvar{E} \ensuremath{\vdash} \coqdocvariable{T'} :* \coqdocvariable{k} and
    \coqdocvariable{e} = \coqdocvar{E}, \coqdocvariable{X} :* \coqdocvariable{k}, \coqdocvar{E'} and \coqdocvariable{e'} = \coqdocvar{E}, (\coqdocvariable{X} $\mapsto$ \coqdocvariable{T'}) \coqdocvar{E'}.  \begin{coqdoccode}
\coqdocemptyline
\coqdocnoindent
\coqdockw{Inductive} \coqdef{Fpred.Metatheory.env subst}{env\_subst}{\coqdocinductive{env\_subst}} : \coqexternalref{nat}{http://coq.inria.fr/distrib/8.5beta2/stdlib/Coq.Init.Datatypes}{\coqdocinductive{nat}} \coqexternalref{:type scope:x '->' x}{http://coq.inria.fr/distrib/8.5beta2/stdlib/Coq.Init.Logic}{\coqdocnotation{\ensuremath{\rightarrow}}} \coqdocinductive{typ} \coqexternalref{:type scope:x '->' x}{http://coq.inria.fr/distrib/8.5beta2/stdlib/Coq.Init.Logic}{\coqdocnotation{\ensuremath{\rightarrow}}} \coqdocinductive{env} \coqexternalref{:type scope:x '->' x}{http://coq.inria.fr/distrib/8.5beta2/stdlib/Coq.Init.Logic}{\coqdocnotation{\ensuremath{\rightarrow}}} \coqdocinductive{env} \coqexternalref{:type scope:x '->' x}{http://coq.inria.fr/distrib/8.5beta2/stdlib/Coq.Init.Logic}{\coqdocnotation{\ensuremath{\rightarrow}}} \coqdockw{Prop} :=\coqdoceol
\coqdocnoindent
\ensuremath{|} \coqdef{Fpred.Metatheory.es here}{es\_here}{\coqdocconstructor{es\_here}} (\coqdocvar{e} : \coqdocinductive{env}) (\coqdocvar{T} : \coqdocinductive{typ}) (\coqdocvar{k} : \coqdocdefinition{kind}) : \coqdocinductive{kinding} \coqdocvariable{e} \coqdocvariable{T} \coqdocvariable{k} \coqexternalref{:type scope:x '->' x}{http://coq.inria.fr/distrib/8.5beta2/stdlib/Coq.Init.Logic}{\coqdocnotation{\ensuremath{\rightarrow}}} \coqref{Fpred.Metatheory.env subst}{\coqdocinductive{env\_subst}} 0 \coqdocvariable{T} (\coqdocconstructor{etvar} \coqdocvariable{e} \coqdocvariable{k}) \coqdocvariable{e}\coqdoceol
\coqdocnoindent
\ensuremath{|} \coqdef{Fpred.Metatheory.es var}{es\_var}{\coqdocconstructor{es\_var}} (\coqdocvar{X} : \coqexternalref{nat}{http://coq.inria.fr/distrib/8.5beta2/stdlib/Coq.Init.Datatypes}{\coqdocinductive{nat}}) (\coqdocvar{T} \coqdocvar{T'} : \coqdocinductive{typ}) (\coqdocvar{e} \coqdocvar{e'} : \coqdocinductive{env}) : \coqdoceol
\coqdocindent{2.00em}
\coqref{Fpred.Metatheory.env subst}{\coqdocinductive{env\_subst}} \coqdocvariable{X} \coqdocvariable{T'} \coqdocvariable{e} \coqdocvariable{e'} \coqexternalref{:type scope:x '->' x}{http://coq.inria.fr/distrib/8.5beta2/stdlib/Coq.Init.Logic}{\coqdocnotation{\ensuremath{\rightarrow}}} \coqref{Fpred.Metatheory.env subst}{\coqdocinductive{env\_subst}} \coqdocvariable{X} \coqdocvariable{T'} (\coqdocconstructor{evar} \coqdocvariable{e} \coqdocvariable{T}) (\coqdocconstructor{evar} \coqdocvariable{e'} (\coqdocdefinition{tsubst} \coqdocvariable{T} \coqdocvariable{X} \coqdocvariable{T'}))\coqdoceol
\coqdocnoindent
\ensuremath{|} \coqdef{Fpred.Metatheory.es kind}{es\_kind}{\coqdocconstructor{es\_kind}} (\coqdocvar{X} : \coqexternalref{nat}{http://coq.inria.fr/distrib/8.5beta2/stdlib/Coq.Init.Datatypes}{\coqdocinductive{nat}}) \coqdocvar{k} (\coqdocvar{T'} : \coqdocinductive{typ}) (\coqdocvar{e} \coqdocvar{e'} : \coqdocinductive{env}) :\coqdoceol
\coqdocindent{2.00em}
\coqref{Fpred.Metatheory.env subst}{\coqdocinductive{env\_subst}} \coqdocvariable{X} \coqdocvariable{T'} \coqdocvariable{e} \coqdocvariable{e'} \coqexternalref{:type scope:x '->' x}{http://coq.inria.fr/distrib/8.5beta2/stdlib/Coq.Init.Logic}{\coqdocnotation{\ensuremath{\rightarrow}}} \coqref{Fpred.Metatheory.env subst}{\coqdocinductive{env\_subst}} (1 \coqexternalref{:nat scope:x '+' x}{http://coq.inria.fr/distrib/8.5beta2/stdlib/Coq.Init.Peano}{\coqdocnotation{+}} \coqdocvariable{X}) (\coqdocdefinition{tshift} 0 \coqdocvariable{T'}) (\coqdocconstructor{etvar} \coqdocvariable{e} \coqdocvariable{k}) (\coqdocconstructor{etvar} \coqdocvariable{e'} \coqdocvariable{k}).\coqdoceol
\coqdocemptyline
\coqdocemptyline
\end{coqdoccode}
\subsection{Typing and well-formedness}

 Actually, both kinding and typing imply well-formedness. In other
   words, it is possible to kind a type \coqdocvariable{T} in an environment \coqdocvariable{e} only
   if both the type and the environment are well-formed. \begin{coqdoccode}
\coqdocemptyline
\coqdocnoindent
\coqdockw{Lemma} \coqdef{Fpred.Metatheory.kinding wf}{kinding\_wf}{\coqdoclemma{kinding\_wf}} (\coqdocvar{e} : \coqdocinductive{env}) (\coqdocvar{T} : \coqdocinductive{typ}) (\coqdocvar{k} : \coqdocdefinition{kind}) : \coqdocinductive{kinding} \coqdocvariable{e} \coqdocvariable{T} \coqdocvariable{k} \coqexternalref{:type scope:x '->' x}{http://coq.inria.fr/distrib/8.5beta2/stdlib/Coq.Init.Logic}{\coqdocnotation{\ensuremath{\rightarrow}}} \coqdocdefinition{wf\_env} \coqdocvariable{e} \coqexternalref{:type scope:x '/x5C' x}{http://coq.inria.fr/distrib/8.5beta2/stdlib/Coq.Init.Logic}{\coqdocnotation{\ensuremath{\land}}} \coqdocdefinition{wf\_typ} \coqdocvariable{e} \coqdocvariable{T}.\coqdoceol
\coqdocemptyline
\coqdocemptyline
\end{coqdoccode}
\subsection{Weakening}

\begin{coqdoccode}
\coqdocemptyline
\coqdocemptyline
\end{coqdoccode}
We only show the main weakening lemma for typing: if \coqdocvariable{e'} results from
  \coqdocvariable{e} by inserting a type variable at position \coqdocvariable{X} with any kind,
  the term and types of a typing derivation can be shifted accordingly
  to give a new typing derivation in the extended environment. \begin{coqdoccode}
\coqdocemptyline
\coqdocnoindent
\coqdockw{Lemma} \coqdef{Fpred.Metatheory.typing weakening kind ind}{typing\_weakening\_kind\_ind}{\coqdoclemma{typing\_weakening\_kind\_ind}} (\coqdocvar{e} \coqdocvar{e'} : \coqdocinductive{env}) (\coqdocvar{X} : \coqexternalref{nat}{http://coq.inria.fr/distrib/8.5beta2/stdlib/Coq.Init.Datatypes}{\coqdocinductive{nat}}) (\coqdocvar{t} : \coqdocinductive{term}) (\coqdocvar{U} : \coqdocinductive{typ}) :\coqdoceol
\coqdocindent{1.00em}
\coqdocinductive{insert\_kind} \coqdocvariable{X} \coqdocvariable{e} \coqdocvariable{e'} \coqexternalref{:type scope:x '->' x}{http://coq.inria.fr/distrib/8.5beta2/stdlib/Coq.Init.Logic}{\coqdocnotation{\ensuremath{\rightarrow}}} \coqdocinductive{typing} \coqdocvariable{e} \coqdocvariable{t} \coqdocvariable{U} \coqexternalref{:type scope:x '->' x}{http://coq.inria.fr/distrib/8.5beta2/stdlib/Coq.Init.Logic}{\coqdocnotation{\ensuremath{\rightarrow}}} \coqdocinductive{typing} \coqdocvariable{e'} (\coqdocdefinition{shift\_typ} \coqdocvariable{X} \coqdocvariable{t}) (\coqdocdefinition{tshift} \coqdocvariable{X} \coqdocvariable{U}).\coqdoceol
\coqdocemptyline
\coqdocemptyline
\end{coqdoccode}
Weakening by a term variable preserves typing as well. \begin{coqdoccode}
\coqdocemptyline
\coqdocnoindent
\coqdockw{Lemma} \coqdef{Fpred.Metatheory.typing weakening var}{typing\_weakening\_var}{\coqdoclemma{typing\_weakening\_var}} (\coqdocvar{e} : \coqdocinductive{env}) (\coqdocvar{t} : \coqdocinductive{term}) (\coqdocvar{U} \coqdocvar{V} : \coqdocinductive{typ}) :\coqdoceol
\coqdocindent{1.00em}
\coqdocdefinition{wf\_typ} \coqdocvariable{e} \coqdocvariable{V} \coqexternalref{:type scope:x '->' x}{http://coq.inria.fr/distrib/8.5beta2/stdlib/Coq.Init.Logic}{\coqdocnotation{\ensuremath{\rightarrow}}} \coqdocinductive{typing} \coqdocvariable{e} \coqdocvariable{t} \coqdocvariable{U} \coqexternalref{:type scope:x '->' x}{http://coq.inria.fr/distrib/8.5beta2/stdlib/Coq.Init.Logic}{\coqdocnotation{\ensuremath{\rightarrow}}} \coqdocinductive{typing} (\coqdocconstructor{evar} \coqdocvariable{e} \coqdocvariable{V}) (\coqdocdefinition{shift} 0 \coqdocvariable{t}) \coqdocvariable{U}.\coqdoceol
\coqdocemptyline
\coqdocemptyline
\end{coqdoccode}
\subsection{Narrowing}

 As the system includes a kind of subtyping relation due to level
  cumulativity, we can prove a narrowing property for derivations. Again
  we define a judgment formalizing that a context \coqdocvariable{e'} is a narrowing of
  a context \coqdocvariable{e} if they are identical but for one type variable
  binding (\coqdocvariable{T} : \coqdocvariable{k'}) in \coqdocvariable{e'} and (\coqdocvariable{T} : \coqdocvariable{k}) in \coqdocvariable{e} with \coqdocvariable{k'} $<$ \coqdocvariable{k}. \begin{coqdoccode}
\coqdocemptyline
\coqdocnoindent
\coqdockw{Inductive} \coqdef{Fpred.Metatheory.narrow}{narrow}{\coqdocinductive{narrow}} : \coqexternalref{nat}{http://coq.inria.fr/distrib/8.5beta2/stdlib/Coq.Init.Datatypes}{\coqdocinductive{nat}} \coqexternalref{:type scope:x '->' x}{http://coq.inria.fr/distrib/8.5beta2/stdlib/Coq.Init.Logic}{\coqdocnotation{\ensuremath{\rightarrow}}} \coqdocinductive{env} \coqexternalref{:type scope:x '->' x}{http://coq.inria.fr/distrib/8.5beta2/stdlib/Coq.Init.Logic}{\coqdocnotation{\ensuremath{\rightarrow}}} \coqdocinductive{env} \coqexternalref{:type scope:x '->' x}{http://coq.inria.fr/distrib/8.5beta2/stdlib/Coq.Init.Logic}{\coqdocnotation{\ensuremath{\rightarrow}}} \coqdockw{Set} :=\coqdoceol
\coqdocindent{2.00em}
\coqdef{Fpred.Metatheory.narrow 0}{narrow\_0}{\coqdocconstructor{narrow\_0}} (\coqdocvar{e} : \coqdocinductive{env}) (\coqdocvar{k} \coqdocvar{k'} : \coqdocdefinition{kind}) : \coqdocvariable{k'} \coqexternalref{:nat scope:x '<' x}{http://coq.inria.fr/distrib/8.5beta2/stdlib/Coq.Init.Peano}{\coqdocnotation{$<$}} \coqdocvariable{k} \coqexternalref{:type scope:x '->' x}{http://coq.inria.fr/distrib/8.5beta2/stdlib/Coq.Init.Logic}{\coqdocnotation{\ensuremath{\rightarrow}}} \coqref{Fpred.Metatheory.narrow}{\coqdocinductive{narrow}} 0 (\coqdocconstructor{etvar} \coqdocvariable{e} \coqdocvariable{k}) (\coqdocconstructor{etvar} \coqdocvariable{e} \coqdocvariable{k'})\coqdoceol
\coqdocindent{1.00em}
\ensuremath{|} \coqdef{Fpred.Metatheory.narrow extend kind}{narrow\_extend\_kind}{\coqdocconstructor{narrow\_extend\_kind}} (\coqdocvar{e} \coqdocvar{e'} : \coqdocinductive{env}) (\coqdocvar{k} : \coqdocdefinition{kind}) (\coqdocvar{X} : \coqexternalref{nat}{http://coq.inria.fr/distrib/8.5beta2/stdlib/Coq.Init.Datatypes}{\coqdocinductive{nat}}) :\coqdoceol
\coqdocindent{3.00em}
\coqref{Fpred.Metatheory.narrow}{\coqdocinductive{narrow}} \coqdocvariable{X} \coqdocvariable{e} \coqdocvariable{e'} \coqexternalref{:type scope:x '->' x}{http://coq.inria.fr/distrib/8.5beta2/stdlib/Coq.Init.Logic}{\coqdocnotation{\ensuremath{\rightarrow}}} \coqref{Fpred.Metatheory.narrow}{\coqdocinductive{narrow}} (1 \coqexternalref{:nat scope:x '+' x}{http://coq.inria.fr/distrib/8.5beta2/stdlib/Coq.Init.Peano}{\coqdocnotation{+}} \coqdocvariable{X}) (\coqdocconstructor{etvar} \coqdocvariable{e} \coqdocvariable{k}) (\coqdocconstructor{etvar} \coqdocvariable{e'} \coqdocvariable{k})\coqdoceol
\coqdocindent{1.00em}
\ensuremath{|} \coqdef{Fpred.Metatheory.narrow extend var}{narrow\_extend\_var}{\coqdocconstructor{narrow\_extend\_var}} (\coqdocvar{e} \coqdocvar{e'} : \coqdocinductive{env}) (\coqdocvar{T} : \coqdocinductive{typ}) (\coqdocvar{X} : \coqexternalref{nat}{http://coq.inria.fr/distrib/8.5beta2/stdlib/Coq.Init.Datatypes}{\coqdocinductive{nat}}) :\coqdoceol
\coqdocindent{3.00em}
\coqdocdefinition{wf\_typ} \coqdocvariable{e'} \coqdocvariable{T} \coqexternalref{:type scope:x '->' x}{http://coq.inria.fr/distrib/8.5beta2/stdlib/Coq.Init.Logic}{\coqdocnotation{\ensuremath{\rightarrow}}} \coqref{Fpred.Metatheory.narrow}{\coqdocinductive{narrow}} \coqdocvariable{X} \coqdocvariable{e} \coqdocvariable{e'} \coqexternalref{:type scope:x '->' x}{http://coq.inria.fr/distrib/8.5beta2/stdlib/Coq.Init.Logic}{\coqdocnotation{\ensuremath{\rightarrow}}} \coqref{Fpred.Metatheory.narrow}{\coqdocinductive{narrow}} \coqdocvariable{X} (\coqdocconstructor{evar} \coqdocvariable{e} \coqdocvariable{T}) (\coqdocconstructor{evar} \coqdocvariable{e'} \coqdocvariable{T}).\coqdoceol
\coqdocemptyline
\coqdocemptyline
\end{coqdoccode}
Before we can show narrowing, we have to show that kinding respects cumulativity:
   If it is provable that a type \coqdocvariable{T} has kind \coqdocvariable{k} in the context \coqdocvariable{e}, then
   we can also prove that it has any kind \coqdocvariable{k'} for \coqdocvariable{k} \ensuremath{\le} \coqdocvariable{k'}. \begin{coqdoccode}
\coqdocemptyline
\coqdocnoindent
\coqdockw{Lemma} \coqdef{Fpred.Metatheory.kinding transitive}{kinding\_transitive}{\coqdoclemma{kinding\_transitive}} \coqdocvar{e} \coqdocvar{T} \coqdocvar{k} \coqdocvar{k'} : \coqdocinductive{kinding} \coqdocvariable{e} \coqdocvariable{T} \coqdocvariable{k} \coqexternalref{:type scope:x '->' x}{http://coq.inria.fr/distrib/8.5beta2/stdlib/Coq.Init.Logic}{\coqdocnotation{\ensuremath{\rightarrow}}} \coqdocvariable{k} \coqexternalref{:nat scope:x '<=' x}{http://coq.inria.fr/distrib/8.5beta2/stdlib/Coq.Init.Peano}{\coqdocnotation{\ensuremath{\le}}} \coqdocvariable{k'} \coqexternalref{:type scope:x '->' x}{http://coq.inria.fr/distrib/8.5beta2/stdlib/Coq.Init.Logic}{\coqdocnotation{\ensuremath{\rightarrow}}} \coqdocinductive{kinding} \coqdocvariable{e} \coqdocvariable{T} \coqdocvariable{k'}.\coqdoceol
\coqdocemptyline
\coqdocemptyline
\end{coqdoccode}
Narrowing is a strong property, in the sense that a type \coqdocvariable{T} can
    have in a narrowing of a context \coqdocvariable{e} any kind that it can have in
    \coqdocvariable{e} itself. \begin{coqdoccode}
\coqdocemptyline
\coqdocnoindent
\coqdockw{Lemma} \coqdef{Fpred.Metatheory.typing narrowing ind}{typing\_narrowing\_ind}{\coqdoclemma{typing\_narrowing\_ind}} (\coqdocvar{e} \coqdocvar{e'} : \coqdocinductive{env}) (\coqdocvar{X} : \coqexternalref{nat}{http://coq.inria.fr/distrib/8.5beta2/stdlib/Coq.Init.Datatypes}{\coqdocinductive{nat}}) (\coqdocvar{t} : \coqdocinductive{term}) (\coqdocvar{U} : \coqdocinductive{typ}) : \coqref{Fpred.Metatheory.narrow}{\coqdocinductive{narrow}} \coqdocvariable{X} \coqdocvariable{e} \coqdocvariable{e'} \coqexternalref{:type scope:x '->' x}{http://coq.inria.fr/distrib/8.5beta2/stdlib/Coq.Init.Logic}{\coqdocnotation{\ensuremath{\rightarrow}}} \coqdocinductive{typing} \coqdocvariable{e} \coqdocvariable{t} \coqdocvariable{U} \coqexternalref{:type scope:x '->' x}{http://coq.inria.fr/distrib/8.5beta2/stdlib/Coq.Init.Logic}{\coqdocnotation{\ensuremath{\rightarrow}}} \coqdocinductive{typing} \coqdocvariable{e'} \coqdocvariable{t} \coqdocvariable{U}.\coqdoceol
\coqdocemptyline
\coqdocemptyline
\end{coqdoccode}
\subsection{Substitution}

 Now, substitution lemmas can be proven for the various
    substitution functions. \begin{coqdoccode}
\coqdocemptyline
\coqdocnoindent
\coqdockw{Lemma} \coqdef{Fpred.Metatheory.subst preserves typing}{subst\_preserves\_typing}{\coqdoclemma{subst\_preserves\_typing}} (\coqdocvar{e} : \coqdocinductive{env}) (\coqdocvar{x} : \coqexternalref{nat}{http://coq.inria.fr/distrib/8.5beta2/stdlib/Coq.Init.Datatypes}{\coqdocinductive{nat}}) (\coqdocvar{t} \coqdocvar{u} : \coqdocinductive{term}) (\coqdocvar{V} \coqdocvar{W} : \coqdocinductive{typ}) :\coqdoceol
\coqdocindent{1.00em}
\coqdocinductive{typing} \coqdocvariable{e} \coqdocvariable{t} \coqdocvariable{V} \coqexternalref{:type scope:x '->' x}{http://coq.inria.fr/distrib/8.5beta2/stdlib/Coq.Init.Logic}{\coqdocnotation{\ensuremath{\rightarrow}}} \coqdocinductive{typing} (\coqdocdefinition{remove\_var} \coqdocvariable{e} \coqdocvariable{x}) \coqdocvariable{u} \coqdocvariable{W} \coqexternalref{:type scope:x '->' x}{http://coq.inria.fr/distrib/8.5beta2/stdlib/Coq.Init.Logic}{\coqdocnotation{\ensuremath{\rightarrow}}} \coqdocdefinition{get\_var} \coqdocvariable{e} \coqdocvariable{x} \coqexternalref{:type scope:x '=' x}{http://coq.inria.fr/distrib/8.5beta2/stdlib/Coq.Init.Logic}{\coqdocnotation{=}} \coqexternalref{Some}{http://coq.inria.fr/distrib/8.5beta2/stdlib/Coq.Init.Datatypes}{\coqdocconstructor{Some}} \coqdocvariable{W} \coqexternalref{:type scope:x '->' x}{http://coq.inria.fr/distrib/8.5beta2/stdlib/Coq.Init.Logic}{\coqdocnotation{\ensuremath{\rightarrow}}} \coqdocinductive{typing} (\coqdocdefinition{remove\_var} \coqdocvariable{e} \coqdocvariable{x}) (\coqdocdefinition{subst} \coqdocvariable{t} \coqdocvariable{x} \coqdocvariable{u}) \coqdocvariable{V}.\coqdoceol
\coqdocemptyline
\coqdocemptyline
\coqdocnoindent
\coqdockw{Lemma} \coqdef{Fpred.Metatheory.subst typ preserves typing}{subst\_typ\_preserves\_typing}{\coqdoclemma{subst\_typ\_preserves\_typing}} (\coqdocvar{e} : \coqdocinductive{env}) (\coqdocvar{t} : \coqdocinductive{term}) (\coqdocvar{U} \coqdocvar{P} : \coqdocinductive{typ}) \coqdocvar{k} :\coqdoceol
\coqdocindent{1.00em}
\coqdocinductive{typing} (\coqdocconstructor{etvar} \coqdocvariable{e} \coqdocvariable{k}) \coqdocvariable{t} \coqdocvariable{U} \coqexternalref{:type scope:x '->' x}{http://coq.inria.fr/distrib/8.5beta2/stdlib/Coq.Init.Logic}{\coqdocnotation{\ensuremath{\rightarrow}}} \coqdocinductive{kinding} \coqdocvariable{e} \coqdocvariable{P} \coqdocvariable{k} \coqexternalref{:type scope:x '->' x}{http://coq.inria.fr/distrib/8.5beta2/stdlib/Coq.Init.Logic}{\coqdocnotation{\ensuremath{\rightarrow}}} \coqdocinductive{typing} \coqdocvariable{e} (\coqdocdefinition{subst\_typ} \coqdocvariable{t} 0 \coqdocvariable{P}) (\coqdocdefinition{tsubst} \coqdocvariable{U} 0 \coqdocvariable{P}).\coqdoceol
\coqdocemptyline
\end{coqdoccode}
Finally, we prove regularity, which is to say that the type of any
    well-typed term is kinded. This is a consequence of the fact that
    any well-formed type is kindable. All these results correspond
    directly to the paper proofs of Stump and Hayes. \begin{coqdoccode}
\coqdocemptyline
\coqdocemptyline
\coqdocnoindent
\coqdockw{Theorem} \coqdef{Fpred.Metatheory.regularity}{regularity}{\coqdoclemma{regularity}} (\coqdocvar{e} : \coqdocinductive{env}) (\coqdocvar{t} : \coqdocinductive{term}) (\coqdocvar{U} : \coqdocinductive{typ}) : \coqdocinductive{typing} \coqdocvariable{e} \coqdocvariable{t} \coqdocvariable{U} \coqexternalref{:type scope:x '->' x}{http://coq.inria.fr/distrib/8.5beta2/stdlib/Coq.Init.Logic}{\coqdocnotation{\ensuremath{\rightarrow}}} \coqexternalref{:type scope:'exists' x '..' x ',' x}{http://coq.inria.fr/distrib/8.5beta2/stdlib/Coq.Init.Logic}{\coqdocnotation{\ensuremath{\exists}}} \coqdocvar{k} : \coqdocdefinition{kind}\coqexternalref{:type scope:'exists' x '..' x ',' x}{http://coq.inria.fr/distrib/8.5beta2/stdlib/Coq.Init.Logic}{\coqdocnotation{,}} \coqdocinductive{kinding} \coqdocvariable{e} \coqdocvariable{U} \coqdocvariable{k}.\coqdoceol
\end{coqdoccode}

\section{Normalization}
\label{sec:normalization}
\begin{coqdoccode}
\coqdocemptyline
\coqdocemptyline
\end{coqdoccode}
To show that hereditary substitution is well-defined, we must
   provide an order of termination. In our case, we will have a
   lexicographic combination of a multiset ordering on kinds.  To
   formalize this, we reuse CoLoR's \cite{BlanquiCoLoR} library of
   multisets and definition of the multiset order. Those are multisets
   on ordered types, here natural numbers with the usual ordering, which
   is well-founded. \begin{coqdoccode}
\coqdocemptyline
\coqdocnoindent
\coqdockw{Notation} \coqdef{Fpred.Normalization.::x '\ltmul' x}{"}{"}X \ltmul Y" := (\coqdocdefinition{MultisetLt} \coqexternalref{gt}{http://coq.inria.fr/distrib/8.5beta2/stdlib/Coq.Init.Peano}{\coqdocdefinition{gt}} \coqdocvar{X} \coqdocvar{Y}) (\coqdoctac{at} \coqdockw{level} 70).\coqdoceol
\coqdocemptyline
\coqdocnoindent
\coqdockw{Definition} \coqdef{Fpred.Normalization.wf multiset order}{wf\_multiset\_order}{\coqdocdefinition{wf\_multiset\_order}} : \coqexternalref{well founded}{http://coq.inria.fr/distrib/8.5beta2/stdlib/Coq.Init.Wf}{\coqdocdefinition{well\_founded}} (\coqdocdefinition{MultisetLt} \coqexternalref{gt}{http://coq.inria.fr/distrib/8.5beta2/stdlib/Coq.Init.Peano}{\coqdocdefinition{gt}}).\coqdoceol
\coqdocemptyline
\end{coqdoccode}
The \coqref{Fpred.Normalization.kinds of}{\coqdocdefinition{kinds\_of}} function computes the multiset of kinds appearing in a type,
  which reduces to the bounds of universal quantifications. \begin{coqdoccode}
\coqdocemptyline
\coqdocnoindent
\coqdockw{Equations} \coqdef{Fpred.Normalization.kinds of}{kinds\_of}{\coqdocdefinition{kinds\_of}} (\coqdocvar{t} : \coqdocinductive{typ}) : \coqdocdefinition{Multiset} :=\coqdoceol
\coqdocnoindent
\coqref{Fpred.Normalization.kinds of}{\coqdocdefinition{kinds\_of}} (\coqdocconstructor{tvar} \coqdocconstructor{\_})    \ensuremath{\Rightarrow} \coqdocdefinition{empty}; \coqref{Fpred.Normalization.kinds of}{\coqdocdefinition{kinds\_of}} (\coqdocconstructor{arrow} \coqdocconstructor{T} \coqdocconstructor{U}) \ensuremath{\Rightarrow} \coqdocdefinition{union} (\coqref{Fpred.Normalization.kinds of}{\coqdocdefinition{kinds\_of}} \coqdocvar{T}) (\coqref{Fpred.Normalization.kinds of}{\coqdocdefinition{kinds\_of}} \coqdocvar{U});\coqdoceol
\coqdocnoindent
\coqref{Fpred.Normalization.kinds of}{\coqdocdefinition{kinds\_of}} (\coqdocconstructor{all} \coqdocconstructor{k} \coqdocconstructor{T})   \ensuremath{\Rightarrow} \coqdocdefinition{union} \coqdocnotation{$\{$}\coqdocvar{k}\coqdocnotation{$\}$} (\coqref{Fpred.Normalization.kinds of}{\coqdocdefinition{kinds\_of}} \coqdocvar{T}).\coqdoceol
\coqdocemptyline
\coqdocemptyline
\end{coqdoccode}
Clearly, the singleton multiset built from any valid kind for \coqdocvariable{T}
  bounds the bag of kinds appearing in \coqdocvariable{T}, according to the kinding
  rules. This is proved by induction on the kinding derivation: \begin{coqdoccode}
\coqdocemptyline
\coqdocnoindent
\coqdockw{Lemma} \coqdef{Fpred.Normalization.kinds of kinded}{kinds\_of\_kinded}{\coqdoclemma{kinds\_of\_kinded}} \coqdocvar{e} \coqdocvar{T} \coqdocvar{k} : \coqdocinductive{kinding} \coqdocvariable{e} \coqdocvariable{T} \coqdocvariable{k} \coqexternalref{:type scope:x '->' x}{http://coq.inria.fr/distrib/8.5beta2/stdlib/Coq.Init.Logic}{\coqdocnotation{\ensuremath{\rightarrow}}} \coqref{Fpred.Normalization.kinds of}{\coqdocdefinition{kinds\_of}} \coqdocvariable{T} \coqref{Fpred.Normalization.::x '\ltmul' x}{\coqdocnotation{$<$}}\coqref{Fpred.Normalization.::x '\ltmul' x}{\coqdocnotation{mul}} \coqdocnotation{$\{$} \coqdocvariable{k} \coqdocnotation{$\}$}.\coqdoceol
\coqdocemptyline
\end{coqdoccode}
Kinds in a type are invariant by shifting or lifting. This is a simple example
  of a proof by functional elimination. The \coqdocvariable{T} argument and result of \coqref{Fpred.Normalization.kinds of}{\coqdocdefinition{kinds\_of}} \coqdocvariable{T} 
  get refined and we just need to simplify the right hand sides according to the
  definitions of \coqref{Fpred.Normalization.kinds of}{\coqdocdefinition{kinds\_of}} and \coqdocdefinition{tshift}, using rewriting not computation, and 
  finish by rewriting with the induction hypotheses.
\begin{coqdoccode}
\coqdocemptyline
\coqdocnoindent
\coqdockw{Lemma} \coqdef{Fpred.Normalization.kinds of tshift}{kinds\_of\_tshift}{\coqdoclemma{kinds\_of\_tshift}} \coqdocvar{X} \coqdocvar{T} : \coqref{Fpred.Normalization.kinds of}{\coqdocdefinition{kinds\_of}} (\coqdocdefinition{tshift} \coqdocvariable{X} \coqdocvariable{T}) \coqexternalref{:type scope:x '=' x}{http://coq.inria.fr/distrib/8.5beta2/stdlib/Coq.Init.Logic}{\coqdocnotation{=}} \coqref{Fpred.Normalization.kinds of}{\coqdocdefinition{kinds\_of}} \coqdocvariable{T}.\coqdoceol
\coqdocnoindent
\coqdockw{Proof}.\coqdoceol
\coqdocindent{1.00em}
\coqdocvar{funelim} (\coqref{Fpred.Normalization.kinds of}{\coqdocdefinition{kinds\_of}} \coqdocvar{T}); \coqdocvar{simp} \coqdocvar{kinds\_of} \coqdocvar{tshift}; \coqdocvar{now} \coqdoctac{rewrite} \coqdocvar{H}, ?\coqdocvar{H0}.\coqdoceol
\coqdocnoindent
\coqdockw{Qed}.\coqdoceol
\coqdocemptyline
\end{coqdoccode}
For type substitution of \coqdocvariable{T} in \coqdocvariable{U} however, an exact arithmetic relation holds.
  We know that the multiset of kinds of the substituted type can appear 
  a finite number of times in the resulting type, along with the original 
  kinds of \coqdocvariable{U}. \begin{coqdoccode}
\coqdocemptyline
\coqdocnoindent
\coqdockw{Lemma} \coqdef{Fpred.Normalization.kinds of tsubst}{kinds\_of\_tsubst}{\coqdoclemma{kinds\_of\_tsubst}} \coqdocvar{e} \coqdocvar{e'} \coqdocvar{X} \coqdocvar{T} \coqdocvar{U} \coqdocvar{k} : \coqdocinductive{env\_subst} \coqdocvariable{X} \coqdocvariable{T} \coqdocvariable{e} \coqdocvariable{e'} \coqexternalref{:type scope:x '->' x}{http://coq.inria.fr/distrib/8.5beta2/stdlib/Coq.Init.Logic}{\coqdocnotation{\ensuremath{\rightarrow}}} \coqdocinductive{kinding} \coqdocvariable{e} \coqdocvariable{U} \coqdocvariable{k} \coqexternalref{:type scope:x '->' x}{http://coq.inria.fr/distrib/8.5beta2/stdlib/Coq.Init.Logic}{\coqdocnotation{\ensuremath{\rightarrow}}}\coqdoceol
\coqdocindent{1.00em}
\coqexternalref{:type scope:'exists' x '..' x ',' x}{http://coq.inria.fr/distrib/8.5beta2/stdlib/Coq.Init.Logic}{\coqdocnotation{\ensuremath{\exists}}} \coqdocvar{n} : \coqexternalref{nat}{http://coq.inria.fr/distrib/8.5beta2/stdlib/Coq.Init.Datatypes}{\coqdocinductive{nat}}\coqexternalref{:type scope:'exists' x '..' x ',' x}{http://coq.inria.fr/distrib/8.5beta2/stdlib/Coq.Init.Logic}{\coqdocnotation{,}} \coqref{Fpred.Normalization.kinds of}{\coqdocdefinition{kinds\_of}} (\coqdocdefinition{tsubst} \coqdocvariable{U} \coqdocvariable{X} \coqdocvariable{T}) \coqdocnotation{=}\coqdocnotation{mul}\coqdocnotation{=} \coqref{Fpred.Normalization.kinds of}{\coqdocdefinition{kinds\_of}} \coqdocvariable{U} \coqdocnotation{+} \coqref{Fpred.Normalization.mul sum}{\coqdocdefinition{mul\_sum}} \coqdocvariable{n} (\coqref{Fpred.Normalization.kinds of}{\coqdocdefinition{kinds\_of}} \coqdocvariable{T}).\coqdoceol
\coqdocemptyline
\end{coqdoccode}
This allows us to derive a general result about kindings of
   universal types: any well-kinded instance substitution produces a
   type with a strictly smaller bag of kinds. This is the central 
   result needed to show termination. In Stump's work, the measure
   considered was solely the depth of types, and only through the stricter
   kinding invariant could the order be shown well-founded. \begin{coqdoccode}
\coqdocemptyline
\coqdocnoindent
\coqdockw{Lemma} \coqdef{Fpred.Normalization.kinds of tsubst all}{kinds\_of\_tsubst\_all}{\coqdoclemma{kinds\_of\_tsubst\_all}} \coqdocvar{e} \coqdocvar{U} \coqdocvar{k} \coqdocvar{k'} \coqdocvar{T} : \coqdocinductive{kinding} \coqdocvariable{e} (\coqdocconstructor{all} \coqdocvariable{k} \coqdocvariable{U}) \coqdocvariable{k'} \coqexternalref{:type scope:x '->' x}{http://coq.inria.fr/distrib/8.5beta2/stdlib/Coq.Init.Logic}{\coqdocnotation{\ensuremath{\rightarrow}}}\coqdoceol
\coqdocindent{1.00em}
\coqdocinductive{kinding} \coqdocvariable{e} \coqdocvariable{T} \coqdocvariable{k} \coqexternalref{:type scope:x '->' x}{http://coq.inria.fr/distrib/8.5beta2/stdlib/Coq.Init.Logic}{\coqdocnotation{\ensuremath{\rightarrow}}} \coqref{Fpred.Normalization.kinds of}{\coqdocdefinition{kinds\_of}} (\coqdocdefinition{tsubst} \coqdocvariable{U} 0 \coqdocvariable{T}) \coqref{Fpred.Normalization.::x '\ltmul' x}{\coqdocnotation{$<$}}\coqref{Fpred.Normalization.::x '\ltmul' x}{\coqdocnotation{mul}} \coqref{Fpred.Normalization.kinds of}{\coqdocdefinition{kinds\_of}} (\coqdocconstructor{all} \coqdocvariable{k} \coqdocvariable{U}).\coqdoceol
\coqdocemptyline
\coqdocemptyline
\end{coqdoccode}
\subsection{Definition of the measure.}

 We first define the depth of a type as being the number of universal
    quantifications and type variables in that type. \begin{coqdoccode}
\coqdocemptyline
\coqdocnoindent
\coqdockw{Equations} \coqdef{Fpred.Normalization.depth}{depth}{\coqdocdefinition{depth}} (\coqdocvar{t} : \coqdocinductive{typ}) : \coqexternalref{nat}{http://coq.inria.fr/distrib/8.5beta2/stdlib/Coq.Init.Datatypes}{\coqdocinductive{nat}} :=\coqdoceol
\coqdocnoindent
\coqref{Fpred.Normalization.depth}{\coqdocdefinition{depth}} (\coqdocconstructor{tvar}  \coqdocconstructor{\_})   \ensuremath{\Rightarrow} 1; \coqref{Fpred.Normalization.depth}{\coqdocdefinition{depth}} (\coqdocconstructor{arrow} \coqdocconstructor{T} \coqdocconstructor{U}) \ensuremath{\Rightarrow} (\coqref{Fpred.Normalization.depth}{\coqdocdefinition{depth}} \coqdocvar{T} \coqexternalref{:nat scope:x '+' x}{http://coq.inria.fr/distrib/8.5beta2/stdlib/Coq.Init.Peano}{\coqdocnotation{+}} \coqref{Fpred.Normalization.depth}{\coqdocdefinition{depth}} \coqdocvar{U})\%\coqdocvar{nat};\coqdoceol
\coqdocnoindent
\coqref{Fpred.Normalization.depth}{\coqdocdefinition{depth}} (\coqdocconstructor{all}   \coqdocconstructor{k} \coqdocconstructor{U}) \ensuremath{\Rightarrow} \coqexternalref{S}{http://coq.inria.fr/distrib/8.5beta2/stdlib/Coq.Init.Datatypes}{\coqdocconstructor{S}} (\coqref{Fpred.Normalization.depth}{\coqdocdefinition{depth}} \coqdocvar{U}).\coqdoceol
\coqdocemptyline
\coqdocemptyline
\end{coqdoccode}
Of course, it cannot be zero, which is useful since it allows
    to have \coqref{Fpred.Normalization.depth}{\coqdocdefinition{depth}} \coqdocvariable{T} $<$ \coqref{Fpred.Normalization.depth}{\coqdocdefinition{depth}} (\coqdocconstructor{arrow} \coqdocvariable{T} \coqdocvariable{U}), which will be needed to
    prove the well-foundedness of the hereditary substitution. \begin{coqdoccode}
\coqdocemptyline
\coqdocnoindent
\coqdockw{Lemma} \coqdef{Fpred.Normalization.depth nz}{depth\_nz}{\coqdoclemma{depth\_nz}} \coqdocvar{t} : 0 \coqexternalref{:nat scope:x '<' x}{http://coq.inria.fr/distrib/8.5beta2/stdlib/Coq.Init.Peano}{\coqdocnotation{$<$}} \coqref{Fpred.Normalization.depth}{\coqdocdefinition{depth}} \coqdocvariable{t}.\coqdoceol
\coqdocemptyline
\coqdocemptyline
\end{coqdoccode}
The order that we will use on types is a lexicographical order on
    the multiset of kinds and the depth. As the first part of the
    lexicographic product is a multiset, and as those should not be
    compared with the Leibniz equality but rather a specific setoid
    equality, we defined a generalized notion of lexicographic product
    up-to an equivalence relation on the first component, here \coqdocdefinition{meq}
    which represents multiset equality. \begin{coqdoccode}
\coqdocemptyline
\coqdocnoindent
\coqdockw{Definition} \coqdef{Fpred.Normalization.relmd}{relmd}{\coqdocdefinition{relmd}} : \coqexternalref{relation}{http://coq.inria.fr/distrib/8.5beta2/stdlib/Coq.Relations.Relation\_Definitions}{\coqdocdefinition{relation}} (\coqdocdefinition{Multiset} \coqexternalref{:type scope:x '*' x}{http://coq.inria.fr/distrib/8.5beta2/stdlib/Coq.Init.Datatypes}{\coqdocnotation{\ensuremath{\times}}} \coqexternalref{nat}{http://coq.inria.fr/distrib/8.5beta2/stdlib/Coq.Init.Datatypes}{\coqdocinductive{nat}}) := \coqref{Fpred.Normalization.lexprod}{\coqdocinductive{lexprod}} (\coqdocdefinition{MultisetLt} \coqexternalref{gt}{http://coq.inria.fr/distrib/8.5beta2/stdlib/Coq.Init.Peano}{\coqdocdefinition{gt}}) \coqdocdefinition{meq} \coqexternalref{lt}{http://coq.inria.fr/distrib/8.5beta2/stdlib/Coq.Init.Peano}{\coqdocdefinition{lt}}.\coqdoceol
\coqdocemptyline
\end{coqdoccode}
It is well-founded, relying ultimately on the well-foundedness of \coqexternalref{lt}{http://coq.inria.fr/distrib/8.5beta2/stdlib/Coq.Init.Peano}{\coqdocdefinition{lt}}. \begin{coqdoccode}
\coqdocemptyline
\coqdocnoindent
\coqdockw{Definition} \coqdef{Fpred.Normalization.wf relmd}{wf\_relmd}{\coqdocdefinition{wf\_relmd}} : \coqexternalref{well founded}{http://coq.inria.fr/distrib/8.5beta2/stdlib/Coq.Init.Wf}{\coqdocdefinition{well\_founded}} \coqref{Fpred.Normalization.relmd}{\coqdocdefinition{relmd}}.\coqdoceol
\coqdocemptyline
\end{coqdoccode}
The actual order is \coqref{Fpred.Normalization.relmd}{\coqdocdefinition{relmd}} \textit{on} the \coqref{Fpred.Normalization.kinds of}{\coqdocdefinition{kinds\_of}} and \coqref{Fpred.Normalization.depth}{\coqdocdefinition{depth}} measures on types. \begin{coqdoccode}
\coqdocemptyline
\coqdocnoindent
\coqdockw{Definition} \coqdef{Fpred.Normalization.order}{order}{\coqdocdefinition{order}} (\coqdocvar{x} \coqdocvar{y} : \coqdocinductive{typ}) := \coqref{Fpred.Normalization.relmd}{\coqdocdefinition{relmd}} \coqexternalref{:core scope:'(' x ',' x ',' '..' ',' x ')'}{http://coq.inria.fr/distrib/8.5beta2/stdlib/Coq.Init.Datatypes}{\coqdocnotation{(}}\coqref{Fpred.Normalization.kinds of}{\coqdocdefinition{kinds\_of}} \coqdocvariable{x}\coqexternalref{:core scope:'(' x ',' x ',' '..' ',' x ')'}{http://coq.inria.fr/distrib/8.5beta2/stdlib/Coq.Init.Datatypes}{\coqdocnotation{,}} \coqref{Fpred.Normalization.depth}{\coqdocdefinition{depth}} \coqdocvariable{x}\coqexternalref{:core scope:'(' x ',' x ',' '..' ',' x ')'}{http://coq.inria.fr/distrib/8.5beta2/stdlib/Coq.Init.Datatypes}{\coqdocnotation{)}} \coqexternalref{:core scope:'(' x ',' x ',' '..' ',' x ')'}{http://coq.inria.fr/distrib/8.5beta2/stdlib/Coq.Init.Datatypes}{\coqdocnotation{(}}\coqref{Fpred.Normalization.kinds of}{\coqdocdefinition{kinds\_of}} \coqdocvariable{y}\coqexternalref{:core scope:'(' x ',' x ',' '..' ',' x ')'}{http://coq.inria.fr/distrib/8.5beta2/stdlib/Coq.Init.Datatypes}{\coqdocnotation{,}} \coqref{Fpred.Normalization.depth}{\coqdocdefinition{depth}} \coqdocvariable{y}\coqexternalref{:core scope:'(' x ',' x ',' '..' ',' x ')'}{http://coq.inria.fr/distrib/8.5beta2/stdlib/Coq.Init.Datatypes}{\coqdocnotation{)}}.\coqdoceol
\coqdocemptyline
\coqdocnoindent
\coqdockw{Definition} \coqdef{Fpred.Normalization.wf order}{wf\_order}{\coqdocdefinition{wf\_order}} : \coqexternalref{well founded}{http://coq.inria.fr/distrib/8.5beta2/stdlib/Coq.Init.Wf}{\coqdocdefinition{well\_founded}} \coqref{Fpred.Normalization.order}{\coqdocdefinition{order}}.\coqdoceol
 \coqdocemptyline
\end{coqdoccode}
It is well-founded and clearly transitive. \begin{coqdoccode}
\coqdocnoindent
\coqdockw{Lemma} \coqdef{Fpred.Normalization.order trans}{order\_trans}{\coqdoclemma{order\_trans}} \coqdocvar{t} \coqdocvar{u} \coqdocvar{v} : \coqref{Fpred.Normalization.order}{\coqdocdefinition{order}} \coqdocvariable{t} \coqdocvariable{u} \coqexternalref{:type scope:x '->' x}{http://coq.inria.fr/distrib/8.5beta2/stdlib/Coq.Init.Logic}{\coqdocnotation{\ensuremath{\rightarrow}}} \coqref{Fpred.Normalization.order}{\coqdocdefinition{order}} \coqdocvariable{u} \coqdocvariable{v} \coqexternalref{:type scope:x '->' x}{http://coq.inria.fr/distrib/8.5beta2/stdlib/Coq.Init.Logic}{\coqdocnotation{\ensuremath{\rightarrow}}} \coqref{Fpred.Normalization.order}{\coqdocdefinition{order}} \coqdocvariable{t} \coqdocvariable{v}.\coqdoceol
\coqdocemptyline
\end{coqdoccode}
As we expected, we can compare a type with an arrow on that type, on the left
    and on the right. \begin{coqdoccode}
\coqdocemptyline
\coqdocnoindent
\coqdockw{Lemma} \coqdef{Fpred.Normalization.order arrow l}{order\_arrow\_l}{\coqdoclemma{order\_arrow\_l}} : \coqdockw{\ensuremath{\forall}} \coqdocvar{A} \coqdocvar{B}, \coqref{Fpred.Normalization.order}{\coqdocdefinition{order}} \coqdocvariable{A} (\coqdocconstructor{arrow} \coqdocvariable{A} \coqdocvariable{B}).\coqdoceol
\coqdocemptyline
\coqdocnoindent
\coqdockw{Lemma} \coqdef{Fpred.Normalization.order arrow r}{order\_arrow\_r}{\coqdoclemma{order\_arrow\_r}} : \coqdockw{\ensuremath{\forall}} \coqdocvar{A} \coqdocvar{B}, \coqref{Fpred.Normalization.order}{\coqdocdefinition{order}} \coqdocvariable{B} (\coqdocconstructor{arrow} \coqdocvariable{A} \coqdocvariable{B}).\coqdoceol
\coqdocemptyline
\end{coqdoccode}
We also define the reflexive closure of this order. It will be useful
    to express the postcondition of the hereditary substitution function,
    as we will explain below. \begin{coqdoccode}
\coqdocemptyline
\coqdocnoindent
\coqdockw{Definition} \coqdef{Fpred.Normalization.ordtyp}{ordtyp}{\coqdocdefinition{ordtyp}} := \coqexternalref{clos refl}{http://coq.inria.fr/distrib/8.5beta2/stdlib/Coq.Relations.Relation\_Operators}{\coqdocinductive{clos\_refl}} \coqref{Fpred.Normalization.order}{\coqdocdefinition{order}}.\coqdoceol
\coqdocemptyline
\end{coqdoccode}
Finally, we define the size of a term as usual. \begin{coqdoccode}
\coqdocemptyline
\coqdocnoindent
\coqdockw{Equations}(\coqdocvar{nocomp}) \coqdef{Fpred.Normalization.term size}{term\_size}{\coqdocdefinition{term\_size}} (\coqdocvar{t} : \coqdocinductive{term}) : \coqexternalref{nat}{http://coq.inria.fr/distrib/8.5beta2/stdlib/Coq.Init.Datatypes}{\coqdocinductive{nat}} :=\coqdoceol
\coqdocnoindent
\coqref{Fpred.Normalization.term size}{\coqdocdefinition{term\_size}} (\coqdocconstructor{var}  \coqdocconstructor{\_})   \ensuremath{\Rightarrow} 0; \coqref{Fpred.Normalization.term size}{\coqdocdefinition{term\_size}} (\coqdocconstructor{abs}  \coqdocconstructor{T} \coqdocconstructor{t}) \ensuremath{\Rightarrow} \coqexternalref{S}{http://coq.inria.fr/distrib/8.5beta2/stdlib/Coq.Init.Datatypes}{\coqdocconstructor{S}} (\coqref{Fpred.Normalization.term size}{\coqdocdefinition{term\_size}} \coqdocvar{t});\coqdoceol
\coqdocnoindent
\coqref{Fpred.Normalization.term size}{\coqdocdefinition{term\_size}} (\coqdocconstructor{app}  \coqdocconstructor{t} \coqdocconstructor{u}) \ensuremath{\Rightarrow} \coqexternalref{S}{http://coq.inria.fr/distrib/8.5beta2/stdlib/Coq.Init.Datatypes}{\coqdocconstructor{S}} (\coqref{Fpred.Normalization.term size}{\coqdocdefinition{term\_size}} \coqdocvar{t} \coqexternalref{:nat scope:x '+' x}{http://coq.inria.fr/distrib/8.5beta2/stdlib/Coq.Init.Peano}{\coqdocnotation{+}} \coqref{Fpred.Normalization.term size}{\coqdocdefinition{term\_size}} \coqdocvar{u});\coqdoceol
\coqdocnoindent
\coqref{Fpred.Normalization.term size}{\coqdocdefinition{term\_size}} (\coqdocconstructor{tabs} \coqdocconstructor{k} \coqdocconstructor{t}) \ensuremath{\Rightarrow} \coqexternalref{S}{http://coq.inria.fr/distrib/8.5beta2/stdlib/Coq.Init.Datatypes}{\coqdocconstructor{S}} (\coqref{Fpred.Normalization.term size}{\coqdocdefinition{term\_size}} \coqdocvar{t}); \coqref{Fpred.Normalization.term size}{\coqdocdefinition{term\_size}} (\coqdocconstructor{tapp} \coqdocconstructor{t} \coqdocconstructor{U}) \ensuremath{\Rightarrow} \coqexternalref{S}{http://coq.inria.fr/distrib/8.5beta2/stdlib/Coq.Init.Datatypes}{\coqdocconstructor{S}} (\coqref{Fpred.Normalization.term size}{\coqdocdefinition{term\_size}} \coqdocvar{t}).\coqdoceol
\coqdocemptyline
\coqdocnoindent
\coqdockw{Definition} \coqdef{Fpred.Normalization.wf term size}{wf\_term\_size}{\coqdocdefinition{wf\_term\_size}} : \coqexternalref{well founded}{http://coq.inria.fr/distrib/8.5beta2/stdlib/Coq.Init.Wf}{\coqdocdefinition{well\_founded}} (\coqref{Fpred.Normalization.MR}{\coqdocdefinition{MR}} \coqref{Fpred.Normalization.term size}{\coqdocdefinition{term\_size}} \coqexternalref{lt}{http://coq.inria.fr/distrib/8.5beta2/stdlib/Coq.Init.Peano}{\coqdocdefinition{lt}}) := \coqexternalref{wf inverse image}{http://coq.inria.fr/distrib/8.5beta2/stdlib/Coq.Wellfounded.Inverse\_Image}{\coqdoclemma{wf\_inverse\_image}} \coqexternalref{lt}{http://coq.inria.fr/distrib/8.5beta2/stdlib/Coq.Init.Peano}{\coqdocdefinition{lt}} \coqref{Fpred.Normalization.term size}{\coqdocdefinition{term\_size}} \coqexternalref{lt wf}{http://coq.inria.fr/distrib/8.5beta2/stdlib/Coq.Arith.Wf\_nat}{\coqdoclemma{lt\_wf}}.\coqdoceol
\coqdocemptyline
\end{coqdoccode}
The hereditary substitution order is a lexicographic combination 
  of the order on the multisets of kinds in the substituted term's type,
  the number of universal quantifiers and type variables in the substituted
  term's type, and the term size of the substituend.
  In other words, with \coqdocvariable{U} the type of the substituted term and \coqdocvariable{t} the
  substituend, we first compare the multiset of kinds in \coqdocvariable{U}, then the
  depth of \coqdocvariable{U}, and ultimately the size of \coqdocvariable{t}. \begin{coqdoccode}
\coqdocemptyline
\coqdocnoindent
\coqdockw{Definition} \coqdef{Fpred.Normalization.her order}{her\_order}{\coqdocdefinition{her\_order}} : \coqexternalref{relation}{http://coq.inria.fr/distrib/8.5beta2/stdlib/Coq.Relations.Relation\_Definitions}{\coqdocdefinition{relation}} (\coqdocinductive{typ} \coqexternalref{:type scope:x '*' x}{http://coq.inria.fr/distrib/8.5beta2/stdlib/Coq.Init.Datatypes}{\coqdocnotation{\ensuremath{\times}}} \coqdocinductive{term}) := \coqdoceol
\coqdocindent{1.00em}
\coqref{Fpred.Normalization.lexprod}{\coqdocinductive{lexprod}} \coqref{Fpred.Normalization.order}{\coqdocdefinition{order}} (\coqdockw{fun} \coqdocvar{x} \coqdocvar{y} \ensuremath{\Rightarrow} \coqref{Fpred.Normalization.kinds of}{\coqdocdefinition{kinds\_of}} \coqdocvariable{x} \coqexternalref{:type scope:x '=' x}{http://coq.inria.fr/distrib/8.5beta2/stdlib/Coq.Init.Logic}{\coqdocnotation{=}} \coqref{Fpred.Normalization.kinds of}{\coqdocdefinition{kinds\_of}} \coqdocvariable{y} \coqexternalref{:type scope:x '/x5C' x}{http://coq.inria.fr/distrib/8.5beta2/stdlib/Coq.Init.Logic}{\coqdocnotation{\ensuremath{\land}}} \coqref{Fpred.Normalization.depth}{\coqdocdefinition{depth}} \coqdocvariable{x} \coqexternalref{:type scope:x '=' x}{http://coq.inria.fr/distrib/8.5beta2/stdlib/Coq.Init.Logic}{\coqdocnotation{=}} \coqref{Fpred.Normalization.depth}{\coqdocdefinition{depth}} \coqdocvariable{y}) (\coqref{Fpred.Normalization.MR}{\coqdocdefinition{MR}} \coqref{Fpred.Normalization.term size}{\coqdocdefinition{term\_size}} \coqexternalref{lt}{http://coq.inria.fr/distrib/8.5beta2/stdlib/Coq.Init.Peano}{\coqdocdefinition{lt}}).\coqdoceol
\coqdocemptyline
\coqdocnoindent
\coqdockw{Instance} \coqdef{Fpred.Normalization.WF her order}{WF\_her\_order}{\coqdocinstance{WF\_her\_order}} : \coqdocclass{WellFounded} \coqref{Fpred.Normalization.her order}{\coqdocdefinition{her\_order}}.\coqdoceol
\coqdocemptyline
\coqdocemptyline
\end{coqdoccode}
\subsection{The model}

  We now turn to the interpretation proper. We characterize the normal forms as a
 subset of the terms using mutually-inductive \coqref{Fpred.Normalization.normal}{\coqdocinductive{normal}} and \coqref{Fpred.Normalization.neutral}{\coqdocinductive{neutral}} judgments.
 The plan is to show that the hereditary substitution function, when given two terms in 
 \coqref{Fpred.Normalization.normal}{\coqdocinductive{normal}} form will produce terms in \coqref{Fpred.Normalization.normal}{\coqdocinductive{normal}} form. We can already expect some complications
 as normal terms also include \coqref{Fpred.Normalization.neutral}{\coqdocinductive{neutral}} ones... \begin{coqdoccode}
\coqdocemptyline
\coqdocnoindent
\coqdockw{Inductive} \coqdef{Fpred.Normalization.normal}{normal}{\coqdocinductive{normal}} : \coqdocinductive{term} \coqexternalref{:type scope:x '->' x}{http://coq.inria.fr/distrib/8.5beta2/stdlib/Coq.Init.Logic}{\coqdocnotation{\ensuremath{\rightarrow}}} \coqdockw{Prop} :=\coqdoceol
\coqdocnoindent
\ensuremath{|} \coqdef{Fpred.Normalization.normal abs}{normal\_abs}{\coqdocconstructor{normal\_abs}} \coqdocvar{T} \coqdocvar{t} : \coqref{Fpred.Normalization.normal}{\coqdocinductive{normal}} \coqdocvariable{t} \coqexternalref{:type scope:x '->' x}{http://coq.inria.fr/distrib/8.5beta2/stdlib/Coq.Init.Logic}{\coqdocnotation{\ensuremath{\rightarrow}}} \coqref{Fpred.Normalization.normal}{\coqdocinductive{normal}} (\coqdocconstructor{abs} \coqdocvariable{T} \coqdocvariable{t})\coqdoceol
\coqdocnoindent
\ensuremath{|} \coqdef{Fpred.Normalization.normal tabs}{normal\_tabs}{\coqdocconstructor{normal\_tabs}} \coqdocvar{k} \coqdocvar{t} : \coqref{Fpred.Normalization.normal}{\coqdocinductive{normal}} \coqdocvariable{t} \coqexternalref{:type scope:x '->' x}{http://coq.inria.fr/distrib/8.5beta2/stdlib/Coq.Init.Logic}{\coqdocnotation{\ensuremath{\rightarrow}}} \coqref{Fpred.Normalization.normal}{\coqdocinductive{normal}} (\coqdocconstructor{tabs} \coqdocvariable{k} \coqdocvariable{t})\coqdoceol
\coqdocnoindent
\ensuremath{|} \coqdef{Fpred.Normalization.normal neutral}{normal\_neutral}{\coqdocconstructor{normal\_neutral}} \coqdocvar{r} : \coqref{Fpred.Normalization.neutral}{\coqdocinductive{neutral}} \coqdocvariable{r} \coqexternalref{:type scope:x '->' x}{http://coq.inria.fr/distrib/8.5beta2/stdlib/Coq.Init.Logic}{\coqdocnotation{\ensuremath{\rightarrow}}} \coqref{Fpred.Normalization.normal}{\coqdocinductive{normal}} \coqdocvariable{r}\coqdoceol
\coqdocnoindent
\coqdockw{with} \coqdef{Fpred.Normalization.neutral}{neutral}{\coqdocinductive{neutral}} : \coqdocinductive{term} \coqexternalref{:type scope:x '->' x}{http://coq.inria.fr/distrib/8.5beta2/stdlib/Coq.Init.Logic}{\coqdocnotation{\ensuremath{\rightarrow}}} \coqdockw{Prop} :=\coqdoceol
\coqdocnoindent
\ensuremath{|} \coqdef{Fpred.Normalization.neutral var}{neutral\_var}{\coqdocconstructor{neutral\_var}} \coqdocvar{i} : \coqref{Fpred.Normalization.neutral}{\coqdocinductive{neutral}} (\coqdocconstructor{var} \coqdocvariable{i})\coqdoceol
\coqdocnoindent
\ensuremath{|} \coqdef{Fpred.Normalization.neutral app}{neutral\_app}{\coqdocconstructor{neutral\_app}} \coqdocvar{t} \coqdocvar{n} : \coqref{Fpred.Normalization.neutral}{\coqdocinductive{neutral}} \coqdocvariable{t} \coqexternalref{:type scope:x '->' x}{http://coq.inria.fr/distrib/8.5beta2/stdlib/Coq.Init.Logic}{\coqdocnotation{\ensuremath{\rightarrow}}} \coqref{Fpred.Normalization.normal}{\coqdocinductive{normal}} \coqdocvariable{n} \coqexternalref{:type scope:x '->' x}{http://coq.inria.fr/distrib/8.5beta2/stdlib/Coq.Init.Logic}{\coqdocnotation{\ensuremath{\rightarrow}}} \coqref{Fpred.Normalization.neutral}{\coqdocinductive{neutral}} (\coqdocconstructor{app} \coqdocvariable{t} \coqdocvariable{n})\coqdoceol
\coqdocnoindent
\ensuremath{|} \coqdef{Fpred.Normalization.neutral tapp}{neutral\_tapp}{\coqdocconstructor{neutral\_tapp}} \coqdocvar{t} \coqdocvar{T} : \coqref{Fpred.Normalization.neutral}{\coqdocinductive{neutral}} \coqdocvariable{t} \coqexternalref{:type scope:x '->' x}{http://coq.inria.fr/distrib/8.5beta2/stdlib/Coq.Init.Logic}{\coqdocnotation{\ensuremath{\rightarrow}}} \coqref{Fpred.Normalization.neutral}{\coqdocinductive{neutral}} (\coqdocconstructor{tapp} \coqdocvariable{t} \coqdocvariable{T}).\coqdoceol
\coqdocemptyline
\coqdocemptyline
\end{coqdoccode}
A term \coqdocvariable{t} is said to be a canonical inhabitant of a type \coqdocvariable{T} in
    environment \coqdocvariable{e} if \coqdocvariable{e} \ensuremath{\vdash} \coqdocvariable{t} : \coqdocvariable{T} and \coqdocvariable{t} is in normal form. Our goal
    will be to show that every typeable term can be normalized to a 
    canonical one. \begin{coqdoccode}
\coqdocemptyline
\coqdocnoindent
\coqdockw{Definition} \coqdef{Fpred.Normalization.canonical}{canonical}{\coqdocdefinition{canonical}} \coqdocvar{e} \coqdocvar{t} \coqdocvar{T} := \coqdocinductive{typing} \coqdocvariable{e} \coqdocvariable{t} \coqdocvariable{T} \coqexternalref{:type scope:x '/x5C' x}{http://coq.inria.fr/distrib/8.5beta2/stdlib/Coq.Init.Logic}{\coqdocnotation{\ensuremath{\land}}} \coqref{Fpred.Normalization.normal}{\coqdocinductive{normal}} \coqdocvariable{t}.\coqdoceol
\coqdocemptyline
\coqdocemptyline
\end{coqdoccode}
We define a relation expressing that \coqdocvariable{n} is the interpretation of some arbitrary term \coqdocvariable{t}
  of type \coqdocvariable{T} and in environment \coqdocvariable{e}. \begin{coqdoccode}
\coqdocnoindent
\coqdockw{Definition} \coqdef{Fpred.Normalization.interp}{interp}{\coqdocdefinition{interp}} \coqdocvar{e} \coqdocvar{t} \coqdocvar{T} \coqdocvar{n} := \coqdocdefinition{reds} \coqdocvariable{t} \coqdocvariable{n} \coqexternalref{:type scope:x '/x5C' x}{http://coq.inria.fr/distrib/8.5beta2/stdlib/Coq.Init.Logic}{\coqdocnotation{\ensuremath{\land}}} \coqref{Fpred.Normalization.canonical}{\coqdocdefinition{canonical}} \coqdocvariable{e} \coqdocvariable{n} \coqdocvariable{T}.\coqdoceol
\coqdocemptyline
\coqdocemptyline
\end{coqdoccode}
\subsection{Hereditary substitution}

  As we said, hereditary substitution takes two terms \coqdocvariable{t} and \coqdocvariable{u} in \coqref{Fpred.Normalization.normal}{\coqdocinductive{normal}}
  form and returns a term which is the result of substituting \coqdocvariable{u} in \coqdocvariable{t} at
  some index. From a purely algorithmic point of view, we only need \coqdocvariable{t}, \coqdocvariable{u}
  and the index \coqdocvariable{X} to compute the result of this function. However, we need
  more to prove its correctness.

  First of all, the well-founded order that we use to justify its termination
  is an order on the type of the substituted and on the substituend, which
  is why the function \coqref{Fpred.Normalization.hsubst}{\coqdocdefinition{hsubst}} also takes as an argument the type
  of the substituted term.

  We then need a typing environment for \coqdocvariable{t} and \coqdocvariable{u}, which is not useful from
  a computational point of view but will serve to prove the termination and
  the correctness of \coqref{Fpred.Normalization.hsubst}{\coqdocdefinition{hsubst}}. To this effect, we will decorate
  the function with a precondition and a postcondition. We define those
  in the universe of propositions to underline the fact that they are not
  useful in a computational way. \begin{coqdoccode}
\coqdocemptyline
\coqdocnoindent
\coqdockw{Definition} \coqdef{Fpred.Normalization.pre}{pre}{\coqdocdefinition{pre}} (\coqdocvar{t} : \coqdocinductive{typ} \coqexternalref{:type scope:x '*' x}{http://coq.inria.fr/distrib/8.5beta2/stdlib/Coq.Init.Datatypes}{\coqdocnotation{\ensuremath{\times}}} \coqdocinductive{term}) (\coqdocvar{u} : \coqdocinductive{term}) (\coqdocvar{X} : \coqexternalref{nat}{http://coq.inria.fr/distrib/8.5beta2/stdlib/Coq.Init.Datatypes}{\coqdocinductive{nat}}) (\coqdocvar{p} : \coqdocinductive{env} \coqexternalref{:type scope:x '*' x}{http://coq.inria.fr/distrib/8.5beta2/stdlib/Coq.Init.Datatypes}{\coqdocnotation{\ensuremath{\times}}} \coqdocinductive{typ}) : \coqdockw{Prop} :=\coqdoceol
\coqdocindent{2.00em}
(\coqdocdefinition{get\_var} (\coqexternalref{fst}{http://coq.inria.fr/distrib/8.5beta2/stdlib/Coq.Init.Datatypes}{\coqdocdefinition{fst}} \coqdocvariable{p}) \coqdocvariable{X} \coqexternalref{:type scope:x '=' x}{http://coq.inria.fr/distrib/8.5beta2/stdlib/Coq.Init.Logic}{\coqdocnotation{=}} \coqexternalref{Some}{http://coq.inria.fr/distrib/8.5beta2/stdlib/Coq.Init.Datatypes}{\coqdocconstructor{Some}} (\coqexternalref{fst}{http://coq.inria.fr/distrib/8.5beta2/stdlib/Coq.Init.Datatypes}{\coqdocdefinition{fst}} \coqdocvariable{t}) \coqexternalref{:type scope:x '/x5C' x}{http://coq.inria.fr/distrib/8.5beta2/stdlib/Coq.Init.Logic}{\coqdocnotation{\ensuremath{\land}}} \coqref{Fpred.Normalization.canonical}{\coqdocdefinition{canonical}} (\coqexternalref{fst}{http://coq.inria.fr/distrib/8.5beta2/stdlib/Coq.Init.Datatypes}{\coqdocdefinition{fst}} \coqdocvariable{p}) (\coqexternalref{snd}{http://coq.inria.fr/distrib/8.5beta2/stdlib/Coq.Init.Datatypes}{\coqdocdefinition{snd}} \coqdocvariable{t}) (\coqexternalref{snd}{http://coq.inria.fr/distrib/8.5beta2/stdlib/Coq.Init.Datatypes}{\coqdocdefinition{snd}} \coqdocvariable{p}) \coqdoceol
\coqdocindent{2.50em}
\coqexternalref{:type scope:x '/x5C' x}{http://coq.inria.fr/distrib/8.5beta2/stdlib/Coq.Init.Logic}{\coqdocnotation{\ensuremath{\land}}} \coqref{Fpred.Normalization.canonical}{\coqdocdefinition{canonical}} (\coqdocdefinition{remove\_var} (\coqexternalref{fst}{http://coq.inria.fr/distrib/8.5beta2/stdlib/Coq.Init.Datatypes}{\coqdocdefinition{fst}} \coqdocvariable{p}) \coqdocvariable{X}) \coqdocvariable{u} (\coqexternalref{fst}{http://coq.inria.fr/distrib/8.5beta2/stdlib/Coq.Init.Datatypes}{\coqdocdefinition{fst}} \coqdocvariable{t})).\coqdoceol
\coqdocemptyline
\coqdocemptyline
\end{coqdoccode}
There is one subtle point in the formulation of the
    postcondition. When we substitute in an application \coqdocconstructor{app} \coqdocvar{t1} \coqdocvar{t2}, it
    may be that the result of substituting in \coqdocvar{t1} is an abstraction
    \coqdocconstructor{abs} \coqdocvariable{T} \coqdocvariable{t}. If that's the case, to preserve the invariant that the
    result of \coqref{Fpred.Normalization.hsubst}{\coqdocdefinition{hsubst}} is in normal form, we have to call
    again \coqref{Fpred.Normalization.hsubst}{\coqdocdefinition{hsubst}} to perform the beta-reduction. However to
    do this, we need to know that the type of \coqdocvar{t1} is smaller than the
    type of the substituted term.  Note that we need to add a
    side-condition to this property, which is not always true: if the
    substituted variable did not appear at all, then there is no reason
    to have any relation between those types. There is a relation only if
    the original term was not an abstraction but the substituted
    term is. \begin{coqdoccode}
\coqdocemptyline
\coqdocnoindent
\coqdockw{Equations} \coqdef{Fpred.Normalization.is abs}{is\_abs}{\coqdocdefinition{is\_abs}} (\coqdocvar{t} : \coqdocinductive{term}) : \coqdockw{Prop} :=\coqdoceol
\coqdocnoindent
\coqref{Fpred.Normalization.is abs}{\coqdocdefinition{is\_abs}} (\coqdocconstructor{abs}  \coqdocconstructor{\_} \coqdocconstructor{\_}) \ensuremath{\Rightarrow} \coqexternalref{True}{http://coq.inria.fr/distrib/8.5beta2/stdlib/Coq.Init.Logic}{\coqdocinductive{True}}; \coqref{Fpred.Normalization.is abs}{\coqdocdefinition{is\_abs}} (\coqdocconstructor{tabs} \coqdocconstructor{\_} \coqdocconstructor{\_}) \ensuremath{\Rightarrow} \coqexternalref{True}{http://coq.inria.fr/distrib/8.5beta2/stdlib/Coq.Init.Logic}{\coqdocinductive{True}}; \coqref{Fpred.Normalization.is abs}{\coqdocdefinition{is\_abs}}  \coqdocvar{\_}         \ensuremath{\Rightarrow} \coqexternalref{False}{http://coq.inria.fr/distrib/8.5beta2/stdlib/Coq.Init.Logic}{\coqdocinductive{False}}.\coqdoceol
\coqdocemptyline
\coqdocemptyline
\coqdocnoindent
\coqdockw{Definition} \coqdef{Fpred.Normalization.post}{post}{\coqdocdefinition{post}} (\coqdocvar{t} : \coqdocinductive{typ} \coqexternalref{:type scope:x '*' x}{http://coq.inria.fr/distrib/8.5beta2/stdlib/Coq.Init.Datatypes}{\coqdocnotation{\ensuremath{\times}}} \coqdocinductive{term}) (\coqdocvar{u} : \coqdocinductive{term}) (\coqdocvar{X} : \coqexternalref{nat}{http://coq.inria.fr/distrib/8.5beta2/stdlib/Coq.Init.Datatypes}{\coqdocinductive{nat}}) (\coqdocvar{r} : \coqdocinductive{term}) (\coqdocvar{p} : \coqdocinductive{env} \coqexternalref{:type scope:x '*' x}{http://coq.inria.fr/distrib/8.5beta2/stdlib/Coq.Init.Datatypes}{\coqdocnotation{\ensuremath{\times}}} \coqdocinductive{typ}) : \coqdockw{Prop} :=\coqdoceol
\coqdocindent{1.00em}
\coqref{Fpred.Normalization.interp}{\coqdocdefinition{interp}} (\coqdocdefinition{remove\_var} (\coqexternalref{fst}{http://coq.inria.fr/distrib/8.5beta2/stdlib/Coq.Init.Datatypes}{\coqdocdefinition{fst}} \coqdocvariable{p}) \coqdocvariable{X}) (\coqdocdefinition{subst} (\coqexternalref{snd}{http://coq.inria.fr/distrib/8.5beta2/stdlib/Coq.Init.Datatypes}{\coqdocdefinition{snd}} \coqdocvariable{t}) \coqdocvariable{X} \coqdocvariable{u}) (\coqexternalref{snd}{http://coq.inria.fr/distrib/8.5beta2/stdlib/Coq.Init.Datatypes}{\coqdocdefinition{snd}} \coqdocvariable{p}) \coqdocvariable{r} \coqexternalref{:type scope:x '/x5C' x}{http://coq.inria.fr/distrib/8.5beta2/stdlib/Coq.Init.Logic}{\coqdocnotation{\ensuremath{\land}}}\coqdoceol
\coqdocindent{1.00em}
\coqexternalref{:type scope:x '/x5C' x}{http://coq.inria.fr/distrib/8.5beta2/stdlib/Coq.Init.Logic}{\coqdocnotation{(}}\coqexternalref{:type scope:'x7E' x}{http://coq.inria.fr/distrib/8.5beta2/stdlib/Coq.Init.Logic}{\coqdocnotation{\ensuremath{\lnot}}} \coqref{Fpred.Normalization.is abs}{\coqdocdefinition{is\_abs}} (\coqexternalref{snd}{http://coq.inria.fr/distrib/8.5beta2/stdlib/Coq.Init.Datatypes}{\coqdocdefinition{snd}} \coqdocvariable{t}) \coqexternalref{:type scope:x '->' x}{http://coq.inria.fr/distrib/8.5beta2/stdlib/Coq.Init.Logic}{\coqdocnotation{\ensuremath{\rightarrow}}} \coqref{Fpred.Normalization.is abs}{\coqdocdefinition{is\_abs}} \coqdocvariable{r} \coqexternalref{:type scope:x '->' x}{http://coq.inria.fr/distrib/8.5beta2/stdlib/Coq.Init.Logic}{\coqdocnotation{\ensuremath{\rightarrow}}} \coqref{Fpred.Normalization.ordtyp}{\coqdocdefinition{ordtyp}} (\coqexternalref{snd}{http://coq.inria.fr/distrib/8.5beta2/stdlib/Coq.Init.Datatypes}{\coqdocdefinition{snd}} \coqdocvariable{p}) (\coqexternalref{fst}{http://coq.inria.fr/distrib/8.5beta2/stdlib/Coq.Init.Datatypes}{\coqdocdefinition{fst}} \coqdocvariable{t})\coqexternalref{:type scope:x '/x5C' x}{http://coq.inria.fr/distrib/8.5beta2/stdlib/Coq.Init.Logic}{\coqdocnotation{)}}.\coqdoceol
\coqdocemptyline
\end{coqdoccode}
The Program mode proposed by \Coq interacts nicely with \Equations, in
    that it allows us to just return a term, and provide later the
    postcondition, thanks to subtyping of subset types. In the same way,
    we can treat a value returned by a call as if it was just the
    term. We use a standard encoding for the ghost \coqdocvariable{p} : \coqdocinductive{env} \ensuremath{\times} \coqdocinductive{typ} variable.
    The (noind) option disables the generation of the graph and elimination principle
    for the function, its type and computational behavior is all we need here. \begin{coqdoccode}
\coqdocemptyline
\coqdocemptyline
\coqdocnoindent
\coqdockw{Equations}(\coqdocvar{noind}) \coqdef{Fpred.Normalization.hsubst}{hsubst}{\coqdocdefinition{hsubst}} (\coqdocvar{t} : \coqdocinductive{typ} \coqexternalref{:type scope:x '*' x}{http://coq.inria.fr/distrib/8.5beta2/stdlib/Coq.Init.Datatypes}{\coqdocnotation{\ensuremath{\times}}} \coqdocinductive{term}) (\coqdocvar{u} : \coqdocinductive{term}) (\coqdocvar{X} : \coqexternalref{nat}{http://coq.inria.fr/distrib/8.5beta2/stdlib/Coq.Init.Datatypes}{\coqdocinductive{nat}}) (\coqdocvar{P} : \coqexternalref{:type scope:'exists' x '..' x ',' x}{http://coq.inria.fr/distrib/8.5beta2/stdlib/Coq.Init.Logic}{\coqdocnotation{\ensuremath{\exists}}} \coqexternalref{:type scope:'exists' x '..' x ',' x}{http://coq.inria.fr/distrib/8.5beta2/stdlib/Coq.Init.Logic}{\coqdocnotation{(}}\coqdocvar{p} : \coqdocinductive{env} \coqexternalref{:type scope:x '*' x}{http://coq.inria.fr/distrib/8.5beta2/stdlib/Coq.Init.Datatypes}{\coqdocnotation{\ensuremath{\times}}} \coqdocinductive{typ}\coqexternalref{:type scope:'exists' x '..' x ',' x}{http://coq.inria.fr/distrib/8.5beta2/stdlib/Coq.Init.Logic}{\coqdocnotation{),}} \coqref{Fpred.Normalization.pre}{\coqdocdefinition{pre}} \coqdocvariable{t} \coqdocvariable{u} \coqdocvariable{X} \coqdocvariable{p}) :\coqdoceol
\coqdocindent{1.00em}
\coqexternalref{:type scope:'x7B' x ':' x '|' x 'x7D'}{http://coq.inria.fr/distrib/8.5beta2/stdlib/Coq.Init.Specif}{\coqdocnotation{\{}}\coqdocvar{r} \coqexternalref{:type scope:'x7B' x ':' x '|' x 'x7D'}{http://coq.inria.fr/distrib/8.5beta2/stdlib/Coq.Init.Specif}{\coqdocnotation{:}} \coqdocinductive{term} \coqexternalref{:type scope:'x7B' x ':' x '|' x 'x7D'}{http://coq.inria.fr/distrib/8.5beta2/stdlib/Coq.Init.Specif}{\coqdocnotation{\ensuremath{|}}} \coqdockw{\ensuremath{\forall}} (\coqdocvar{p} : \coqdocinductive{env} \coqexternalref{:type scope:x '*' x}{http://coq.inria.fr/distrib/8.5beta2/stdlib/Coq.Init.Datatypes}{\coqdocnotation{\ensuremath{\times}}} \coqdocinductive{typ}), \coqref{Fpred.Normalization.pre}{\coqdocdefinition{pre}} \coqdocvar{t} \coqdocvar{u} \coqdocvar{X} \coqdocvariable{p} \coqexternalref{:type scope:x '->' x}{http://coq.inria.fr/distrib/8.5beta2/stdlib/Coq.Init.Logic}{\coqdocnotation{\ensuremath{\rightarrow}}} \coqref{Fpred.Normalization.post}{\coqdocdefinition{post}} \coqdocvar{t} \coqdocvar{u} \coqdocvar{X} \coqdocvar{r} \coqdocvariable{p}\coqexternalref{:type scope:'x7B' x ':' x '|' x 'x7D'}{http://coq.inria.fr/distrib/8.5beta2/stdlib/Coq.Init.Specif}{\coqdocnotation{\}}} :=\coqdoceol
\coqdocnoindent
\coqref{Fpred.Normalization.hsubst}{\coqdocdefinition{hsubst}} \coqdocvar{t} \coqdocvar{u} \coqdocvar{X} \coqdocvar{P} \coqdoctac{by} \coqdocvar{rec} \coqdocvar{t} \coqref{Fpred.Normalization.her order}{\coqdocdefinition{her\_order}} \ensuremath{\Rightarrow}\coqdoceol
\coqdocnoindent
\coqref{Fpred.Normalization.hsubst}{\coqdocdefinition{hsubst}} (\coqexternalref{pair}{http://coq.inria.fr/distrib/8.5beta2/stdlib/Coq.Init.Datatypes}{\coqdocconstructor{pair}} \coqexternalref{pair}{http://coq.inria.fr/distrib/8.5beta2/stdlib/Coq.Init.Datatypes}{\coqdocconstructor{U}} \coqexternalref{pair}{http://coq.inria.fr/distrib/8.5beta2/stdlib/Coq.Init.Datatypes}{\coqdocconstructor{t}}) \coqdocvar{u} \coqdocvar{X} \coqdocvar{P} $\Leftarrow$ \coqdocvar{t} \ensuremath{\Rightarrow} \{\coqdoceol
\coqdocindent{1.00em}
\ensuremath{|} \coqdocconstructor{var} \coqdocconstructor{i} $\Leftarrow$ \coqexternalref{lt eq lt dec}{http://coq.inria.fr/distrib/8.5beta2/stdlib/Coq.Arith.Compare\_dec}{\coqdocdefinition{lt\_eq\_lt\_dec}} \coqdocvar{i} \coqdocvar{X} \ensuremath{\Rightarrow} \{\coqdoceol
\coqdocindent{2.00em}
\ensuremath{|} \coqexternalref{inleft}{http://coq.inria.fr/distrib/8.5beta2/stdlib/Coq.Init.Specif}{\coqdocconstructor{inleft}} \coqexternalref{inleft}{http://coq.inria.fr/distrib/8.5beta2/stdlib/Coq.Init.Specif}{\coqdocconstructor{(}}\coqexternalref{inleft}{http://coq.inria.fr/distrib/8.5beta2/stdlib/Coq.Init.Specif}{\coqdocconstructor{right}} \coqexternalref{inleft}{http://coq.inria.fr/distrib/8.5beta2/stdlib/Coq.Init.Specif}{\coqdocconstructor{p}}\coqexternalref{inleft}{http://coq.inria.fr/distrib/8.5beta2/stdlib/Coq.Init.Specif}{\coqdocconstructor{)}} \ensuremath{\Rightarrow} \coqdocvar{u}; \ensuremath{|} \coqexternalref{inleft}{http://coq.inria.fr/distrib/8.5beta2/stdlib/Coq.Init.Specif}{\coqdocconstructor{inleft}} \coqexternalref{inleft}{http://coq.inria.fr/distrib/8.5beta2/stdlib/Coq.Init.Specif}{\coqdocconstructor{(}}\coqexternalref{inleft}{http://coq.inria.fr/distrib/8.5beta2/stdlib/Coq.Init.Specif}{\coqdocconstructor{left}} \coqexternalref{inleft}{http://coq.inria.fr/distrib/8.5beta2/stdlib/Coq.Init.Specif}{\coqdocconstructor{p}}\coqexternalref{inleft}{http://coq.inria.fr/distrib/8.5beta2/stdlib/Coq.Init.Specif}{\coqdocconstructor{)}} \ensuremath{\Rightarrow} \coqdocconstructor{var} \coqdocvar{i};\coqdoceol
\coqdocindent{2.00em}
\ensuremath{|} \coqexternalref{inright}{http://coq.inria.fr/distrib/8.5beta2/stdlib/Coq.Init.Specif}{\coqdocconstructor{inright}} \coqexternalref{inright}{http://coq.inria.fr/distrib/8.5beta2/stdlib/Coq.Init.Specif}{\coqdocconstructor{p}} \ensuremath{\Rightarrow} \coqdocconstructor{var} (\coqexternalref{pred}{http://coq.inria.fr/distrib/8.5beta2/stdlib/Coq.Init.Peano}{\coqdocabbreviation{pred}} \coqdocvar{i}) \};\coqdoceol
\coqdocindent{1.00em}
\ensuremath{|} \coqdocconstructor{abs} \coqdocconstructor{T} \coqdocconstructor{t} \ensuremath{\Rightarrow} \coqdocconstructor{abs} \coqdocvar{T} (\coqref{Fpred.Normalization.hsubst comp proj}{\coqdocdefinition{hsubst}} \coqexternalref{:core scope:'(' x ',' x ',' '..' ',' x ')'}{http://coq.inria.fr/distrib/8.5beta2/stdlib/Coq.Init.Datatypes}{\coqdocnotation{(}}\coqdocvar{U}\coqexternalref{:core scope:'(' x ',' x ',' '..' ',' x ')'}{http://coq.inria.fr/distrib/8.5beta2/stdlib/Coq.Init.Datatypes}{\coqdocnotation{,}} \coqdocvar{t}\coqexternalref{:core scope:'(' x ',' x ',' '..' ',' x ')'}{http://coq.inria.fr/distrib/8.5beta2/stdlib/Coq.Init.Datatypes}{\coqdocnotation{)}} (\coqdocdefinition{shift} 0 \coqdocvar{u}) (\coqexternalref{S}{http://coq.inria.fr/distrib/8.5beta2/stdlib/Coq.Init.Datatypes}{\coqdocconstructor{S}} \coqdocvar{X}) \coqdocvar{\_}) ;\coqdoceol
\coqdocindent{1.00em}
\ensuremath{|} \coqdocconstructor{tabs} \coqdocconstructor{k} \coqdocconstructor{t} \ensuremath{\Rightarrow} \coqdocconstructor{tabs} \coqdocvar{k} (\coqref{Fpred.Normalization.hsubst comp proj}{\coqdocdefinition{hsubst}} \coqexternalref{:core scope:'(' x ',' x ',' '..' ',' x ')'}{http://coq.inria.fr/distrib/8.5beta2/stdlib/Coq.Init.Datatypes}{\coqdocnotation{(}}\coqdocdefinition{tshift} 0 \coqdocvar{U}\coqexternalref{:core scope:'(' x ',' x ',' '..' ',' x ')'}{http://coq.inria.fr/distrib/8.5beta2/stdlib/Coq.Init.Datatypes}{\coqdocnotation{,}} \coqdocvar{t}\coqexternalref{:core scope:'(' x ',' x ',' '..' ',' x ')'}{http://coq.inria.fr/distrib/8.5beta2/stdlib/Coq.Init.Datatypes}{\coqdocnotation{)}} (\coqdocdefinition{shift\_typ} 0 \coqdocvar{u}) \coqdocvar{X} \coqdocvar{\_}) ;\coqdoceol
\coqdocindent{1.00em}
\ensuremath{|} \coqdocconstructor{tapp} \coqdocconstructor{t} \coqdocconstructor{T} $\Leftarrow$ \coqref{Fpred.Normalization.hsubst comp proj}{\coqdocdefinition{hsubst}} \coqexternalref{:core scope:'(' x ',' x ',' '..' ',' x ')'}{http://coq.inria.fr/distrib/8.5beta2/stdlib/Coq.Init.Datatypes}{\coqdocnotation{(}}\coqdocvar{U}\coqexternalref{:core scope:'(' x ',' x ',' '..' ',' x ')'}{http://coq.inria.fr/distrib/8.5beta2/stdlib/Coq.Init.Datatypes}{\coqdocnotation{,}} \coqdocvar{t}\coqexternalref{:core scope:'(' x ',' x ',' '..' ',' x ')'}{http://coq.inria.fr/distrib/8.5beta2/stdlib/Coq.Init.Datatypes}{\coqdocnotation{)}} \coqdocvar{u} \coqdocvar{X} \coqdocvar{\_} \ensuremath{\Rightarrow} \{\coqdoceol
\coqdocindent{2.00em}
\ensuremath{|} \coqexternalref{exist}{http://coq.inria.fr/distrib/8.5beta2/stdlib/Coq.Init.Specif}{\coqdocconstructor{exist}} \coqexternalref{exist}{http://coq.inria.fr/distrib/8.5beta2/stdlib/Coq.Init.Specif}{\coqdocconstructor{(}}\coqexternalref{exist}{http://coq.inria.fr/distrib/8.5beta2/stdlib/Coq.Init.Specif}{\coqdocconstructor{tabs}} \coqexternalref{exist}{http://coq.inria.fr/distrib/8.5beta2/stdlib/Coq.Init.Specif}{\coqdocconstructor{k}} \coqexternalref{exist}{http://coq.inria.fr/distrib/8.5beta2/stdlib/Coq.Init.Specif}{\coqdocconstructor{t'}}\coqexternalref{exist}{http://coq.inria.fr/distrib/8.5beta2/stdlib/Coq.Init.Specif}{\coqdocconstructor{)}} \coqexternalref{exist}{http://coq.inria.fr/distrib/8.5beta2/stdlib/Coq.Init.Specif}{\coqdocconstructor{P}} \ensuremath{\Rightarrow} \coqdocdefinition{subst\_typ} \coqdocvar{t'} 0 \coqdocvar{T};\coqdoceol
\coqdocindent{2.00em}
\ensuremath{|} \coqexternalref{exist}{http://coq.inria.fr/distrib/8.5beta2/stdlib/Coq.Init.Specif}{\coqdocconstructor{exist}} \coqexternalref{exist}{http://coq.inria.fr/distrib/8.5beta2/stdlib/Coq.Init.Specif}{\coqdocconstructor{r}} \coqexternalref{exist}{http://coq.inria.fr/distrib/8.5beta2/stdlib/Coq.Init.Specif}{\coqdocconstructor{P}} \ensuremath{\Rightarrow} \coqdocconstructor{tapp} \coqdocvar{r} \coqdocvar{T} \};\coqdoceol
\coqdocindent{1.00em}
\ensuremath{|} \coqdocconstructor{app} \coqdocconstructor{t1} \coqdocconstructor{t2} $\Leftarrow$ \coqref{Fpred.Normalization.hsubst comp proj}{\coqdocdefinition{hsubst}} \coqexternalref{:core scope:'(' x ',' x ',' '..' ',' x ')'}{http://coq.inria.fr/distrib/8.5beta2/stdlib/Coq.Init.Datatypes}{\coqdocnotation{(}}\coqdocvar{U}\coqexternalref{:core scope:'(' x ',' x ',' '..' ',' x ')'}{http://coq.inria.fr/distrib/8.5beta2/stdlib/Coq.Init.Datatypes}{\coqdocnotation{,}} \coqdocvar{t2}\coqexternalref{:core scope:'(' x ',' x ',' '..' ',' x ')'}{http://coq.inria.fr/distrib/8.5beta2/stdlib/Coq.Init.Datatypes}{\coqdocnotation{)}} \coqdocvar{u} \coqdocvar{X} \coqdocvar{\_} \ensuremath{\Rightarrow} \{\coqdoceol
\coqdocindent{2.00em}
\ensuremath{|} \coqexternalref{exist}{http://coq.inria.fr/distrib/8.5beta2/stdlib/Coq.Init.Specif}{\coqdocconstructor{exist}} \coqexternalref{exist}{http://coq.inria.fr/distrib/8.5beta2/stdlib/Coq.Init.Specif}{\coqdocconstructor{r2}} \coqexternalref{exist}{http://coq.inria.fr/distrib/8.5beta2/stdlib/Coq.Init.Specif}{\coqdocconstructor{P2}} $\Leftarrow$ \coqref{Fpred.Normalization.hsubst comp proj}{\coqdocdefinition{hsubst}} \coqexternalref{:core scope:'(' x ',' x ',' '..' ',' x ')'}{http://coq.inria.fr/distrib/8.5beta2/stdlib/Coq.Init.Datatypes}{\coqdocnotation{(}}\coqdocvar{U}\coqexternalref{:core scope:'(' x ',' x ',' '..' ',' x ')'}{http://coq.inria.fr/distrib/8.5beta2/stdlib/Coq.Init.Datatypes}{\coqdocnotation{,}} \coqdocvar{t1}\coqexternalref{:core scope:'(' x ',' x ',' '..' ',' x ')'}{http://coq.inria.fr/distrib/8.5beta2/stdlib/Coq.Init.Datatypes}{\coqdocnotation{)}} \coqdocvar{u} \coqdocvar{X} \coqdocvar{\_} \ensuremath{\Rightarrow} \{\coqdoceol
\coqdocindent{3.00em}
\ensuremath{|} \coqexternalref{exist}{http://coq.inria.fr/distrib/8.5beta2/stdlib/Coq.Init.Specif}{\coqdocconstructor{exist}} \coqexternalref{exist}{http://coq.inria.fr/distrib/8.5beta2/stdlib/Coq.Init.Specif}{\coqdocconstructor{(}}\coqexternalref{exist}{http://coq.inria.fr/distrib/8.5beta2/stdlib/Coq.Init.Specif}{\coqdocconstructor{abs}} \coqexternalref{exist}{http://coq.inria.fr/distrib/8.5beta2/stdlib/Coq.Init.Specif}{\coqdocconstructor{T'}} \coqexternalref{exist}{http://coq.inria.fr/distrib/8.5beta2/stdlib/Coq.Init.Specif}{\coqdocconstructor{t'}}\coqexternalref{exist}{http://coq.inria.fr/distrib/8.5beta2/stdlib/Coq.Init.Specif}{\coqdocconstructor{)}} \coqexternalref{exist}{http://coq.inria.fr/distrib/8.5beta2/stdlib/Coq.Init.Specif}{\coqdocconstructor{P1}} \ensuremath{\Rightarrow} \coqref{Fpred.Normalization.hsubst comp proj}{\coqdocdefinition{hsubst}} \coqexternalref{:core scope:'(' x ',' x ',' '..' ',' x ')'}{http://coq.inria.fr/distrib/8.5beta2/stdlib/Coq.Init.Datatypes}{\coqdocnotation{(}}\coqdocvar{T'}\coqexternalref{:core scope:'(' x ',' x ',' '..' ',' x ')'}{http://coq.inria.fr/distrib/8.5beta2/stdlib/Coq.Init.Datatypes}{\coqdocnotation{,}} \coqdocvar{t'}\coqexternalref{:core scope:'(' x ',' x ',' '..' ',' x ')'}{http://coq.inria.fr/distrib/8.5beta2/stdlib/Coq.Init.Datatypes}{\coqdocnotation{)}} \coqdocvar{r2} 0 \coqdocvar{\_};\coqdoceol
\coqdocindent{3.00em}
\ensuremath{|} \coqexternalref{exist}{http://coq.inria.fr/distrib/8.5beta2/stdlib/Coq.Init.Specif}{\coqdocconstructor{exist}} \coqexternalref{exist}{http://coq.inria.fr/distrib/8.5beta2/stdlib/Coq.Init.Specif}{\coqdocconstructor{r1}} \coqexternalref{exist}{http://coq.inria.fr/distrib/8.5beta2/stdlib/Coq.Init.Specif}{\coqdocconstructor{P1}} \ensuremath{\Rightarrow} \coqdocconstructor{app} \coqdocvar{r1} \coqdocvar{r2} \} \} \}.\coqdoceol
\coqdocemptyline
\coqdocemptyline
\coqdocemptyline
\end{coqdoccode}
With \coqref{Fpred.Normalization.hsubst}{\coqdocdefinition{hsubst}} defined, it is now easy to implement a
    \coqref{Fpred.Normalization.normalize}{\coqdocdefinition{normalize}} function which takes a term and returns its normal form.
    As for \coqref{Fpred.Normalization.hsubst}{\coqdocdefinition{hsubst}}, we add a precondition and a postcondition
    which allow to show correctness by construction. \begin{coqdoccode}
\coqdocemptyline
\coqdocnoindent
\coqdockw{Definition} \coqdef{Fpred.Normalization.pre'}{pre'}{\coqdocdefinition{pre'}} (\coqdocvar{t} : \coqdocinductive{term}) (\coqdocvar{p} : \coqdocinductive{env} \coqexternalref{:type scope:x '*' x}{http://coq.inria.fr/distrib/8.5beta2/stdlib/Coq.Init.Datatypes}{\coqdocnotation{\ensuremath{\times}}} \coqdocinductive{typ}) : \coqdockw{Prop} :=\coqdoceol
\coqdocindent{1.00em}
\coqdocinductive{typing} (\coqexternalref{fst}{http://coq.inria.fr/distrib/8.5beta2/stdlib/Coq.Init.Datatypes}{\coqdocdefinition{fst}} \coqdocvariable{p}) \coqdocvariable{t} (\coqexternalref{snd}{http://coq.inria.fr/distrib/8.5beta2/stdlib/Coq.Init.Datatypes}{\coqdocdefinition{snd}} \coqdocvariable{p}).\coqdoceol
\coqdocnoindent
\coqdockw{Definition} \coqdef{Fpred.Normalization.post'}{post'}{\coqdocdefinition{post'}} (\coqdocvar{t} : \coqdocinductive{term}) (\coqdocvar{n} : \coqdocinductive{term}) (\coqdocvar{p} : \coqdocinductive{env} \coqexternalref{:type scope:x '*' x}{http://coq.inria.fr/distrib/8.5beta2/stdlib/Coq.Init.Datatypes}{\coqdocnotation{\ensuremath{\times}}} \coqdocinductive{typ}) : \coqdockw{Prop} :=\coqdoceol
\coqdocindent{1.00em}
\coqref{Fpred.Normalization.interp}{\coqdocdefinition{interp}} (\coqexternalref{fst}{http://coq.inria.fr/distrib/8.5beta2/stdlib/Coq.Init.Datatypes}{\coqdocdefinition{fst}} \coqdocvariable{p}) \coqdocvariable{t} (\coqexternalref{snd}{http://coq.inria.fr/distrib/8.5beta2/stdlib/Coq.Init.Datatypes}{\coqdocdefinition{snd}} \coqdocvariable{p}) \coqdocvariable{n}.\coqdoceol
\coqdocemptyline
\coqdocnoindent
\coqdockw{Equations}(\coqdocvar{noind}) \coqdef{Fpred.Normalization.normalize}{normalize}{\coqdocdefinition{normalize}} (\coqdocvar{t} : \coqdocinductive{term}) (\coqdocvar{P} : \coqexternalref{:type scope:'exists' x '..' x ',' x}{http://coq.inria.fr/distrib/8.5beta2/stdlib/Coq.Init.Logic}{\coqdocnotation{\ensuremath{\exists}}} \coqexternalref{:type scope:'exists' x '..' x ',' x}{http://coq.inria.fr/distrib/8.5beta2/stdlib/Coq.Init.Logic}{\coqdocnotation{(}}\coqdocvar{p} : \coqdocinductive{env} \coqexternalref{:type scope:x '*' x}{http://coq.inria.fr/distrib/8.5beta2/stdlib/Coq.Init.Datatypes}{\coqdocnotation{\ensuremath{\times}}} \coqdocinductive{typ}\coqexternalref{:type scope:'exists' x '..' x ',' x}{http://coq.inria.fr/distrib/8.5beta2/stdlib/Coq.Init.Logic}{\coqdocnotation{),}} \coqref{Fpred.Normalization.pre'}{\coqdocdefinition{pre'}} \coqdocvariable{t} \coqdocvariable{p}) :\coqdoceol
\coqdocindent{1.00em}
\coqexternalref{:type scope:'x7B' x ':' x '|' x 'x7D'}{http://coq.inria.fr/distrib/8.5beta2/stdlib/Coq.Init.Specif}{\coqdocnotation{\{}}\coqdocvar{n} \coqexternalref{:type scope:'x7B' x ':' x '|' x 'x7D'}{http://coq.inria.fr/distrib/8.5beta2/stdlib/Coq.Init.Specif}{\coqdocnotation{:}} \coqdocinductive{term} \coqexternalref{:type scope:'x7B' x ':' x '|' x 'x7D'}{http://coq.inria.fr/distrib/8.5beta2/stdlib/Coq.Init.Specif}{\coqdocnotation{\ensuremath{|}}} \coqdockw{\ensuremath{\forall}} (\coqdocvar{p} : \coqdocinductive{env} \coqexternalref{:type scope:x '*' x}{http://coq.inria.fr/distrib/8.5beta2/stdlib/Coq.Init.Datatypes}{\coqdocnotation{\ensuremath{\times}}} \coqdocinductive{typ}), \coqref{Fpred.Normalization.pre'}{\coqdocdefinition{pre'}} \coqdocvar{t} \coqdocvariable{p} \coqexternalref{:type scope:x '->' x}{http://coq.inria.fr/distrib/8.5beta2/stdlib/Coq.Init.Logic}{\coqdocnotation{\ensuremath{\rightarrow}}} \coqref{Fpred.Normalization.post'}{\coqdocdefinition{post'}} \coqdocvar{t} \coqdocvar{n} \coqdocvariable{p}\coqexternalref{:type scope:'x7B' x ':' x '|' x 'x7D'}{http://coq.inria.fr/distrib/8.5beta2/stdlib/Coq.Init.Specif}{\coqdocnotation{\}}} :=\coqdoceol
\coqdocnoindent
\coqref{Fpred.Normalization.normalize}{\coqdocdefinition{normalize}} (\coqdocconstructor{var} \coqdocconstructor{i}) \coqdocvar{P} \ensuremath{\Rightarrow} \coqdocconstructor{var} \coqdocvar{i};\coqdoceol
\coqdocnoindent
\coqref{Fpred.Normalization.normalize}{\coqdocdefinition{normalize}} (\coqdocconstructor{abs} \coqdocconstructor{T1} \coqdocconstructor{t}) \coqdocvar{P} \ensuremath{\Rightarrow} \coqdocconstructor{abs} \coqdocvar{T1} (\coqref{Fpred.Normalization.normalize}{\coqdocdefinition{normalize}} \coqdocvar{t} \coqdocvar{\_}) ;\coqdoceol
\coqdocnoindent
\coqref{Fpred.Normalization.normalize}{\coqdocdefinition{normalize}} (\coqdocconstructor{app} \coqdocconstructor{t1} \coqdocconstructor{t2}) \coqdocvar{P} $\Leftarrow$ \coqref{Fpred.Normalization.normalize}{\coqdocdefinition{normalize}} \coqdocvar{t2} \coqdocvar{\_} \ensuremath{\Rightarrow} \{\coqdoceol
\coqdocindent{1.00em}
\ensuremath{|} \coqexternalref{exist}{http://coq.inria.fr/distrib/8.5beta2/stdlib/Coq.Init.Specif}{\coqdocconstructor{exist}} \coqexternalref{exist}{http://coq.inria.fr/distrib/8.5beta2/stdlib/Coq.Init.Specif}{\coqdocconstructor{t2'}} \coqexternalref{exist}{http://coq.inria.fr/distrib/8.5beta2/stdlib/Coq.Init.Specif}{\coqdocconstructor{P2'}} $\Leftarrow$ \coqref{Fpred.Normalization.normalize}{\coqdocdefinition{normalize}} \coqdocvar{t1} \coqdocvar{\_} \ensuremath{\Rightarrow} \{\coqdoceol
\coqdocindent{2.00em}
\ensuremath{|} \coqexternalref{exist}{http://coq.inria.fr/distrib/8.5beta2/stdlib/Coq.Init.Specif}{\coqdocconstructor{exist}} \coqexternalref{exist}{http://coq.inria.fr/distrib/8.5beta2/stdlib/Coq.Init.Specif}{\coqdocconstructor{(}}\coqexternalref{exist}{http://coq.inria.fr/distrib/8.5beta2/stdlib/Coq.Init.Specif}{\coqdocconstructor{abs}} \coqexternalref{exist}{http://coq.inria.fr/distrib/8.5beta2/stdlib/Coq.Init.Specif}{\coqdocconstructor{T}} \coqexternalref{exist}{http://coq.inria.fr/distrib/8.5beta2/stdlib/Coq.Init.Specif}{\coqdocconstructor{t}}\coqexternalref{exist}{http://coq.inria.fr/distrib/8.5beta2/stdlib/Coq.Init.Specif}{\coqdocconstructor{)}} \coqexternalref{exist}{http://coq.inria.fr/distrib/8.5beta2/stdlib/Coq.Init.Specif}{\coqdocconstructor{P1'}} \ensuremath{\Rightarrow} \coqref{Fpred.Normalization.hsubst}{\coqdocdefinition{hsubst}} \coqexternalref{:core scope:'(' x ',' x ',' '..' ',' x ')'}{http://coq.inria.fr/distrib/8.5beta2/stdlib/Coq.Init.Datatypes}{\coqdocnotation{(}}\coqdocvar{T}\coqexternalref{:core scope:'(' x ',' x ',' '..' ',' x ')'}{http://coq.inria.fr/distrib/8.5beta2/stdlib/Coq.Init.Datatypes}{\coqdocnotation{,}} \coqdocvar{t}\coqexternalref{:core scope:'(' x ',' x ',' '..' ',' x ')'}{http://coq.inria.fr/distrib/8.5beta2/stdlib/Coq.Init.Datatypes}{\coqdocnotation{)}} \coqdocvar{t2'} 0 \coqdocvar{\_} ;\coqdoceol
\coqdocindent{2.00em}
\ensuremath{|} \coqexternalref{exist}{http://coq.inria.fr/distrib/8.5beta2/stdlib/Coq.Init.Specif}{\coqdocconstructor{exist}} \coqexternalref{exist}{http://coq.inria.fr/distrib/8.5beta2/stdlib/Coq.Init.Specif}{\coqdocconstructor{t1'}} \coqexternalref{exist}{http://coq.inria.fr/distrib/8.5beta2/stdlib/Coq.Init.Specif}{\coqdocconstructor{P1'}} \ensuremath{\Rightarrow} \coqdocconstructor{app} \coqdocvar{t1'} \coqdocvar{t2'} \} \};\coqdoceol
\coqdocnoindent
\coqref{Fpred.Normalization.normalize}{\coqdocdefinition{normalize}} (\coqdocconstructor{tabs} \coqdocconstructor{k} \coqdocconstructor{t}) \coqdocvar{P} \ensuremath{\Rightarrow} \coqdocconstructor{tabs} \coqdocvar{k} (\coqref{Fpred.Normalization.normalize}{\coqdocdefinition{normalize}} \coqdocvar{t} \coqdocvar{\_}) ;\coqdoceol
\coqdocnoindent
\coqref{Fpred.Normalization.normalize}{\coqdocdefinition{normalize}} (\coqdocconstructor{tapp} \coqdocconstructor{t} \coqdocconstructor{T}) \coqdocvar{P} $\Leftarrow$ \coqref{Fpred.Normalization.normalize}{\coqdocdefinition{normalize}} \coqdocvar{t} \coqdocvar{\_} \ensuremath{\Rightarrow} \{\coqdoceol
\coqdocindent{1.00em}
\ensuremath{|} \coqexternalref{exist}{http://coq.inria.fr/distrib/8.5beta2/stdlib/Coq.Init.Specif}{\coqdocconstructor{exist}} \coqexternalref{exist}{http://coq.inria.fr/distrib/8.5beta2/stdlib/Coq.Init.Specif}{\coqdocconstructor{(}}\coqexternalref{exist}{http://coq.inria.fr/distrib/8.5beta2/stdlib/Coq.Init.Specif}{\coqdocconstructor{tabs}} \coqexternalref{exist}{http://coq.inria.fr/distrib/8.5beta2/stdlib/Coq.Init.Specif}{\coqdocconstructor{k}} \coqexternalref{exist}{http://coq.inria.fr/distrib/8.5beta2/stdlib/Coq.Init.Specif}{\coqdocconstructor{t'}}\coqexternalref{exist}{http://coq.inria.fr/distrib/8.5beta2/stdlib/Coq.Init.Specif}{\coqdocconstructor{)}} \coqexternalref{exist}{http://coq.inria.fr/distrib/8.5beta2/stdlib/Coq.Init.Specif}{\coqdocconstructor{P'}} \ensuremath{\Rightarrow} \coqdocdefinition{subst\_typ} \coqdocvar{t'} 0 \coqdocvar{T} ;\coqdoceol
\coqdocindent{1.00em}
\ensuremath{|} \coqexternalref{exist}{http://coq.inria.fr/distrib/8.5beta2/stdlib/Coq.Init.Specif}{\coqdocconstructor{exist}} \coqexternalref{exist}{http://coq.inria.fr/distrib/8.5beta2/stdlib/Coq.Init.Specif}{\coqdocconstructor{t'}} \coqexternalref{exist}{http://coq.inria.fr/distrib/8.5beta2/stdlib/Coq.Init.Specif}{\coqdocconstructor{P'}} \ensuremath{\Rightarrow} \coqdocconstructor{tapp} \coqdocvar{t'} \coqdocvar{T} \}.\coqdoceol
\coqdocemptyline
\coqdocemptyline
\end{coqdoccode}
The existence of the \coqref{Fpred.Normalization.normalize}{\coqdocdefinition{normalize}} function is in itself a proof of the strong
    normalization of Leivant's Predicative System F. \begin{coqdoccode}
\coqdocemptyline
\coqdocnoindent
\coqdockw{Theorem} \coqdef{Fpred.Normalization.normalization}{normalization}{\coqdoclemma{normalization}} \coqdocvar{e} \coqdocvar{t} \coqdocvar{T} : \coqdocinductive{typing} \coqdocvariable{e} \coqdocvariable{t} \coqdocvariable{T} \coqexternalref{:type scope:x '->' x}{http://coq.inria.fr/distrib/8.5beta2/stdlib/Coq.Init.Logic}{\coqdocnotation{\ensuremath{\rightarrow}}} \coqexternalref{:type scope:'exists' x '..' x ',' x}{http://coq.inria.fr/distrib/8.5beta2/stdlib/Coq.Init.Logic}{\coqdocnotation{\ensuremath{\exists}}} \coqdocvar{n}\coqexternalref{:type scope:'exists' x '..' x ',' x}{http://coq.inria.fr/distrib/8.5beta2/stdlib/Coq.Init.Logic}{\coqdocnotation{,}} \coqdocdefinition{reds} \coqdocvariable{t} \coqdocvariable{n} \coqexternalref{:type scope:x '/x5C' x}{http://coq.inria.fr/distrib/8.5beta2/stdlib/Coq.Init.Logic}{\coqdocnotation{\ensuremath{\land}}} \coqdocinductive{typing} \coqdocvariable{e} \coqdocvariable{n} \coqdocvariable{T} \coqexternalref{:type scope:x '/x5C' x}{http://coq.inria.fr/distrib/8.5beta2/stdlib/Coq.Init.Logic}{\coqdocnotation{\ensuremath{\land}}} \coqref{Fpred.Normalization.normal}{\coqdocinductive{normal}} \coqdocvariable{n}.\coqdoceol
\coqdocemptyline
\coqdocemptyline
\end{coqdoccode}
\subsection{Consistency}

   It is easy to show consistency based on the normalization function.
   We just need lemmas showing that neutral terms cannot inhabit any
   type in an environment with only a type variable, by inversion on the
   neutrality derivation.  \begin{coqdoccode}
\coqdocemptyline
\coqdocnoindent
\coqdockw{Lemma} \coqdef{Fpred.Normalization.neutral tvar}{neutral\_tvar}{\coqdoclemma{neutral\_tvar}} \coqdocvar{t} \coqdocvar{k} \coqdocvar{T} : \coqref{Fpred.Normalization.neutral}{\coqdocinductive{neutral}} \coqdocvariable{t} \coqexternalref{:type scope:x '->' x}{http://coq.inria.fr/distrib/8.5beta2/stdlib/Coq.Init.Logic}{\coqdocnotation{\ensuremath{\rightarrow}}} \coqdocinductive{typing} (\coqdocconstructor{etvar} \coqref{Fpred.Normalization.empty env}{\coqdocdefinition{empty\_env}} \coqdocvariable{k}) \coqdocvariable{t} \coqdocvariable{T} \coqexternalref{:type scope:x '->' x}{http://coq.inria.fr/distrib/8.5beta2/stdlib/Coq.Init.Logic}{\coqdocnotation{\ensuremath{\rightarrow}}} \coqexternalref{False}{http://coq.inria.fr/distrib/8.5beta2/stdlib/Coq.Init.Logic}{\coqdocinductive{False}}.\coqdoceol
\coqdocemptyline
\coqdocemptyline
\end{coqdoccode}
Consistency is then proved using case analysis on an assumed typing
 derivation of falsehood at any universe level \coqdocvariable{k}.  Informally, it is
 showing that ∀ \coqdocvariable{X} :* \coqdocvariable{k}, \coqdocvariable{X} is not inhabited for any \coqdocvariable{k}. \begin{coqdoccode}
\coqdocemptyline
\coqdocnoindent
\coqdockw{Corollary} \coqdef{Fpred.Normalization.consistency}{consistency}{\coqdoclemma{consistency}} \coqdocvar{k} : \coqexternalref{:type scope:'x7E' x}{http://coq.inria.fr/distrib/8.5beta2/stdlib/Coq.Init.Logic}{\coqdocnotation{\ensuremath{\lnot}}} \coqexternalref{:type scope:'exists' x '..' x ',' x}{http://coq.inria.fr/distrib/8.5beta2/stdlib/Coq.Init.Logic}{\coqdocnotation{\ensuremath{\exists}}} \coqdocvar{t}\coqexternalref{:type scope:'exists' x '..' x ',' x}{http://coq.inria.fr/distrib/8.5beta2/stdlib/Coq.Init.Logic}{\coqdocnotation{,}} \coqdocinductive{typing} \coqref{Fpred.Normalization.empty env}{\coqdocdefinition{empty\_env}} \coqdocvariable{t} (\coqdocconstructor{all} \coqdocvariable{k} (\coqdocconstructor{tvar} 0)).\coqdoceol
\coqdocemptyline
\end{coqdoccode}

\section{Related Work and Conclusion}
\label{sec:relat-work-concl}

There are many formalizations of similar calculi, and we do not claim
any originality there. However, to our knowledge, the multiset ordering
used to show normalization is original. The point of this paper is more
to show that the \Equations plugin is ready to handle more consequent
developments and showcase its features. Is has similar expressivity
w.r.t. Agda and Idris, but derives more principles, and everything is
compiled down to vanilla \Coq terms, so it does not change the trusted
code base except for the use of K, which we are hopeful we can get rid of
by the time of the workshop.

It would be interesting to study extensions of the language with type
recursion. As shown by Leivant, this would allow to type terms that are
not typeable in second-order lambda calculus. We will also need to
extend the language with existentials, pairs and a minimal notion of
inductive types to be able to handle a larger class of programs. One of
the possible venues for generalization is to extend the work of Malecha
et al \cite{DBLP:conf/itp/MalechaCB14} to reflect a larger fragment of
\textsc{Gallina}, the language of \Coq.

\bibliographystyle{eptcs}
\bibliography{fpred}

\newpage

\appendix
\section{Extracted code}
\begin{lstlisting}[language=Caml,firstline=155,lastline=199]
let hereditary_subst t u x =
  let rec fix_F x0 =
    let h = fun y -> fix_F y in
    (fun u0 x1 _ ->
    let Pair (t0, t1) = x0 in
    (match t1 with
     | Var n ->
       (match lt_eq_lt_dec n x1 with
        | Inleft s ->
          (match s with
           | Left -> Var n
           | Right -> u0)
        | Inright -> Var (pred n))
     | Abs (t2, refine) ->
       Abs (t2, (h (Pair (t0, refine)) (shift O u0) (S x1) __))
     | App (refine1, refine2) ->
       (match h (Pair (t0, refine1)) u0 x1 __ with
        | Abs (t2, x2) ->
          h (Pair (t2, x2)) (h (Pair (t0, refine2)) u0 x1 __) O __
        | x2 -> App (x2, (h (Pair (t0, refine2)) u0 x1 __)))
     | Tabs (k, refine) ->
       Tabs (k, (h (Pair ((tshift O t0), refine)) (shift_typ O u0) x1 __))
     | Tapp (refine, t2) ->
       (match h (Pair (t0, refine)) u0 x1 __ with
        | Tabs (k, x2) -> subst_typ x2 O t2
        | x2 -> Tapp (x2, t2))))
  in fix_F t u x __

type normalize_comp = term

(** val normalize : term -> normalize_comp **)

let rec normalize = function
| Var n -> Var n
| Abs (t0, t1) -> Abs (t0, (normalize t1))
| App (t1, t2) ->
  (match normalize t1 with
   | Abs (t0, x0) -> hereditary_subst (Pair (t0, x0)) (normalize t2) O
   | x -> App (x, (normalize t2)))
| Tabs (k, t0) -> Tabs (k, (normalize t0))
| Tapp (t0, t1) ->
  (match normalize t0 with
   | Tabs (k, x) -> subst_typ x O t1
   | x -> Tapp (x, t1))
\end{lstlisting}

\end{document}